\documentclass[aps,prd,showpacs,eqsecnum,twocolumn,superscriptaddress,a4paper,nofootinbib]
{revtex4-1}
\usepackage{amsmath,amssymb,graphicx,color}
\usepackage{bm}
\usepackage{epsf}

\usepackage{ulem}

\begin{document}

\title{Viscous evolution of a massive disk surrounding stellar-mass black holes in full general relativity}

\author{Sho Fujibayashi}
\affiliation{Max Planck Institute for
  Gravitational Physics (Albert Einstein Institute), Am M\"uhlenberg 1,
  Potsdam-Golm 14476, Germany}

\author{Masaru Shibata} 
\affiliation{Max Planck Institute for
  Gravitational Physics (Albert Einstein Institute), Am M\"uhlenberg 1,
  Potsdam-Golm 14476, Germany}
\affiliation{Center for Gravitational Physics, Yukawa Institute for Theoretical
  Physics, Kyoto University, Kyoto, 606-8502, Japan}

\author{Shinya Wanajo} 
\affiliation{Max Planck Institute for
  Gravitational Physics (Albert Einstein Institute), Am M\"uhlenberg 1,
  Potsdam-Golm 14476, Germany}
  \affiliation{Interdisciplinary Theoretical and Mathematical Sciences Program (iTHEMS), RIKEN,
Wako, Saitama 351-0198, Japan}

\author{Kenta Kiuchi}
\affiliation{Max Planck Institute for
  Gravitational Physics (Albert Einstein Institute), Am M\"uhlenberg 1,
  Potsdam-Golm 14476, Germany}
\affiliation{Center for Gravitational Physics, Yukawa Institute for Theoretical
  Physics, Kyoto University, Kyoto, 606-8502, Japan}

\author{Koutarou Kyutoku} 
\affiliation{Department of Physics, Kyoto University, Kyoto 606-8502, Japan}
\affiliation{Center for Gravitational Physics, Yukawa Institute for Theoretical Physics, Kyoto University, Kyoto, 606-8502, Japan}
\affiliation{Interdisciplinary Theoretical and Mathematical Sciences Program (iTHEMS), RIKEN, Wako, Saitama 351-0198, Japan}

\author{Yuichiro Sekiguchi} \affiliation{Department of Physics, Toho
  University, Funabashi, Chiba 274-8510, Japan}

\date{\today}
\newcommand{\beq}{\begin{equation}}
\newcommand{\eeq}{\end{equation}}
\newcommand{\beqn}{\begin{eqnarray}}
\newcommand{\eeqn}{\end{eqnarray}}
\newcommand{\pa}{\partial}
\newcommand{\vp}{\varphi}
\newcommand{\varep}{\varepsilon}
\newcommand{\ep}{\epsilon}
\newcommand{\comp}{(M/R)_\infty}
\begin{abstract}

Long-term viscous neutrino-radiation hydrodynamics simulations in full
general relativity are performed for a massive disk surrounding
spinning stellar-mass black holes with mass $M_{\rm BH}=4$, 6, and
$10M_\odot$ and initial dimensionless spin $\chi \approx 0.8$.  The
initial disk is chosen to have mass $M_{\rm disk}\approx 0.1$ or
$3M_\odot$ as plausible models of the remnants for the merger of black hole-neutron star binaries or the stellar core collapse from a rapidly rotating progenitor, respectively.
For $M_{\rm disk} \approx 0.1M_\odot$ with the outer disk
edge initially located at $r_{\rm out} \sim 200$\,km, we find that
$15$\%--20\% of $M_{\rm disk}$ is ejected and the average electron
fraction of the ejecta is $\langle Y_e \rangle = 0.30$--0.35 as
found in the previous study.
For $M_{\rm disk} \approx 3M_\odot$, we find that $\approx 10$\%--20\% of $M_{\rm
  disk}$ is ejected for $r_{\rm out}\approx 200$--1000\,km.  In
addition, $\langle Y_e \rangle$ of the ejecta can be enhanced to be
$\agt 0.4$ because the electron fraction is increased significantly
during the long-term viscous expansion of the disk with high neutrino
luminosity until the mass ejection sets in. Our results suggest that
not heavy $r$-process elements but light trans-iron elements would be
synthesized in the matter ejected from a massive torus surrounding
stellar-mass black holes.  We also find that the outcomes of the
viscous evolution for the high-mass disk case is composed of a rapidly
spinning black hole surrounded by a torus with a narrow funnel, which
appears to be suitable for generating gamma-ray bursts.

\end{abstract}


\maketitle

\section{Introduction}\label{sec1}

A stellar-mass spinning black hole surrounded by a massive disk is
believed to be a frequent remnant for the merger of neutron star
binaries (binary neutron stars and black hole-neutron star binaries),
and for the stellar core collapse of massive and rapidly rotating
progenitors. Such remnants have been speculated to be the central
engines of gamma-ray
bursts~\cite{Eichler,Woosley93,Meszaros2004,Piran2005} and
kilonovae~\cite{kilonova,kilonova1,Kasen2013,TH2013}.  This fact
motivates the community to explore in detail the formation process of
the black hole and disk surrounding it (see, e.g.,
Refs.~\cite{SS2011,Shibata16} for a review of numerical simulations
for the formation of black holes surrounded by disks) and subsequent
long-term evolution of the disk by
magnetohydrodynamics/viscous-hydrodynamics processes with various
sophisticated levels (e.g.,
Refs.~\cite{Seti2004,LRP2005,SST2007,MF2013,MF2014,MF2015,Just2015,SM17,FTQFK18,Janiuk19,FTQFK19,Miller19,Fujiba20,FFL20}).
In our previous paper~\cite{Fujiba20}, we performed a long-term
viscous radiation hydrodynamics simulation for the system of a spinning black hole
with low mass $M_{\rm BH}=3M_\odot$ and of a disk with mass $M_{\rm
  disk}=0.03$--$0.5M_\odot$ in full general relativity, paying
particular attention to the mass ejection in the post-merger evolution
of neutron-star binaries.

In this paper, we extend our previous work~\cite{Fujiba20} and explore the dependence of the viscous evolution process of the disk and resulting mass ejection on the black-hole mass, $M_{\rm BH}$, and disk mass, $M_{\rm disk}$.
We consider the systems of $M_{\rm BH}=4$, 6, $10M_\odot$ and of $M_{\rm disk} \approx 0.1$ and $3M_\odot$ with the initial dimensionless spin of the black hole $\chi \approx 0.8$.
These are plausible remnants for the merger of black hole-neutron star binaries~\cite{Kyutoku2015,Kyutoku2011,Foucart2012,Foucart2013,Foucart2014}
or the stellar core collapse from a rapidly rotating progenitor to a black hole~\cite{Woosley1993,MacFadyen1999a}.
The purpose of this study is to explore the quantitative dependence of the disk evolution and subsequent mass
ejection as well as nucleosynthesis on the mass of the black hole and disk. In particular, we pay
attention to the evolution of the high-mass disk model in this paper. 

The system of a spinning black hole and a high-mass disk can be a
plausible model as remnants for the collapse of rapidly rotating
progenitor stars to a black hole.  The evolution of high-mass disks
surrounding a black hole with relevant microphysics (such as realistic
equation of state and neutrino cooling) has not been studied
self-consistently in previous work because not only general
relativistic gravity for the black hole but also the self-gravity of
the disk have to be taken into account. In the present study, we
employ the framework of full general relativity, and thus, we can
explore the high-mass disk model with no assumption (except for the
assumption of axial symmetry).  We will show that in the presence of
a high-mass disk, the neutrino luminosity is enhanced, and as a
result, the viscous evolution timescale of the disk is increased
significantly. This effect modifies the property of the ejected
matter and resulting elements produced in the nucleosynthesis. In
addition, we show that the outcome of the evolution of the system is 
a rapidly spinning black hole with a
geometrically-thick torus and a narrow funnel. This system appears to
be suitable for generating gamma-ray bursts, i.e., for driving a
collimated jet toward the direction of the rotation axis.

The paper is organized as follows. In Sec.~\ref{sec2}, we briefly
summarize the basic equations employed in the present numerical
simulation and initial conditions employed in this work.
Section~\ref{sec3} presents numerical results for the simulations.
We present the evolution process of the disk and black hole, the
properties of the ejecta and nucleosynthesis in the matter ejected
from the disks, and the final outcome of the disk evolution. 
Section~\ref{sec4} is devoted to a summary.  Throughout this paper,
$G$, $c$, and $k$ denote the gravitational constant, speed of light,
and Boltzmann's constant, respectively. 

\section{Summary for the method of numerical computations}\label{sec2}

We perform viscous neutrino-radiation hydrodynamics simulations for
systems of a stellar-mass black hole and a massive disk in the
framework of full general relativity with the assumption of axial
symmetry using the same method and implementation as in
Ref.~\cite{Fujiba20}: We numerically solve Einstein's equation, the
viscous-hydrodynamics equations, the evolution equation for the
viscous tensor, the evolution equations for the lepton fractions
including the electron fraction, and neutrino-radiation transfer
equations.  Einstein's equation is solved using the original version
of the Baumgarte-Shapiro-Shibata-Nakamura formalism~\cite{BSSN}
together with the puncture formulation~\cite{puncture}, Z4c constraint
propagation prescription~\cite{Z4c}, and 5th-order Kreiss-Oliger
dissipation.  The axial symmetry for the geometric variables is
imposed using a cartoon method with the 4th-order accuracy in
space~\cite{cartoon,cartoon2}.  The viscous-hydrodynamics equations and
evolution equations for the viscous tensor are solved by the method
described in Ref.~\cite{SKS17}.  For evolving the lepton fractions, we
take into account electron and positron capture, electron-positron
pair annihilation, nucleon-nucleon bremsstrahlung, and plasmon
decay~\cite{Fujiba2018,Fujiba2019}.  The quantities of black holes
(mass and spin) are determined from their area and circumferential
radii of apparent horizons~\cite{K10}, assuming that these quantities are written as
functions of the mass and spin in the same formulation as in the
vacuum case.

As in our previous work~\cite{Fujiba20}, we employ a tabulated
equation of state based on the DD2 equation of state~\cite{DD2} for a
relatively high-density part and the Timmes equation of state for the
low-density part~\cite{Timmes}.  In this tabulated equation of state,
thermodynamics quantities such as $\varep$, $P$, and $h$ are written
as functions of $\rho$, $Y_e$, and $T$ where $\varep$, $P$,
$h(=c^2+\varep+P/\rho)$, $\rho$, $Y_e$ and $T$ are the specific
internal energy, pressure, specific enthalpy, rest-mass density,
electron fraction, and matter temperature, respectively.  We choose
the lowest rest-mass density to be $0.1\,{\rm g/cm^3}$ in the table,
and the atmosphere density for $\rho_*:=\rho u^t \sqrt{-g}$ in the
hydrodynamics simulation is chosen to be $10\,{\rm g/cm^3}$ in the
central region which is smoothly decreased to $1\,{\rm g/cm^3}$ in the
outer region. Here $u^\mu$ and $g$ denote the four velocity and the
determinant of the spacetime metric, respectively.


For viscous hydrodynamics, we need to input the viscous coefficient $\nu$.
Following our previous work~\cite{Fujiba20}, we set $\nu=\alpha_\nu h c_s H/c^2$ where $\alpha_\nu$ is the dimensionless viscous coefficient (the so-called alpha parameter~\cite{SS73}), $c_s$ is the sound velocity, and $H$ is a scale height.
We basically employ $\alpha_\nu=0.05$ as in our previous work~\cite{Fujiba20} supposing that the origin of the effective viscosity is the turbulence induced hypothetically by the magneto-rotational instability~\cite{BH98,Hawley11,Kiuchi18}.
For comparison, we also employ $\alpha_\nu=0.10$ and 0.15 for a model with $M_{\rm BH}=10M_\odot$ and $M_{\rm disk} \approx 3M_\odot$.
We set $H$ to a constant value that is approximately equal to the radius at the innermost stable circular orbit around the Kerr black hole of $\chi=0.8$, i.e., $H=30\,{\rm km}(M_{\rm BH}/10M_\odot) \approx 2GM_{\rm BH}/c^2$. This setting is also the same as in Ref.~\cite{Fujiba20}.
Note that $H$ is a constant determined by the initial state of the black hole.

The viscous timescale (for heating and angular momentum transport) is written approximately as 
\beqn
\tau_{\rm vis}:={R^2 \over \nu} 
&\approx& 1.0\,{\rm s}(hc^{-2})^{-1}\left({\alpha_\nu \over 0.05}\right)^{-1}
\left({c_s \over 0.05c}\right)^{-1}
\nonumber \\
&&\times 
\left({H \over 30\,{\rm km}}\right)^{-1}
\left({R \over 150\,{\rm km}}\right)^{2},\label{tvis}
\eeqn
where $R$ denotes the cylindrical radius in the disk and the reference values are chosen for $M_{\rm BH}=10M_\odot$. As we show in Sec.~\ref{sec3}, the evolution timescale for our choice of the viscous coefficient is indeed of the order of $\sim 1$\,s, which is much longer than the dynamical timescale (i.e., rotational period) of the disk approximately written as
\beqn
\tau_{\rm dyn}&:=&2\pi\sqrt{{R^{3} \over GM_{\rm BH}}} \nonumber \\
& \approx &10\,{\rm ms} \left({R \over 150\,{\rm km}}\right)^{3/2}
\left({M \over 10M_\odot}\right)^{-1/2}.\label{tdyn}
\eeqn
Thus, in the viscous hydrodynamics of the disks, the evolution should proceed in a quasi-steady manner.


Axisymmetric equilibrium states for black hole-disk systems are
prepared as the initial condition~\cite{Fujiba20} using the method 
shown in Ref.~\cite{Shibata2007}. As in
the previous study, we determine the angular velocity from the
relation
\beq
j \propto \Omega^{-n}, \label{eqj}
\eeq
where $j=c^{-1}h u_\varphi$ is the specific angular momentum.
$\Omega$ is the angular velocity defined by $u^\varphi/u^t$ and $n$ is
a constant that determines the profile of the angular velocity  (see
Ref.~\cite{Poland} for more careful choice for deriving disks with
``Keplerian profile'').  In this paper, we fiducially choose $n=1/7$
to align the value with that chosen in Ref.~\cite{Fujiba20}. For two
models for which the disk has a geometrically thin and less compact
structure, we employ $n=1/5$ and $1/4$ (cf. Table~\ref{table1}). 
These models are used to explore the dependence of the results on the
initial disk compactness (cf. Secs.~\ref{sec3-3} and \ref{sec3-5}).

For obtaining the initial condition, we assume a relation between
$\rho$ and $Y_e$ in the same form of $\rho(Y_e)$ as in
Ref.~\cite{Fujiba20}.  For this model, the value of $Y_e$ is 0.07 in a
high density region and approaches $0.5$ for low density for which
the effect of the electron degeneracy is weak.  In addition, we assume
that the specific entropy, $s$, is initially constant (in order to
obtain the first integral of the Euler equation easily).  We always
choose $s=6k$ in this paper for simplicity.

The initial black-hole mass is chosen to be $M_{\rm BH,0}=4$, 6, and
$10M_\odot$ with the disk mass being $M_{\rm disk}\approx 0.1$ and
$3M_\odot$. For most of the cases, we set the inner edge of the disk
to be $r_{\rm in}=2GM_{\rm BH,0}/c^2$.  For this case, the outer edge of
the disk is located at $r_{\rm out} \sim 200$--250\,km.  For two models
we choose less compact disks for which $r_{\rm out} \sim 500$ and 1000\,km,
$M_{\rm BH,0} \approx 10M_\odot$, and $M_{\rm disk} \approx 3M_\odot$ with
$s/k=6$, $n=1/5$ or 1/4 and $r_{\rm in}=3.5GM_{\rm BH,0}/c^2$ (see Table~\ref{table1}).

In our formalism for obtaining the equilibrium state composed of a
black hole and a disk~\cite{Shibata2007}, we initially prepare a
black-hole geometry as a seed. In this work, we set the dimensionless
spin of the seed black hole to be $\chi=0.8$. This implies that only
in the limit of $M_{\rm disk}\rightarrow 0$, $\chi$ becomes 0.8.  As
already mentioned, the black-hole spin is measured from the area,
$A_{\rm AH}$, and circumferential radii around the equatorial and
meridian planes, $c_e$ and $c_p$, for the black-hole
horizon~\cite{Shibata2007}, assuming that $A_{\rm AH}$, $c_e$, and
$c_p$ are written as functions of the mass, $M_{\rm BH}$, and
dimensionless spin, $\chi$, of Kerr black holes as in the vacuum
black-hole case.  In the presence of matter outside the black hole,
its geometry is modified, and for the high-mass disk case with $M_{\rm
  disk} \approx 3M_\odot$, the dimensionless spin becomes slightly
smaller than 0.8 (see Table~\ref{table1}).


\begin{table*}[t]
\caption{Initial equilibrium models for the numerical simulation. Described
  are the model name, black-hole mass, rest mass of the disk,
  dimensionless spin of the black hole, the coordinate radii at the
  inner and outer edges of the disk ($r_{\rm in}$ and $r_{\rm out}$),
  entropy per baryon ($s/k$), the value of $n$ (that determines the
  rotational profile), and electron fraction ($Y_e$) for the disk, the
  viscous coefficient ($\alpha_{\nu}$), the scale height ($H$), and
  the location of outer boundaries along each axis in units of
  $10^3$\,km ($L$).  The units of the mass are $M_\odot$ and the units
  of $r_{\rm in}$ and $r_{\rm out}$ are $GM_{\rm BH}/c^2 \approx
  14.77(M_{\rm BH}/10M_\odot)$\,km. The last column shows whether the
  neutrino absorption/irradiation is switched on or off. Note that
  model K8 was used in Ref.~\cite{Fujiba20} as a fiducial model.
}
\begin{tabular}{ccccccccccccc} \hline
~Model~ & ~$M_{\rm BH}$~ & ~$M_{\rm disk}$~ & ~~$\chi$~~ 
& ~~$r_{\rm in}$~~ & ~~$r_{\rm out}$~~ & ~$s/k$~
& ~~~~$n$~~~~ & ~~$Y_e$~~& ~~$\alpha_\nu$~~ & ~~$H$\,(km)~~ & $L$\,($10^3$\,km) & neutrino abs\\
 \hline \hline
M04L05 &  4.0 & 0.10 & 0.80 & 2.0 & 30 & 6 & 1/7  & 0.07--0.5 & 0.05 & 12 & 6.56  &yes\\
M04H05 &  4.0 & 3.00 & 0.77 & 2.0 & 32 & 6 & 1/7  & 0.07--0.5 & 0.05 & 12 & 6.56  &yes\\
M06L05 &  6.0 & 0.10 & 0.80 & 2.0 & 22 & 6 & 1/7  & 0.07--0.5 & 0.05 & 18 & 8.39  &yes\\
M06H05 &  6.0 & 3.01 & 0.78 & 2.0 & 24 & 6 & 1/7  & 0.07--0.5 & 0.05 & 18 & 8.39  &yes \\
M10L05 & 10.0 & 0.10 & 0.80 & 2.0 & 16 & 6 & 1/7  & 0.07--0.5 & 0.05 & 30 & 11.01 &yes\\
M10H05 & 10.0 & 3.03 & 0.78 & 2.0 & 17 & 6 & 1/7  & 0.07--0.5 & 0.05 & 30 & 11.01 &yes\\
M10H10 & 10.0 & 3.03 & 0.78 & 2.0 & 17 & 6 & 1/7  & 0.07--0.5 & 0.10 & 30 & 11.01 &yes\\
M10H15 & 10.0 & 3.03 & 0.78 & 2.0 & 17 & 6 & 1/7  & 0.07--0.5 & 0.15 & 30 & 11.01 &yes\\
M10H05n& 10.0 & 3.03 & 0.78 & 2.0 & 17 & 6 & 1/7  & 0.07--0.5 & 0.05 & 30 & 11.01 &no\\
M10H05w& 10.0 & 3.00 & 0.79 & 3.5 & 35 & 6 & 1/5  & 0.07--0.5 & 0.05 & 30 & 11.01 &yes\\
M10H05x& 10.0 & 3.01 & 0.79 & 3.5 & 70 & 6 & 1/4  & 0.07--0.5 & 0.05 & 30 & 11.01 &yes\\\hline
K8     &  3.0 & 0.10 & 0.80 & 2.0 & 41 & 6 & 1/7  & 0.07--0.5 & 0.05 & 9  & 6.27  &yes\\
 \hline
 \end{tabular}
 \label{table1}
\end{table*}


Because the mass ratio, $M_{\rm disk}/M_{\rm BH,0}$, of the initial
conditions employed for the high-mass disk case is very large (3/10--3/4),
the self-gravity of the disk can play an important role for its evolution.  It is
well-known that the self-gravitating disk is subject to
non-axisymmetric deformation and fragmentation (e.g.,
Refs.~\cite{Hachisu90,Hawley91,Zink07,Oleg11,Kiuchi11}).  In our
axisymmetric simulation, this effect cannot be taken into account.
For the high-mass disk, the so-called $m=1$ mode is often non-linearly
excited, and as a result, a significant fraction of the initial disk
mass is likely to be swallowed by the black hole due to the angular
momentum transport in the dynamical timescale of the system.  The
remnant after the non-axisymmetric instability proceeds will be a
black hole surrounded by a disk that satisfies $M_{\rm BH} \gg M_{\rm
  disk}$.  As we show in Sec.~III, in the viscous hydrodynamics, a
significant fraction of the disk is also swallowed by the black hole
after the viscous angular momentum transport sets in, and the resulting remnant
is a black hole and a disk with $M_{\rm BH} \gg M_{\rm disk}$ as
well. Thus, if we focus on the process only after a large fraction of
the disk is swallowed by the black hole, the viscous hydrodynamics
approach is acceptable (that is, the effect of the non-axisymmetric
deformation is substituted by the viscous effect).  However, we should
keep in mind that we need a non-axisymmetric simulation to clarify the
evolution of the high-mass disk system quantitatively.
For the non-axisymmetric simulation, high-amplitude gravitational waves are likely to be emitted during the non-axisymemtric instability proceeds.
Thus, exploring the feature of gravitational waves will be also an interesting topic in the non-axisymmetric simulations.

\section{Numerical Results}\label{sec3}

\subsection{Models, setting, and diagnostics}\label{sec3-1}

Numerical computations are performed for the black hole-disk systems
summarized in the previous section (see also Table~\ref{table1}).  For
most models in this paper, the simulations are performed taking into
account the neutrino absorption/irradiation effect. For one model,
M10H05n, we switch off this neutrino effect to examine the importance of the
neutrino absorption/irradiation effect for the
high-mass disk case. In the present work, we do not incorporate a
heating effect by the neutrino pair annihilation~\cite{Fujiba2018} for
simplicity (thus the mass and kinetic energy of the ejecta might be
underestimated). 

The viscous tensor is set to be zero initially; we prepare an
equilibrium state assuming the ideal fluid with no viscosity. Because the viscous
tensor changes to a non-zero profile during the very early evolution,
the disk profile is modified for all the cases. For the high-mass disk
models, this (artificial) modification is significant in particular for
the relatively low-mass black-hole models.  Specifically, the disk radially
oscillates with a high amplitude for $t \alt 0.2$\,s for such cases, but subsequently, it
relaxes to a quasi-steady state, which is evolved by a long-term viscous
process. Thus in this paper, we pay particular attention to this later phase. 

Following our previous work~\cite{Fujiba20}, we employ a nonuniform grid for the two dimensional coordinates $(x, z)$ in the simulation:
For $x \leq x_\mathrm{uni}=0.9GM_{\rm BH}/c^2$, a uniform grid is used with the grid spacing $\Delta x=0.015GM_{\rm BH}/c^2$, and for $x > x_\mathrm{uni}$, $\Delta x$ is increased uniformly as $\Delta x_{i+1}=1.01\Delta x_i$ where the subscript $i$ denotes the $i$th grid with $\Delta x_i := x_{i+1}-x_i$.
For $z$, the same grid structure as for $x$ is used.
The black-hole horizon is always located in the uniform grid zone.
The location of the outer boundaries along each axis, $L$, is 6500--11000\,km; for larger values of $M_{\rm BH}$, $L$ is larger (see Table~\ref{table1}).

In the analysis of the simulation results, we always derive the
following quantities: average cylindrical radius $R_{\rm mat}$,
average specific entropy $\langle s \rangle$, and average electron
fraction $\langle Y_e \rangle$ both for the matter located outside the
black hole and for the ejecta (see the method for identifying the
ejecta below). Here, these average quantities are defined by
\beqn
R_{\rm mat}&:=&\sqrt{I \over M_{\rm mat}}, \\
\langle s \rangle &:=&{1 \over M_{\rm mat}} \int_{\rm out} \rho_* s \,d^3x, \\
\langle Y_e \rangle &:=&{1 \over M_{\rm mat}} \int_{\rm out} \rho_* Y_e \,d^3x, 
\eeqn
where $I$ and $M_{\rm mat}$ denote a moment of inertia and rest mass of the 
matter located outside the black hole defined, respectively, by 
\beqn
I&:=& \int_{\rm out} \rho_* (x^2+y^2) \,d^3x, \\
M_{\rm mat}&:=& \int_{\rm out} \rho_* \,d^3x. 
\eeqn
$\int_{\rm out}$ implies that the volume integral is performed for the matter located outside the black hole. The integration is practically performed for the $y=0$ plane with $d^3x=2\pi xdxdz$.

The ejecta component is identified using the same criterion as in
Ref.~\cite{Fujiba20}. First, we identify a matter component with $|h
u_t| > h_{\rm min}c^2$ as the ejecta. Here $h_{\rm min}$ denotes the
minimum value of the specific enthalpy in the adopted
equation-of-state table, which is $\approx 0.9987c^2$.  For the matter
escaping from a sphere of $r=r_{\rm ext}$, we define the ejection
rates of the rest mass and energy (kinetic energy plus internal
energy) at a given radius and time by
\beqn
\dot M_{\rm eje}&:=&\oint \rho \sqrt{-g} u^i dS_i, \label{ejectrate}\\
\dot E_{\rm eje}&:=&\oint \rho \hat e \sqrt{-g} u^i dS_i,
\eeqn
where $\hat e:=h \alpha u^t-P/(\rho \alpha u^t)$. The surface integral
is performed on a sphere of $r=r_{\rm ext}$ with $dS_i=\delta_{ir}r_{\rm
  ext}^2d\theta d\varphi$ for the ejecta component. $r_{\rm ext}$ is
chosen to be 4000--5000\,km in the present work. 

Here, $\rho \sqrt{-g} u^t(=\rho_*)$ obeys the continuity equation for the rest mass, and thus, the time integration for it gives a conserved quantity.
In the absence of gravity, $\rho \hat e \sqrt{-g} u^t$ also obeys the source-free energy-conservation equation, 
and far from the central region, the sum of its time integration and the 
gravitational potential energy of the escaped component are approximately conserved.  
Thus, the total rest mass and energy (excluding the gravitational potential
energy) of the ejecta (which escape away from a sphere of $r=r_{\rm
  ext}$) are calculated by
\beqn
M_{\rm eje,esc}(t)&:=&\int^t \dot M_{\rm eje} dt,\\
E_{\rm eje,esc}(t)&:=&\int^t \dot E_{\rm eje} dt. 
\eeqn
In addition, we add the rest mass for the ejecta component located
inside a sphere of $r=r_{\rm ext}$, $M_{\rm eje,in}(t)$, giving the
total ejecta mass on each time slice, $M_{\rm eje}=M_{\rm eje,esc}+M_{\rm eje,in}$.

Far from the central object, $E_{\rm eje,esc}$ is approximated by
\beq
E_{\rm eje,esc}\approx M_{\rm eje,esc} c^2 + U + T_{\rm kin}+{GM_{\rm BH} M_{\rm
  eje,esc} \over r_{\rm ext}}, \label{Eeje}
\eeq
where $U$ and $T_{\rm kin}$ are the values of the internal energy and
kinetic energy of the ejecta at $r_{\rm ext}\rightarrow \infty$,
respectively. The last term of Eq.~(\ref{Eeje}) approximately denotes
the contribution of the gravitational potential energy of the matter
at $r=r_{\rm ext}$~\cite{Fujiba20}.  Since the ratio of the internal
energy to the kinetic energy of the ejecta decreases with its
expansion, we may approximate $U/T_{\rm kin} \approx 0$, and hence,
$E_{\rm eje,esc}$ by $E_{\rm eje,esc} \approx M_{\rm eje,esc}c^2 +
T_{\rm kin}+GM_{\rm BH} M_{\rm eje,esc}/r_{\rm ext}$ for the region
far from the central object.  We then define the average velocity of
the ejecta (for the component that escapes from a sphere of $r=r_{\rm
  ext}$) by
\beq
v_{\rm eje}:=\sqrt{{2(E_{\rm eje,esc}-M_{\rm eje,esc}c^2-GM_{\rm BH} M_{\rm
    eje,esc}/r_{\rm ext}) \over M_{\rm eje,esc}}}. 
\eeq

\begin{figure*}[t]
\includegraphics[width=86mm]{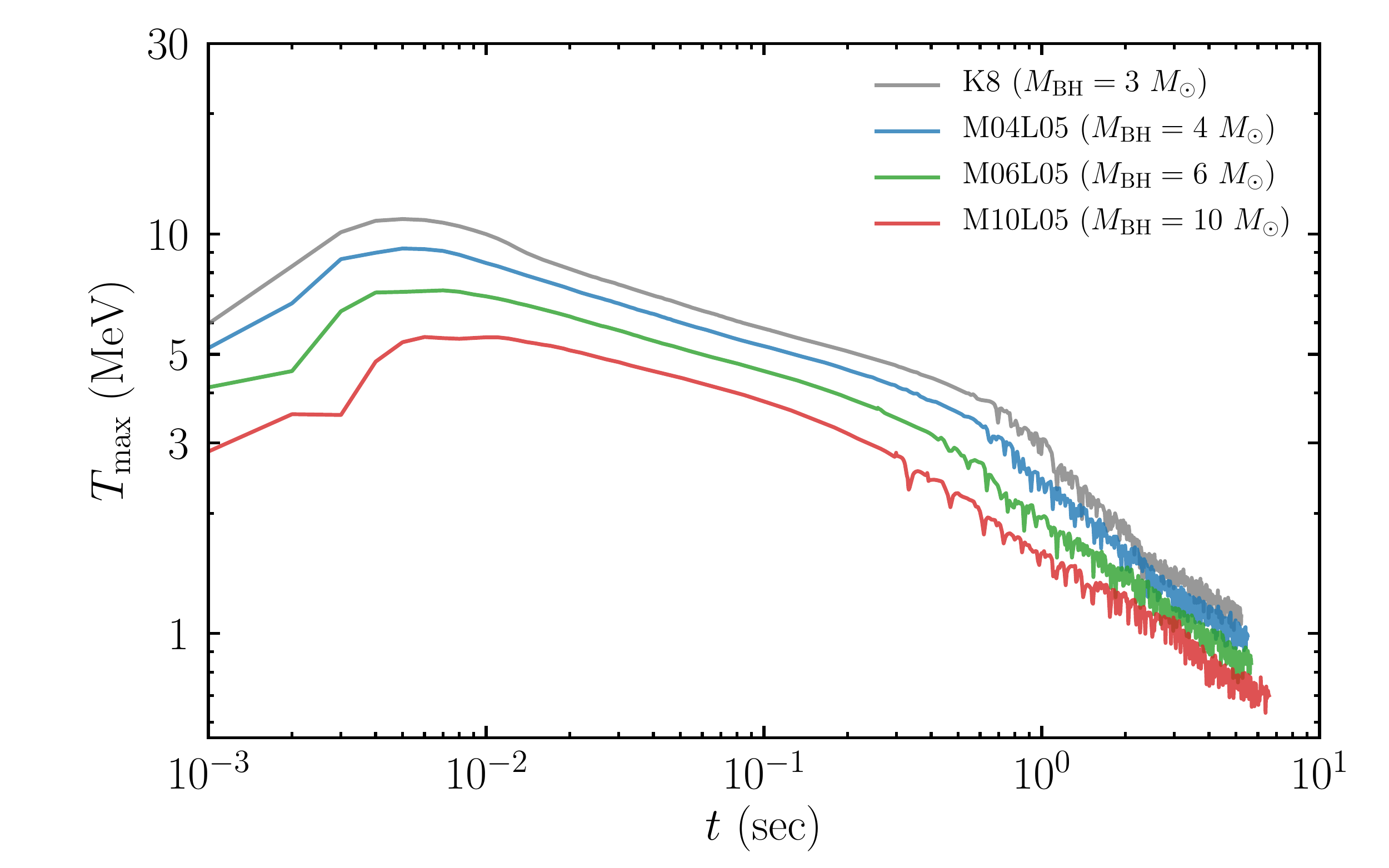} 
\includegraphics[width=86mm]{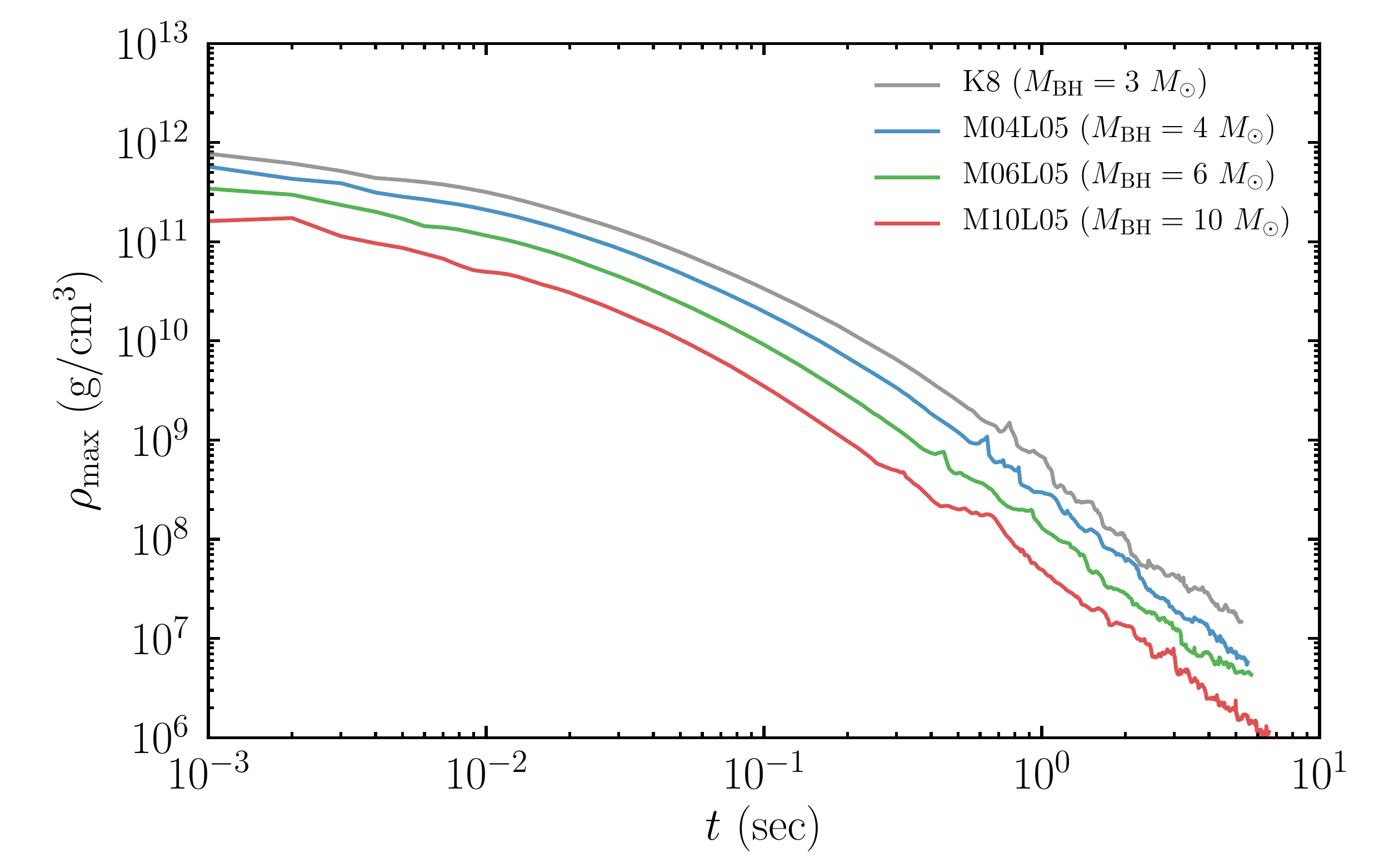} \\
\includegraphics[width=86mm]{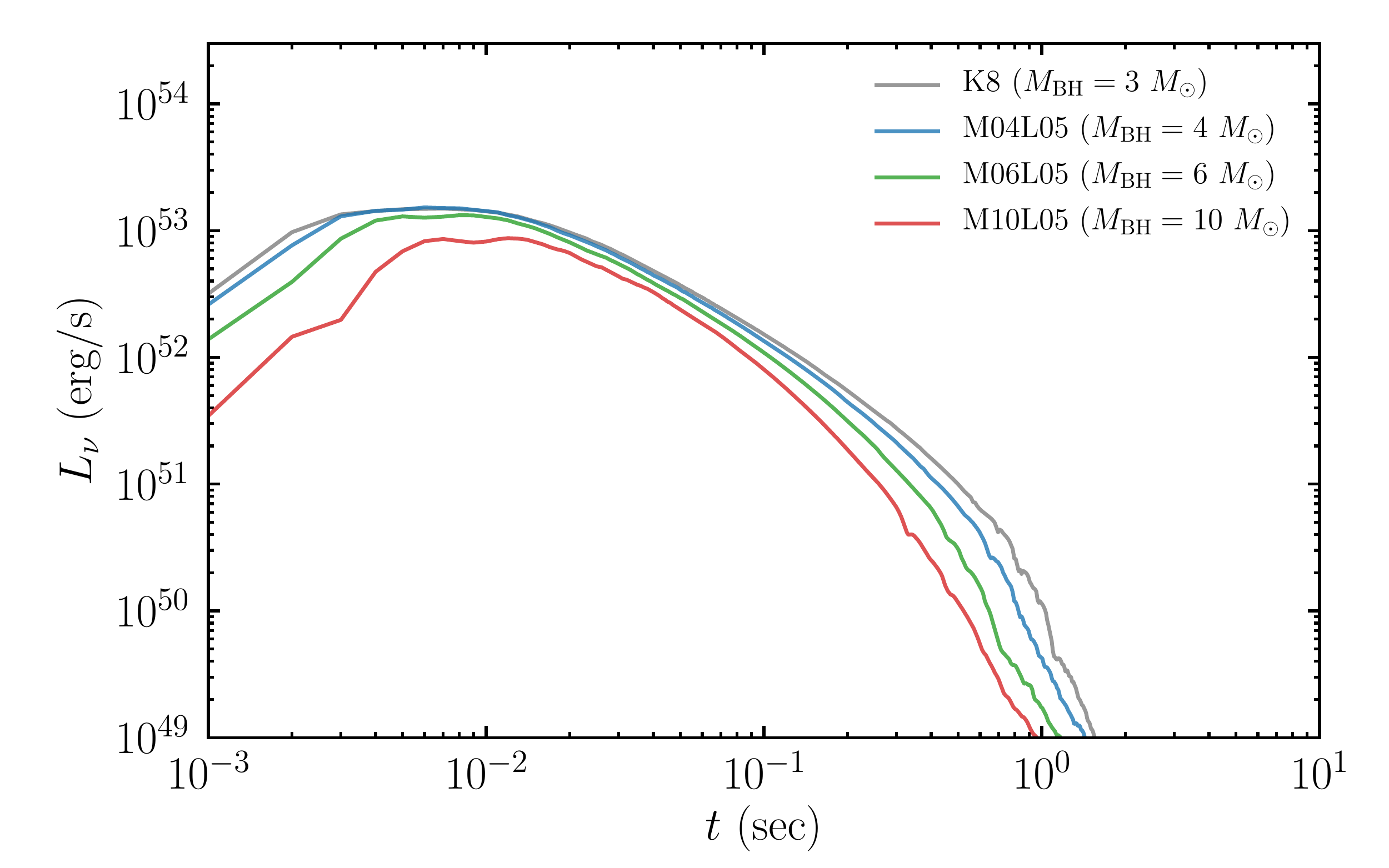} 
\includegraphics[width=86mm]{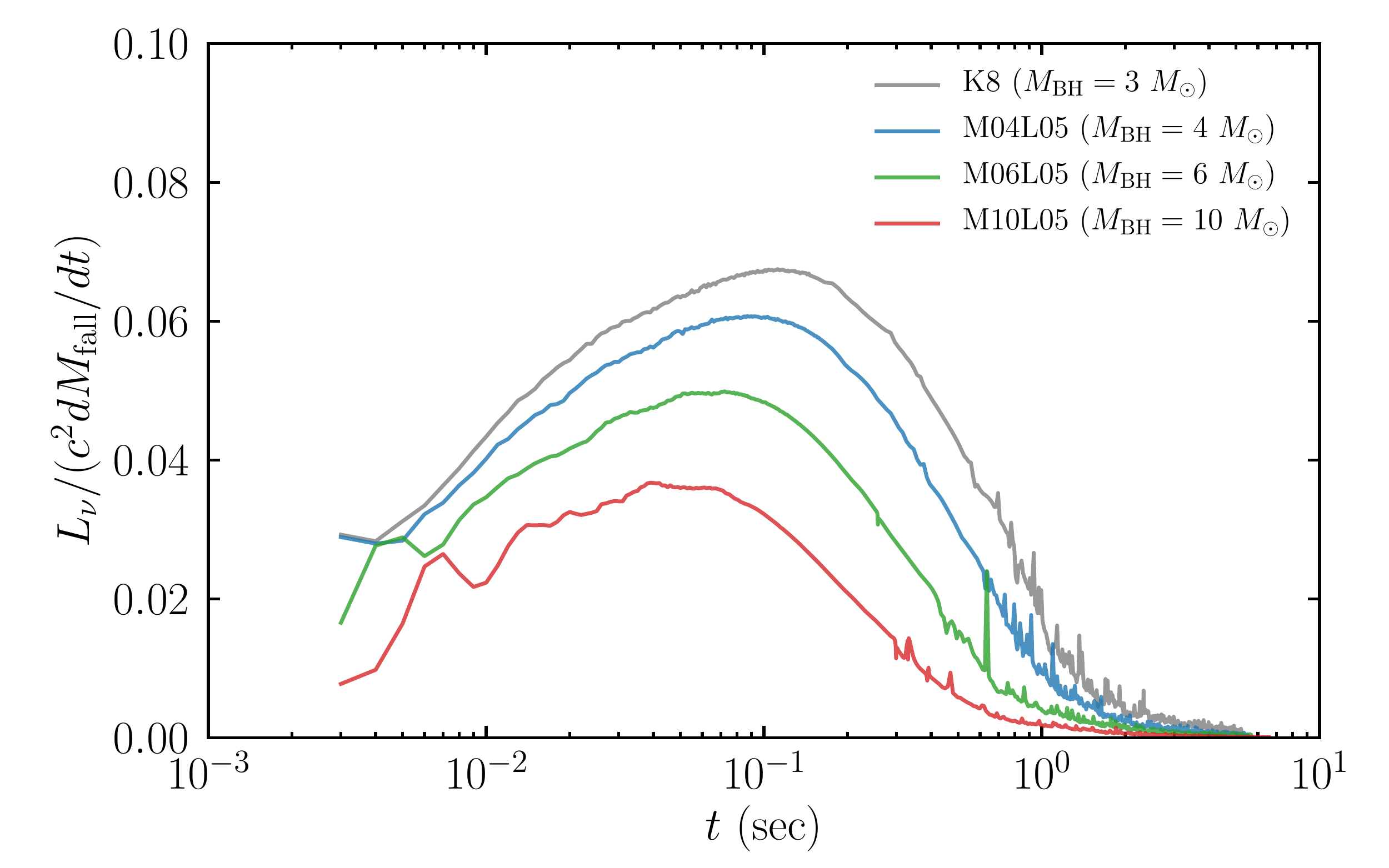} 
\caption{Upper panels: Evolution of the maximum temperature (left) and
  the maximum rest-mass density (right) of the disk for models with
  $M_{\rm disk} \approx 0.1M_\odot$ and $M_{\rm BH}=3$, 4, 6,
  $10M_\odot$ (models K8, M04L05, M06L05, and M10L05). Lower panels:
  The neutrino luminosity and efficiency of the neutrino
  emission. Here the efficiency is defined by the total neutrino
  luminosity, $L_\nu$, divided by the rest-mass energy accretion rate
  of the matter into the black hole, $c^2 dM_{\rm fall}/dt$.  Note
  that the short-timescale oscillation in the curves of $T_{\rm max}$,
  $\rho_{\rm max}$, and neutrino emission efficiency for the late time
  is due to the fact that convective motion is activated, and thus,
  the disk is disturbed.
\label{fig1}}
\end{figure*}

\subsection{Viscous hydrodynamics of disks for $M_{\rm disk}\approx 0.1M_\odot$}
\label{sec3-2}

\begin{figure*}[t]
\includegraphics[width=86mm]{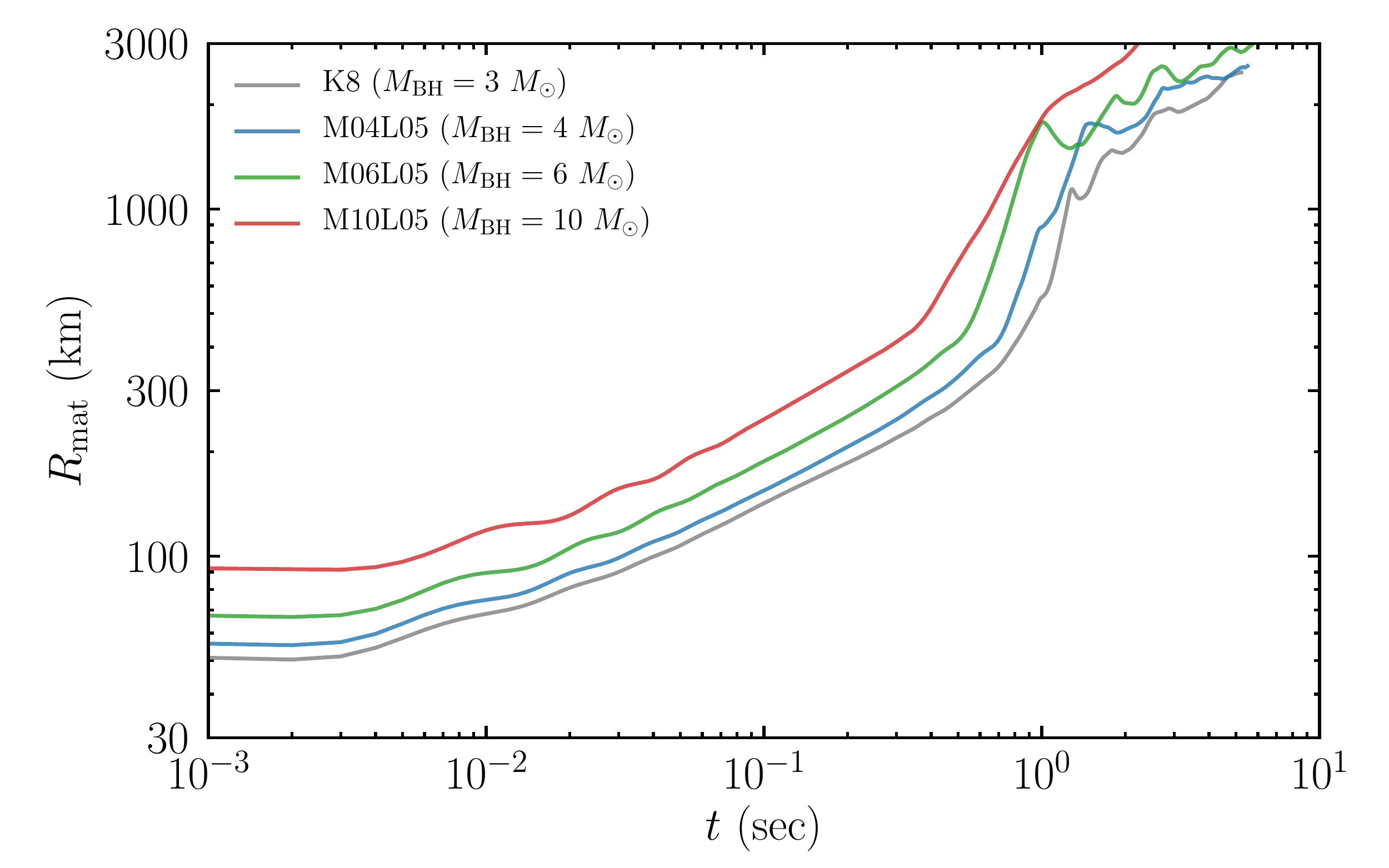} 
\includegraphics[width=86mm]{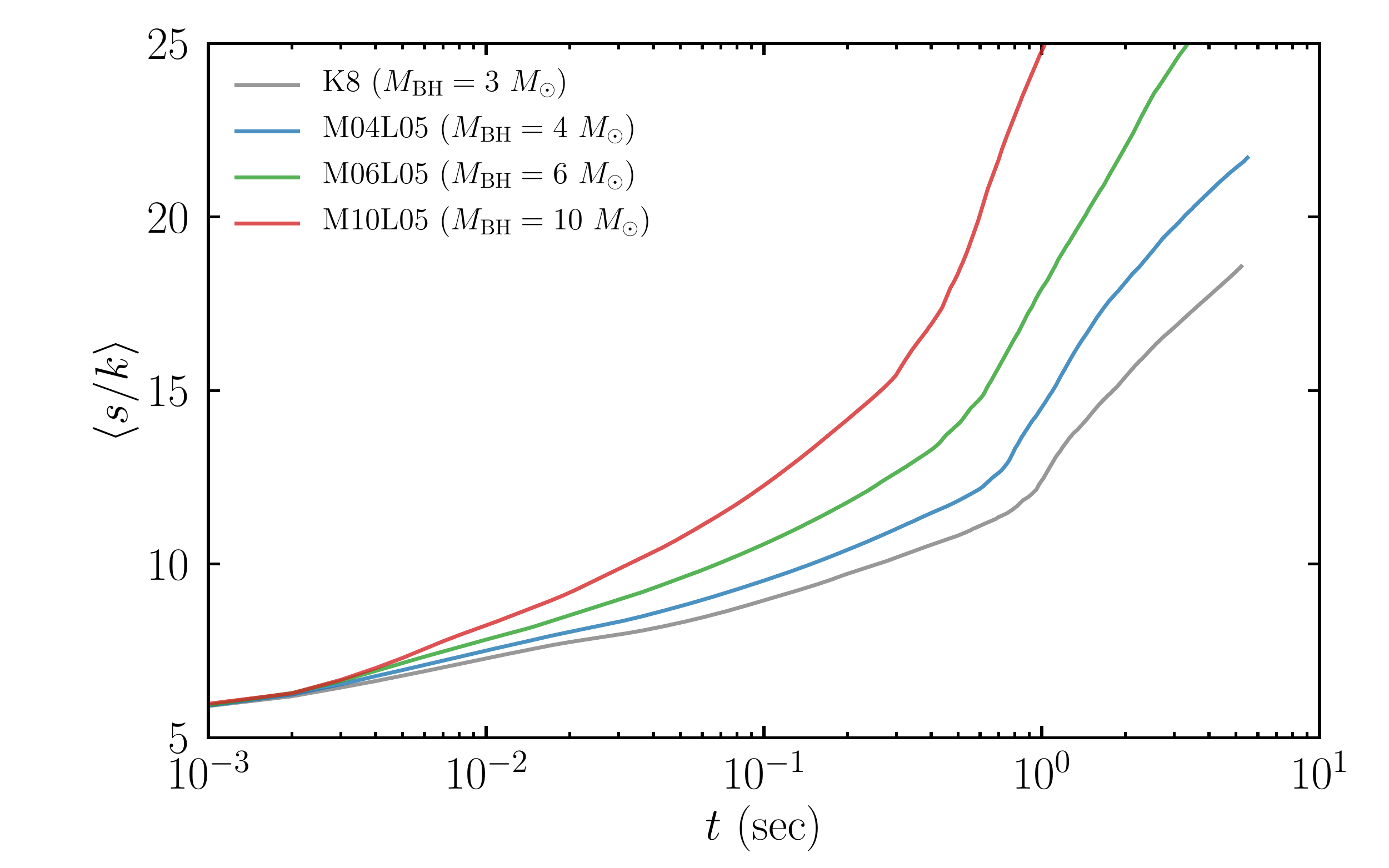} \\
\includegraphics[width=86mm]{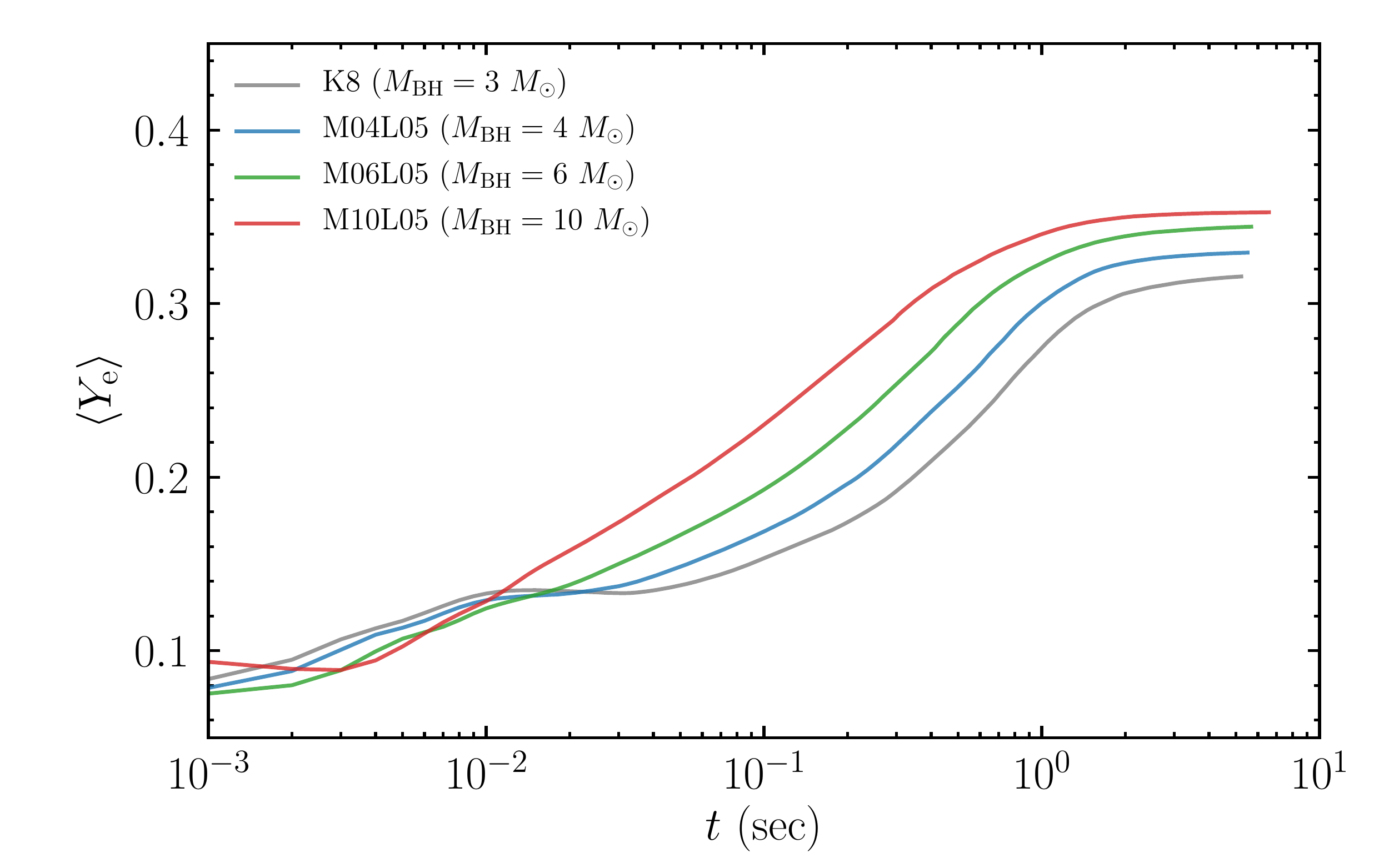} 
\includegraphics[width=86mm]{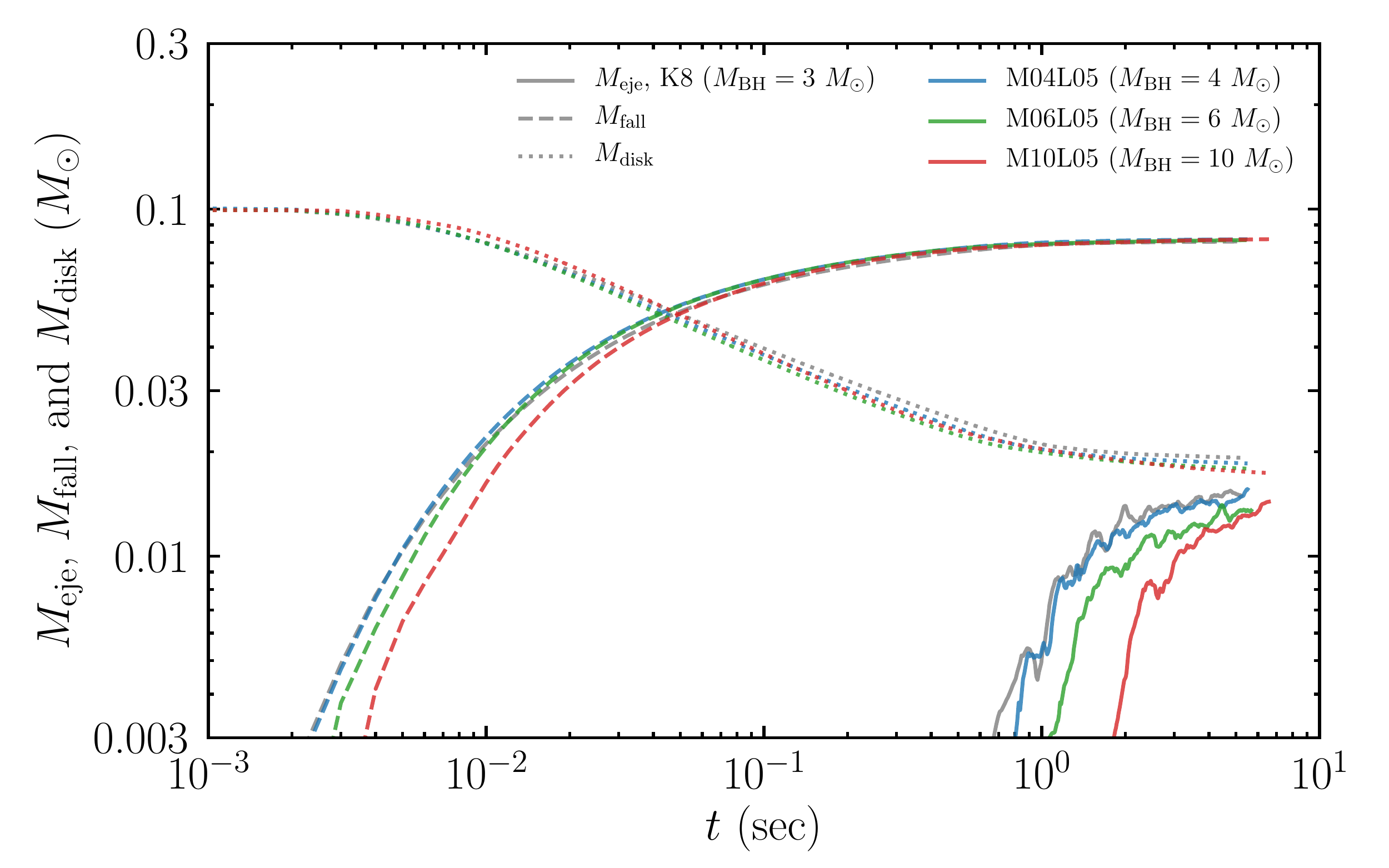} 
\caption{Several quantities for the matter located outside the black hole for models with $M_{\rm disk} \approx 0.1M_\odot$
  and $M_{\rm BH}=3$, 4, 6, and $10M_\odot$ (models K8, M04L05,
  M06L05, and M10L05).  Average cylindrical radius (top left), average
  specific entropy (top right), and average value of $Y_e$ (bottom
  left) of the matter located outside the black hole as functions of
  time.  The bottom right panel shows the total mass swallowed into
  the black hole (dashed curves), disk mass (mass of the matter outside the black
  hole; dotted curves), and the total ejecta mass $M_{\rm eje}$ (solid curves) as functions of
  time. 
\label{fig2}}
\end{figure*}

In this subsection, we compare the results of the low-mass disk 
models with $M_{\rm BH}=3$, 4, 6, and $10M_\odot$ and $M_{\rm disk}=0.1M_\odot$, to clarify the quantitative dependence of the disk evolution and mass ejection on the black-hole mass.
Here, the results with $M_{\rm BH}=3M_\odot$ were already derived in Ref.~\cite{Fujiba20}.
For all these models, the initial condition has the same specific entropy ($s=6k$), the same angular velocity profile ($n=1/7$).
For the higher-mass black holes, the maximum density of the disk is lower because of the larger radius of the outer edge, and thus, the larger volume of the disk.


First we summarize general processes for the viscous evolution of the disk.
We note that the mass and spin of the black hole do not change much because the disk mass is much smaller than the black-hole mass, and thus, the accretion of the matter into the black hole is a minor effect for the models with $M_{\rm disk}=0.1M_\odot$.

In the very early stage of the evolution of the system (for $t \alt 100$\,ms), 60\%--70\% of the disk in rest mass falls into the black hole from its inner region due to the viscous angular momentum transport process (and partly due to the transition process associated with the growth of the viscous tensor).
The maximum rest-mass accretion rate is $\approx 3M_\odot$/s and it monotonically decreases for $t \agt 10$\,ms leading to $< 0.1M_\odot$/s for $t \agt 100$\,ms.
Subsequently, the remaining part of the disk component expands gradually due to the viscous heating and angular momentum transport processes.
By these, the maximum temperature and rest-mass density decrease monotonically with time: see $t > 10$\,ms part of Fig.~\ref{fig1}.
In the relatively early stage with $t \alt 0.5$--2\,s, the maximum temperature of the disk is high, i.e., $kT \agt 3$\,MeV and the maximum rest-mass density is also larger than $\rho \sim 10^{8.5}\,{\rm g/cm^3}$ (see the top panels of Fig.~\ref{fig1}).
As a result, the neutrino emissivity is preserved to be high, $L_\nu \agt 10^{51}$\,erg/s, due to the electron and positron capture on nucleons (see the bottom-left panel of Fig.~\ref{fig1}).
For this stage, the thermal energy generated by the viscous heating is consumed primarily by the neutrino cooling, and hence, the viscous heating cannot be efficiently used for the outward expansion of the disk.
Indeed, the viscous heating rate of the order of $\sim \nu M_{\rm disk} \Omega^2$ is comparable to $L_\nu$ in this stage.
However, for the later stage in which the typical temperature of the disk decreases below $\sim 3$\,MeV$/k$~\cite{Fujiba20}, the neutrino cooling becomes inefficient, because the neutrino emissivity depends strongly on the matter temperature $T$ 
approximately as $\propto T^6$ (e.g., Refs.~\cite{Seti2004,SST2007}).
Then, the viscous heating is mostly used for the outward expansion of the disk.


\begin{figure*}[t]
\includegraphics[width=85mm]{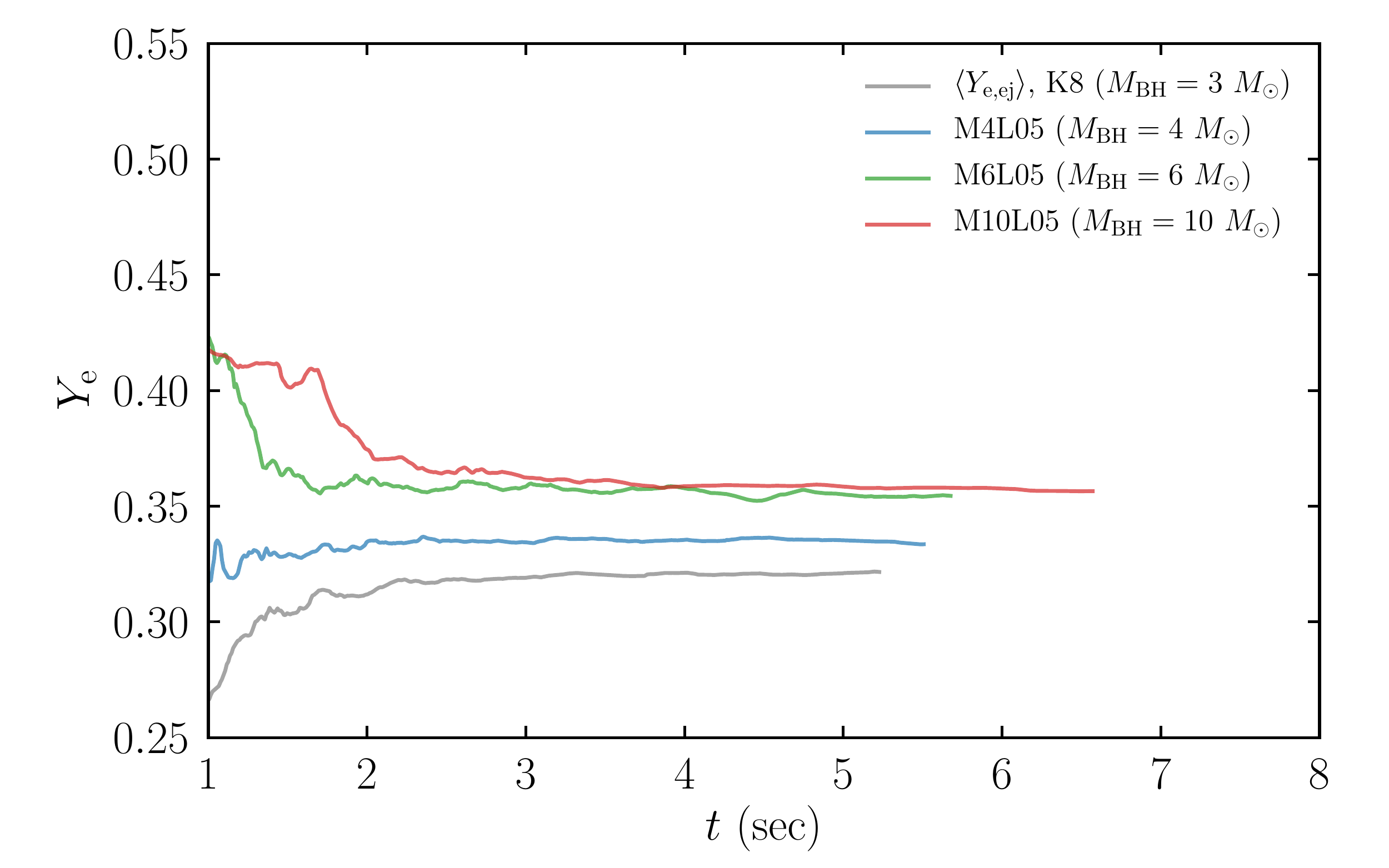} 
\includegraphics[width=85mm]{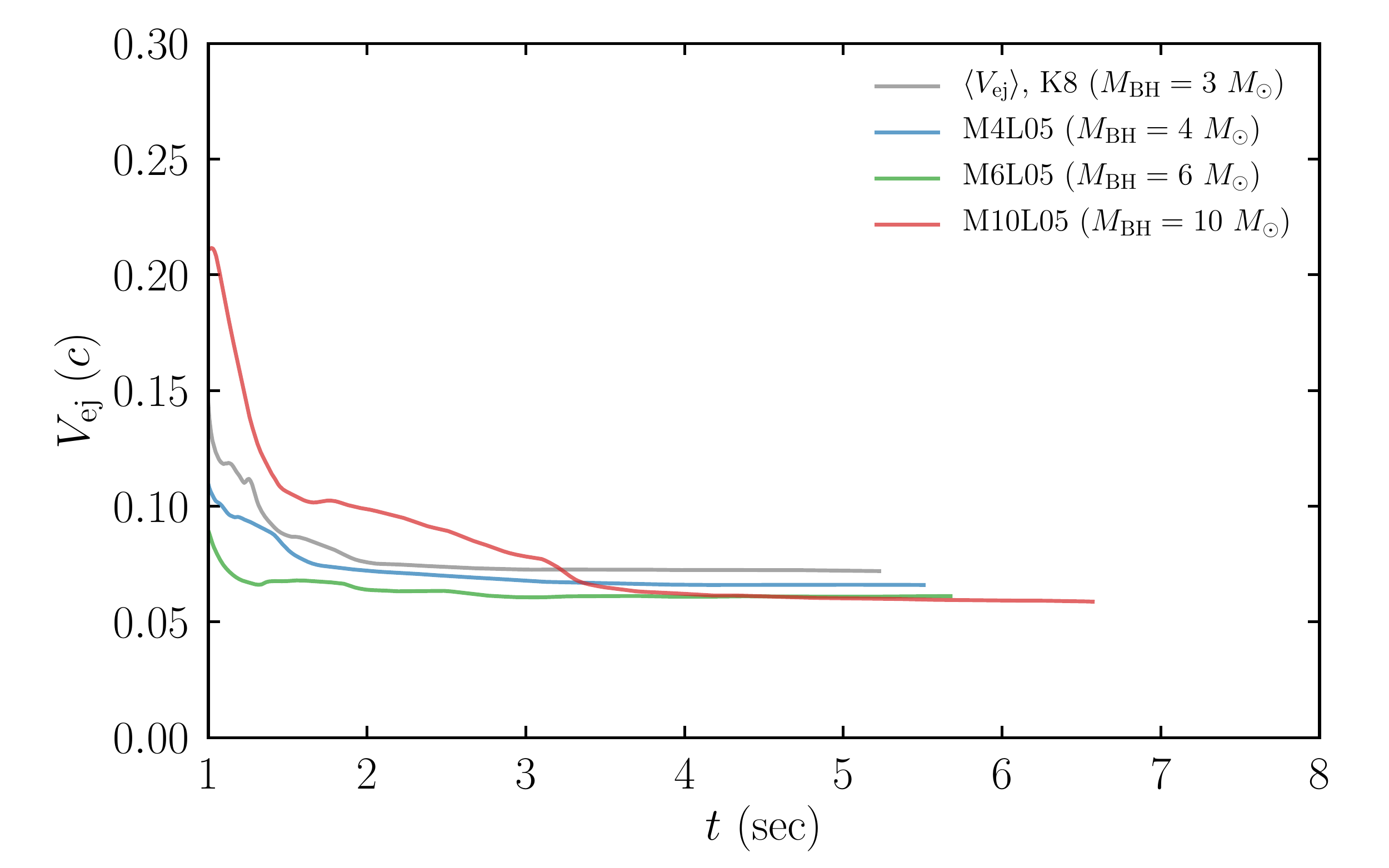}
\caption{The average values of $Y_e$ (left) and velocity (right) of
  the ejecta component for the system of $M_{\rm disk} \approx
  0.1M_\odot$ and $M_{\rm BH,0}=3$, 4, 6, and $10M_\odot$ (models K8,
  M04L05, M06L05, and M10L05).
\label{fig3}}
\end{figure*}

In particular, for the late stage for which the neutrino cooling is inefficient, the convective motion is excited and blobs of the matter viscously heated in the vicinity of the black hole are moved toward the outer region of the disk.
Here, the onset of the convective motion is reflected in a short-timescale variation in the curves of $T_{\rm max}$, 
$\rho_{\rm max}$, and $L_\nu$.  This motion activates the mass ejection of the
matter from the outer part of the disk. All these features are
qualitatively the same as those found in our previous
paper~\cite{Fujiba20}. However, the evolution process and the property
of the ejecta depends quantitatively on the black-hole mass as we
state in the following.

As we described above, the mass ejection sets in when the typical temperature of the disk decreases below $\sim 3$\,MeV$/k$.
Figure~\ref{fig1} shows that for the higher-mass black holes, 
the temperature is lower than that in the lower-mass models at given time.
As already mentioned, this is due to the fact that for the higher-mass black holes, the inner-edge radius of the disk has to be larger due to the larger horizon radius, and hence, the maximum temperature and rest-mass density are smaller from the beginning of the simulation.
As a consequence, the neutrino luminosity as well as the efficiency of the neutrino emission (see the bottom panels of Fig.~\ref{fig1}) is always lower for the higher-mass black holes.
Thus, the acceleration of the disk expansion due to the viscous heating in the inefficient neutrino cooling stage sets in earlier for the higher-mass black holes.

The top-left panel of Fig.~\ref{fig2} shows the evolution of the average cylindrical radius $R_{\rm mat}$.
In the early evolution stage of the disk during which $kT_{\rm max} \agt 3$\,MeV (see Fig.~\ref{fig1}), $R_{\rm mat}$ gradually increases with time
primarily due to the viscous angular momentum transport.
On the other hand, for the later stage of $kT_{\rm max} \alt 3$\,MeV, the increase of $R_{\rm mat}$ is accelerated.
This signals the onset of the mass ejection as we described in our previous paper~\cite{Fujiba20}.
In this mass ejection stage, the increase of the average specific entropy due to the viscous heating is also accelerated (see the top-right panel of Fig.~\ref{fig2}).
For the higher-mass black holes, the mass ejection sets in earlier.
This mass ejection is triggered primarily by the onset of the convective motion.
This is found from the fact that the time from which the short-timescale variation is found in the curves of $T_{\rm max}$ and $\rho_{\rm max}$ agrees approximately with the time at which the steep increases in $R_{\rm mat}$ and $\langle s \rangle$ are found.

The dependence of the evolution timescale of the disk on the black-hole mass is reflected in the composition variation of the disk.
In the early stage of the disk evolution prior to the onset of the mass ejection, electrons are relativistic and in a degenerate state because the rest-mass density is higher than $\rho_{\rm max} \sim 10^{8.5}$--$10^{9}\,{\rm g/cm^3}$.
As a result, the average electron fraction $\langle Y_e \rangle$ is relatively low.
After a significant fraction of the disk falls into the black hole and the disk relaxes to a quasi-steady evolution stage, the average value of $Y_e$ increases with the disk expansion (i.e., with the decrease of the rest-mass density). 
After the typical temperature of the disk decreases below $\sim 3$\,MeV$/k$, not only the neutrino luminosity drops significantly, but also the weak interaction process freezes out.
As a result, the average value of $Y_e$ relaxes to constants.
Our simulation clearly shows this property irrespective of the black-hole mass 
(see the bottom-left panel of Fig.~\ref{fig2}).


For the higher-mass black holes, the average value of $Y_e$ for given time is always higher for $t\gtrsim 10^{-2}$\,s due to the weaker electron degeneracy.
The reason for this is that the efficiency of the neutrino emission (see the bottom-right panel of Fig.~\ref{fig1}) is lower due to the lower temperature and density of the disk, and hence, the viscous heating is more efficient for higher-mass black holes.

On the other hand, the freezeout of $Y_e$ occurs earlier for the higher-mass black holes 
due to the earlier drop of the temperature and density, and 
hence, it is not trivial whether the final average value of $Y_e$ is higher or lower for the higher-mass black holes.
Our present result shows that the final relaxed average value of $Y_e$ becomes slightly higher for the higher-mass black holes.
Our interpretation for this is that the electron degeneracy in the disk is weaker when the weak interaction processes freeze out, and thus, the value of $Y_e$ in the equilibrium of the electron/positron capture at the freeze out is higher for higher-mass black holes.


The average entropy is also always higher for the higher-mass black holes.
The reason for this is the neutrino cooling efficiency is lower, and thus, viscous heating is more efficiently used for enhancing its entropy for the higher-mass black holes.
After the freezeout of the weak interaction at $t\sim 1$\,s, the intermittent production of the high-energy blobs, which eventually become ejecta, also contributes to the 
increase of the entropy. 

We note that this conclusion would hold only for the comparison among the models with the same initial radius of the disk (i.e., with the same physical values of $r_{\rm out}$).
If we compare the models with different values of $r_{\rm out}$, the results could be modified  (see, e.g., Ref.~\cite{FFL20}).

The bottom-right panel of Fig.~\ref{fig2} shows that $\sim 15$\%--20\%
of the initial disk mass becomes the ejecta. Here, for the higher-mass
black holes, the disk is more compact in the sense that $r_{\rm
  out}/M_{\rm BH}$ is smaller.  As a result, the ejecta mass is
smaller. This agrees qualitatively with the results of
Ref.~\cite{FFL20}.  The fraction of the ejecta mass agrees
quantitatively with that in our previous paper~\cite{Fujiba20}. We
note here that all the matter located outside the black hole in the late time
eventually becomes the ejecta as this figure illustrates.  One
non-trivial point is that the increase of the ejecta mass is slower
for the higher-mass black holes. Our interpretation for this is
that although the mass ejection sets in earlier, the viscous heating
timescale, which should be proportional to $R_{\rm disk}^2/\nu \propto
M_{\rm BH}$, is always longer for the higher-mass black holes. This
effect should be reflected in the curve of $M_{\rm eje}$.

Figure~\ref{fig3} displays the evolution for the average values of $Y_e$ and velocity of the ejecta component.
Reflecting the evolution of $Y_e$ in the disk during its viscous expansion, the average value of $Y_e$ for the ejecta becomes 0.30--0.35 and is higher for the higher-mass black holes.
This property also agrees qualitatively with a recent report in Ref.~\cite{FFL20}. 
The velocity of the ejecta is $\approx 0.05$--$0.07c$ depending only weakly on the black-hole mass.


After the viscous evolution of the disk and subsequent mass ejection, the system 
settles to a spinning black hole surrounded by a geometrically-thick torus and a narrow funnel in the vicinity of the rotation axis.
Such an outcome appears to be suitable for generating gamma-ray bursts, i.e., for driving a collimated jet. This topic is discussed in Sec.~\ref{sec3-7}.

\begin{figure*}[t]
\includegraphics[width=85mm]{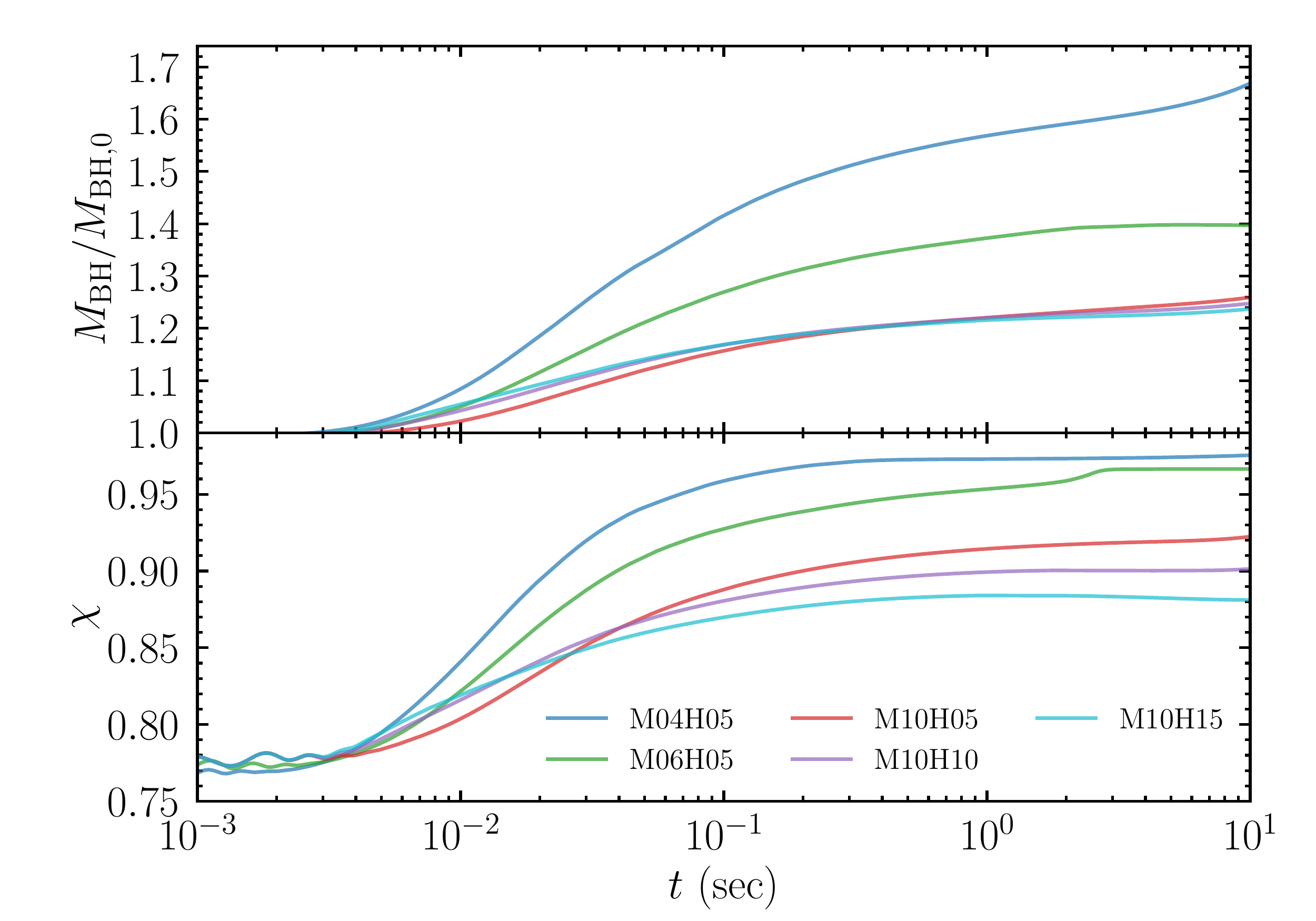} 
\includegraphics[width=85mm]{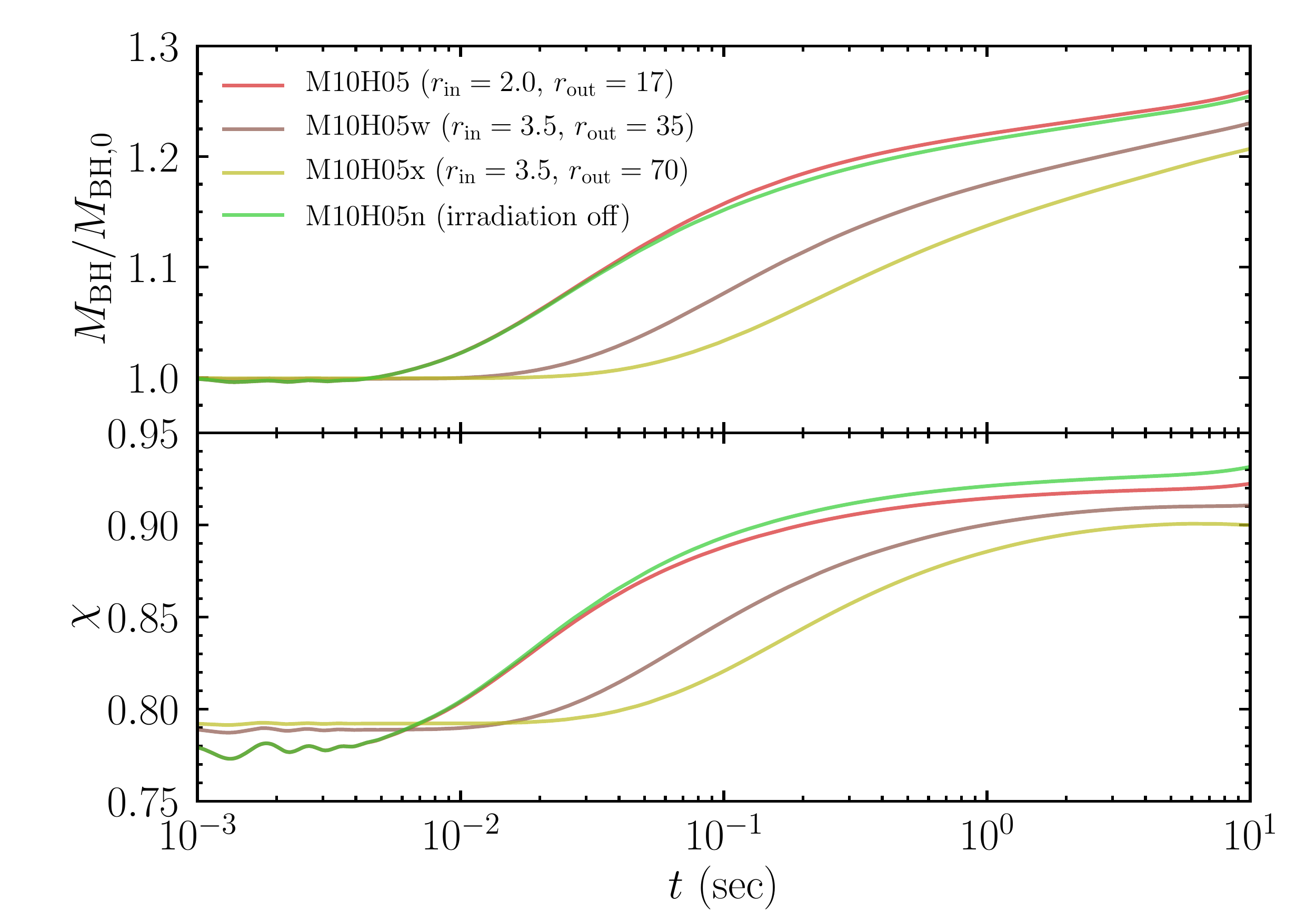} \\
\includegraphics[width=85mm]{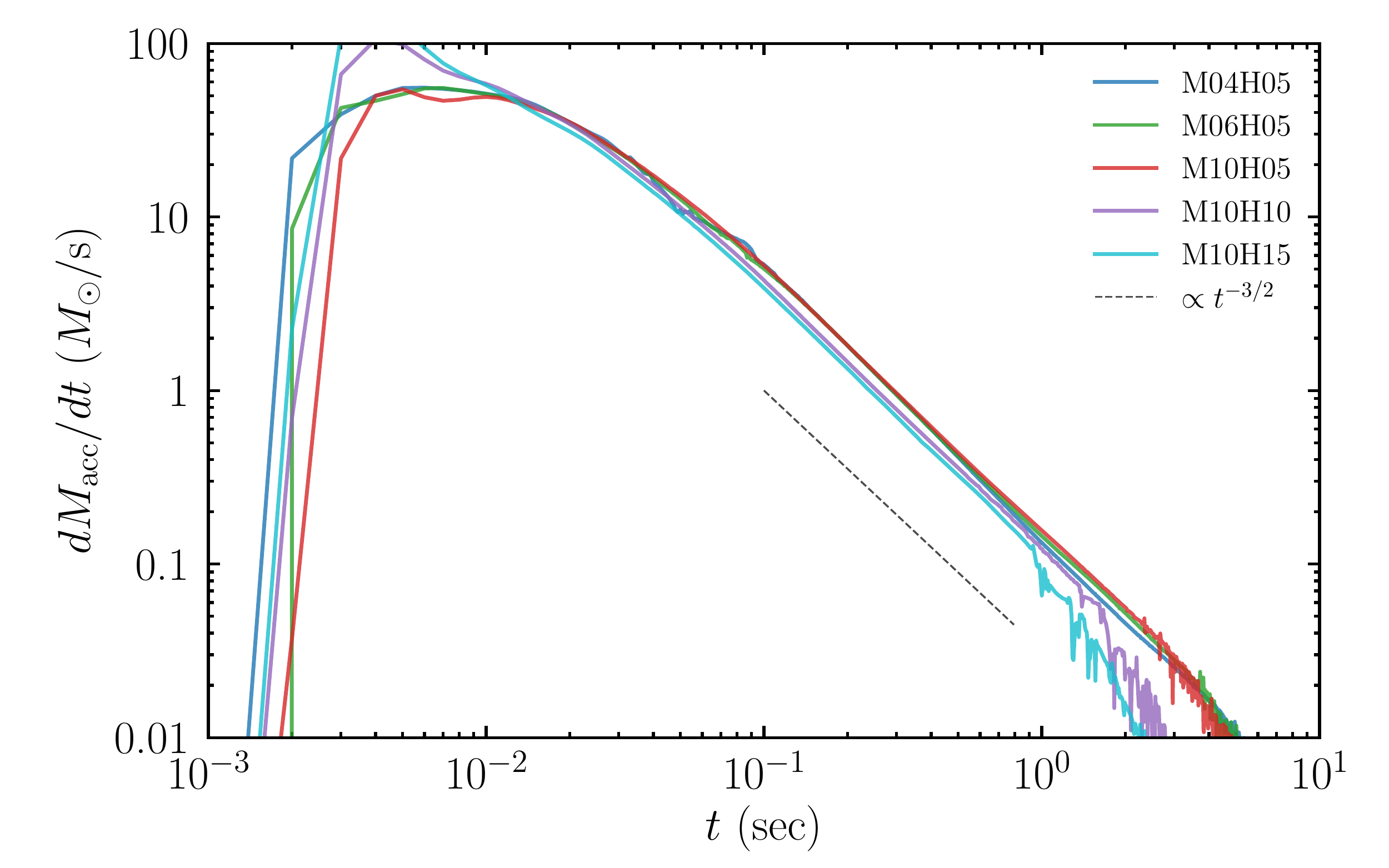} 
\includegraphics[width=85mm]{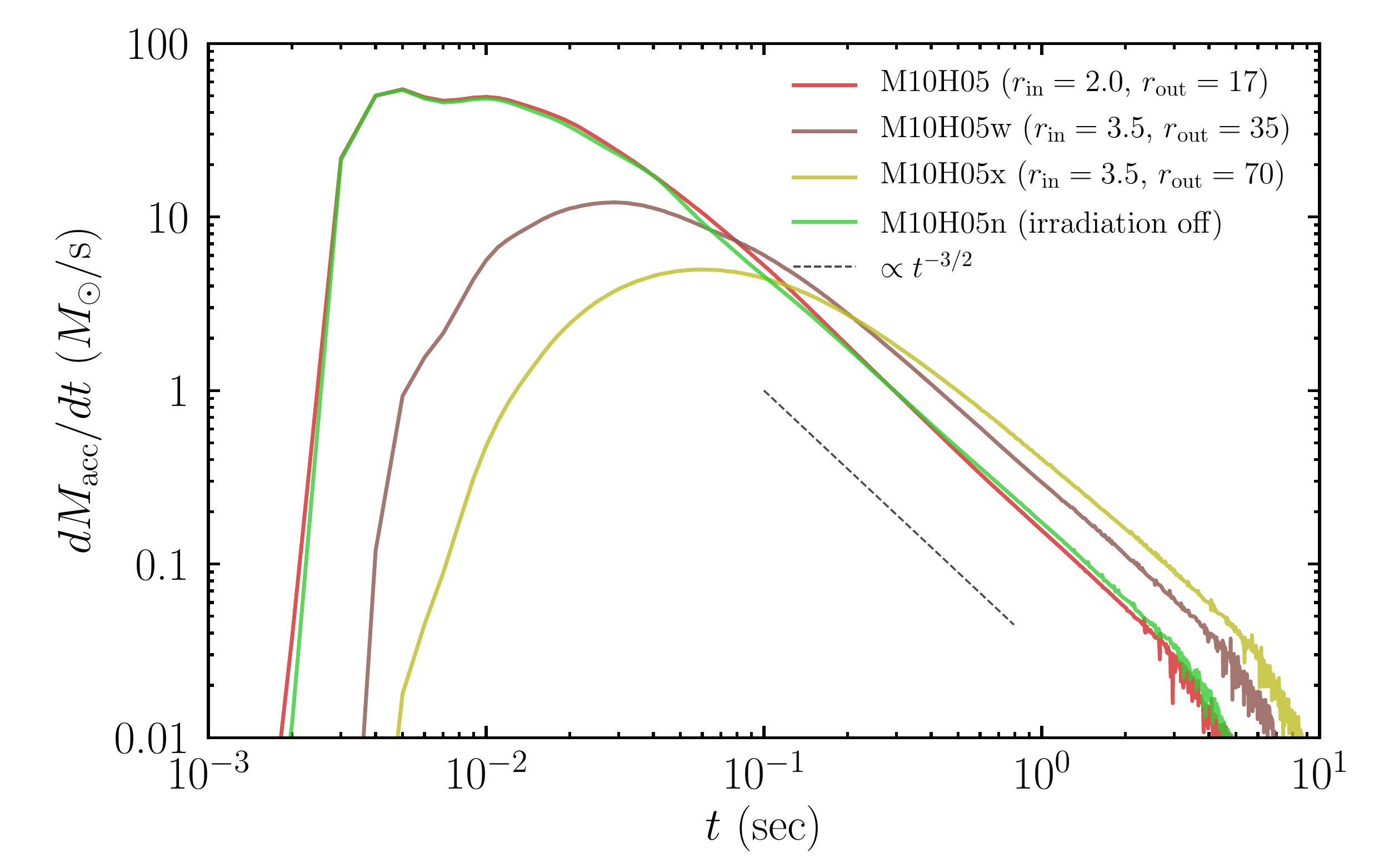} 
\caption{Top panels: The evolution of the mass and dimensionless spin of black holes for the models with $M_{\rm disk} \approx 3M_\odot$.
Bottom panels: The evolution of the rest-mass infall rate into the black holes.
The left panels show the results for $M_{\rm BH,0}=4$, 6, and $10M_\odot$ with $\alpha_\nu=0.05$ and for $M_{\rm BH,0}=10M_\odot$ with different values of $\alpha_\nu$ (models M10H10 and M10H15).
The right panels show the results for $M_{\rm BH,0}=10M_\odot$ with different initial disk radii (M10H05w and M10H05x) and with and without neutrino irradiation (M10H05n).
\label{fig4}}
\end{figure*}

\subsection{Evolution of black holes for high-mass disk models}\label{sec3-3}

In the following three subsections, we describe the results for the high-mass disk models of $M_{\rm disk} \approx 3M_\odot$ with the initial black-hole mass, $M_{\rm BH,0}=4$, 6, and $10M_\odot$.
We first summarize the evolution of the black holes as a result of matter accretion.
Figure~\ref{fig4} shows the evolution of the mass and dimensionless spin of the black holes (top panels) together with the rest-mass infall rate into the black holes (bottom panels) for all the models with $M_{\rm disk} \approx 3M_\odot$. 
The left panels compare the results for $M_{\rm BH,0}=4$, 6, and $10M_\odot$ with $\alpha_\nu=0.05$ and for $M_{\rm BH,0}=10M_\odot$ with different values of the viscous coefficient.
The right ones compare the results for $M_{\rm BH,0}=10M_\odot$ with different initial disk radii and with and without neutrino absorption/irradiation.

\begin{figure*}[t]
\includegraphics[width=85mm]{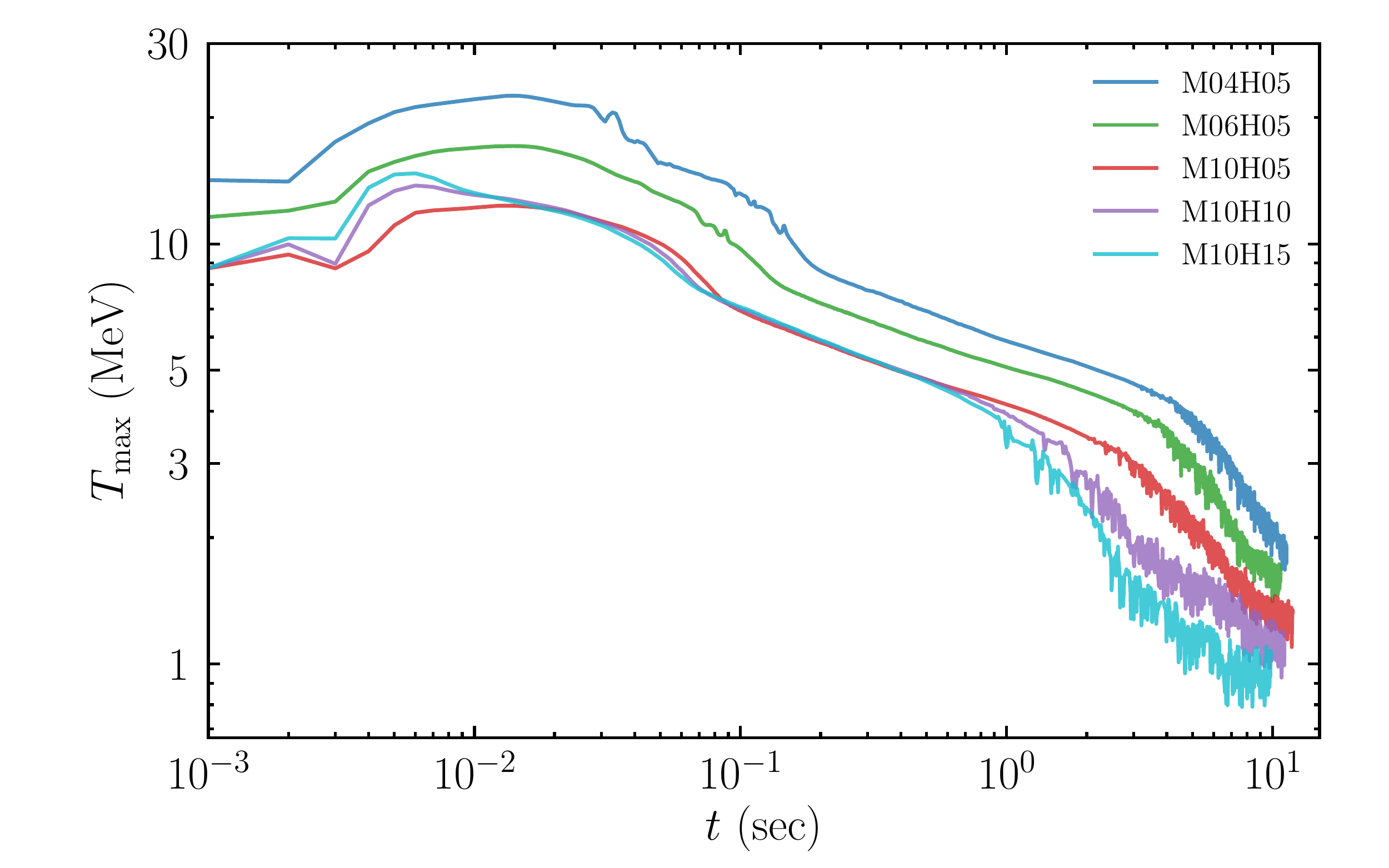} 
\includegraphics[width=85mm]{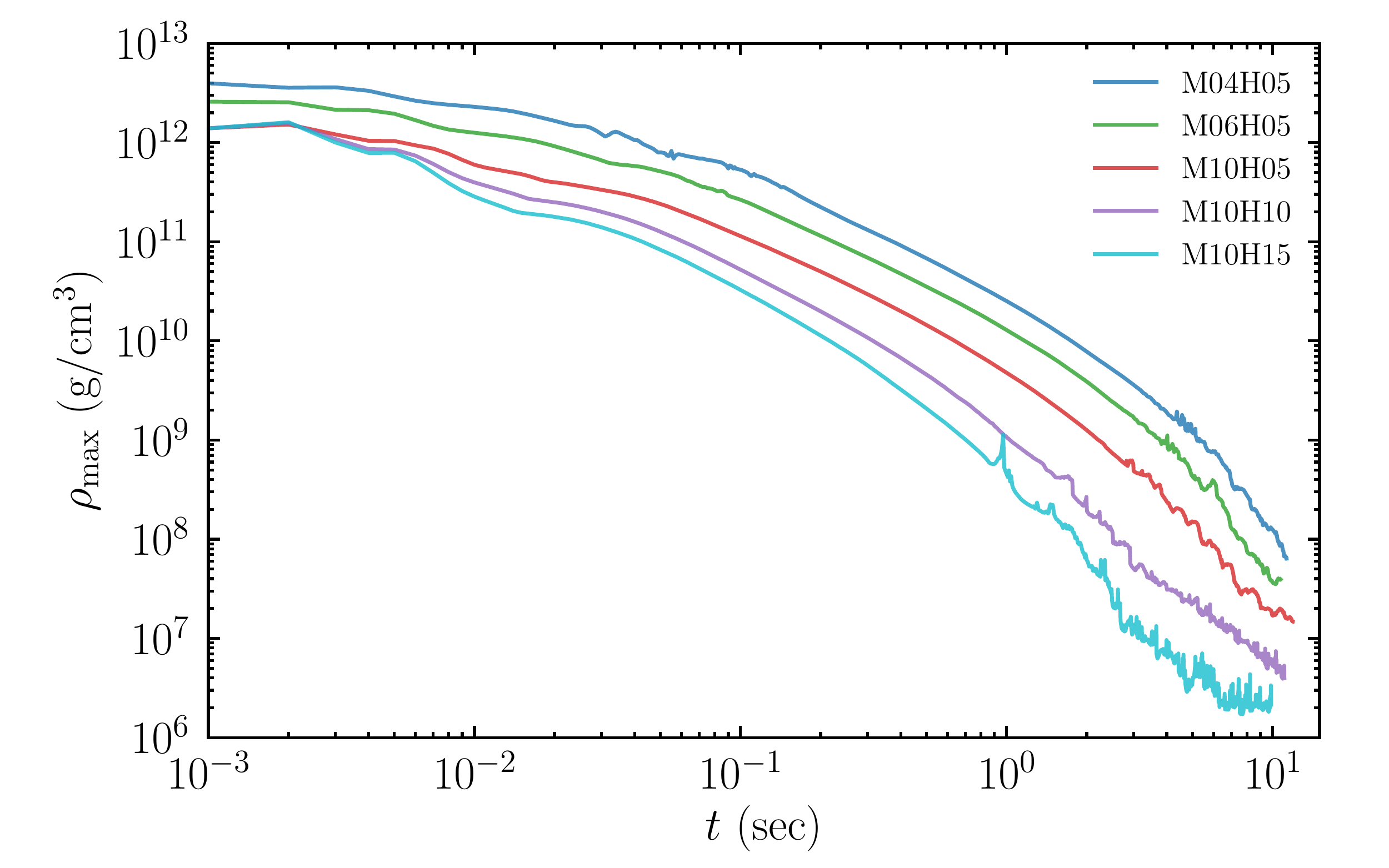} \\
\includegraphics[width=85mm]{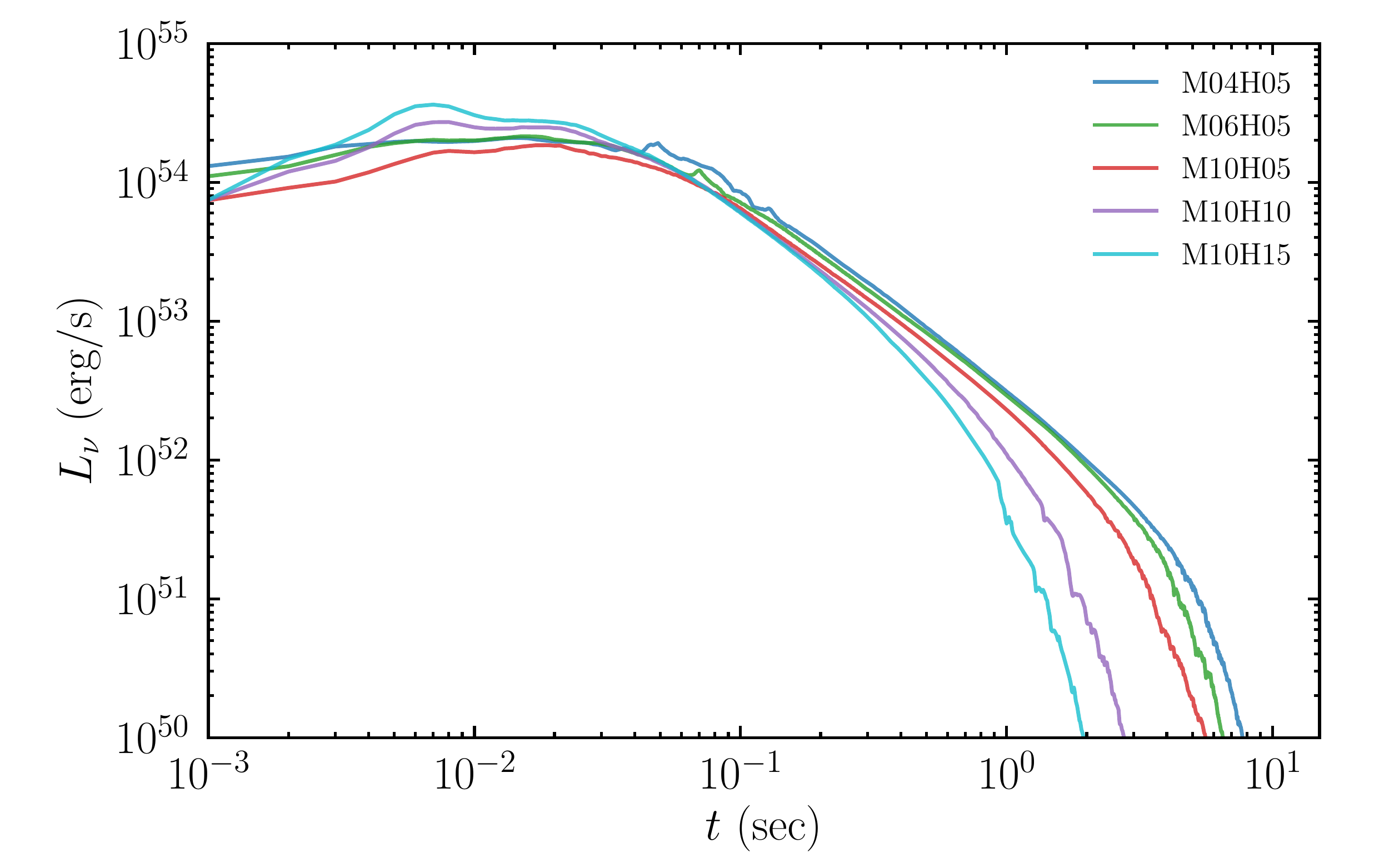} 
\includegraphics[width=85mm]{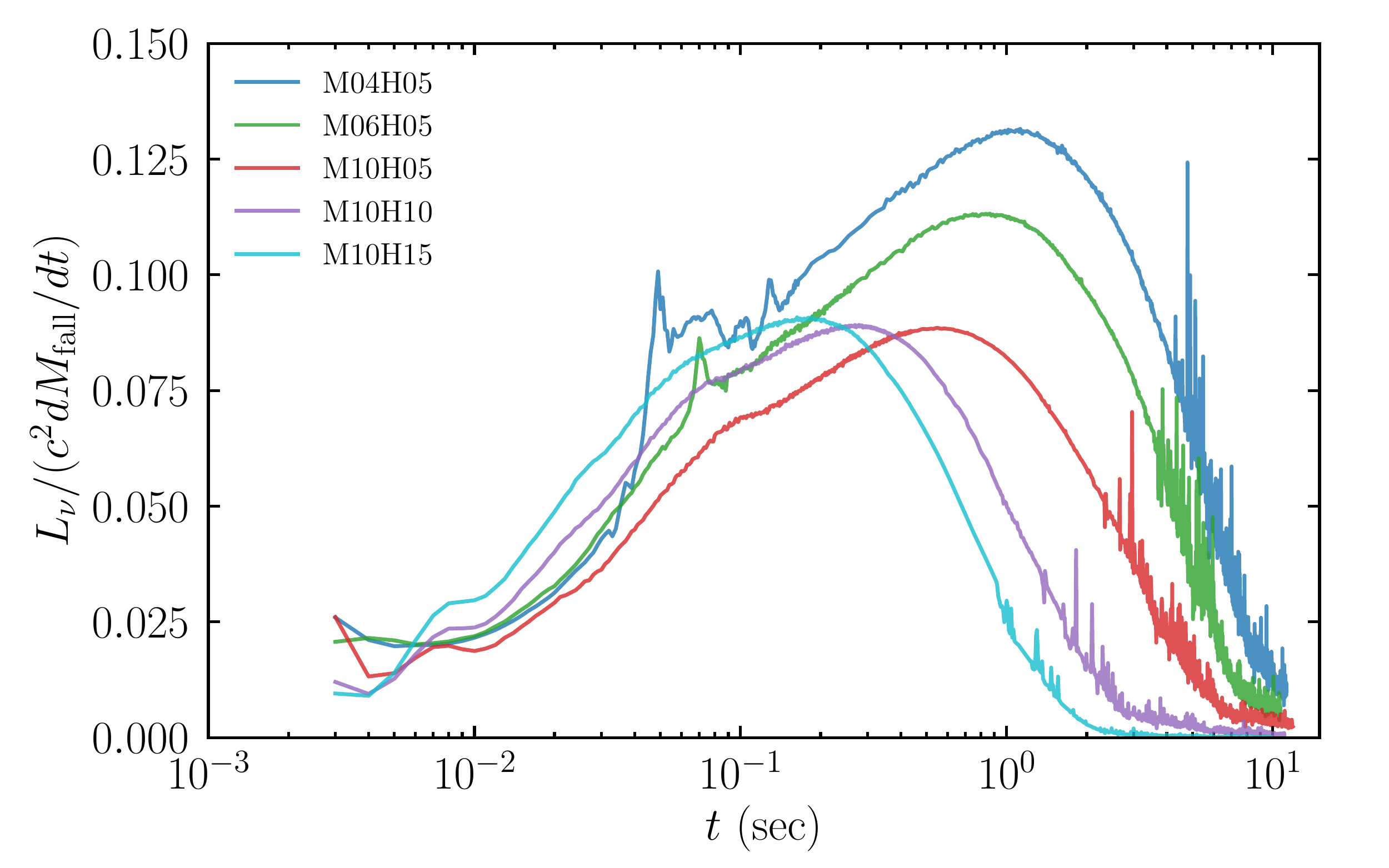} 
\caption{The same as Fig.~\ref{fig1} but for the system of $M_{\rm
    disk}\approx 3M_\odot$ and $M_{\rm BH,0}=4$, 6, and $10M_\odot$ with
  $\alpha_\nu=0.05$ (models M04H05, M06H05, and M10H05) and 
for $M_{\rm BH,0}=10M_\odot$ with $\alpha_\nu=0.10$ and 0.15 
(models M10H10 and M10H15). Note again that the short-timescale variation
  in the curves of $T_{\rm max}$, $\rho_{\rm max}$, and neutrino
  emission efficiency for the late time is due to the fact that 
  convective motion is activated, and thus, the disk is disturbed significantly.
\label{fig5}}
\end{figure*}

\begin{figure*}[t]
\includegraphics[width=85mm]{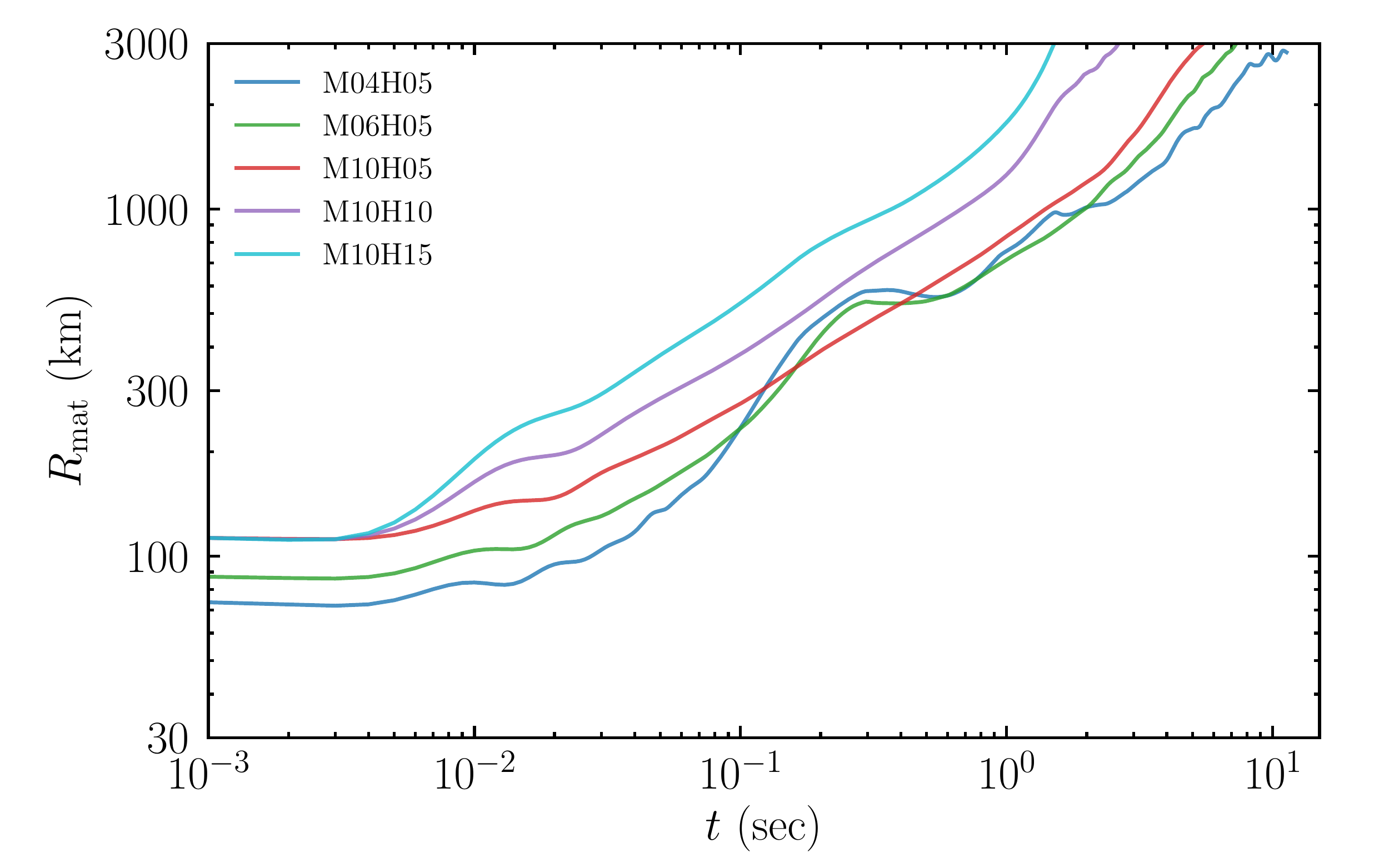} 
\includegraphics[width=85mm]{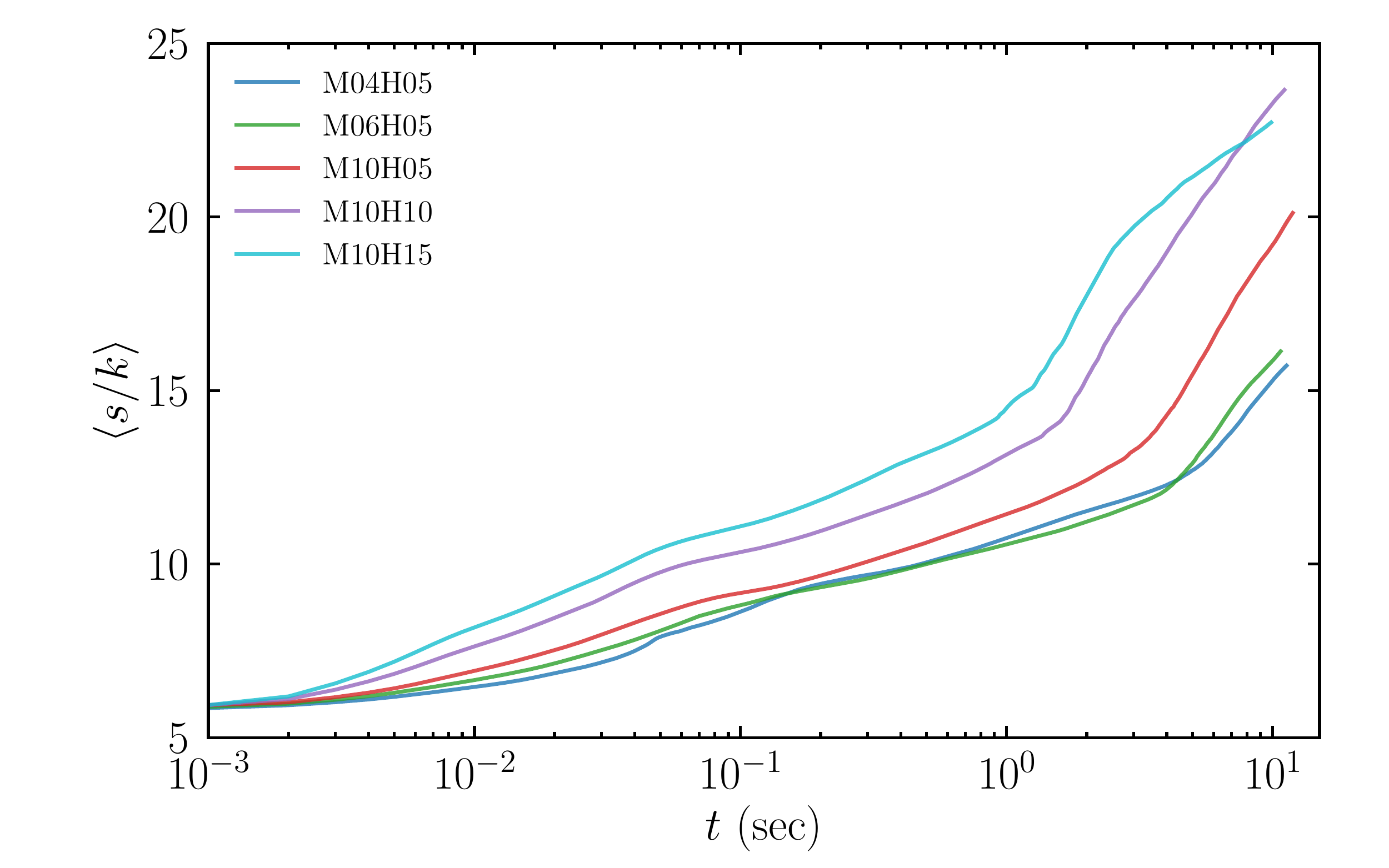} \\
\includegraphics[width=85mm]{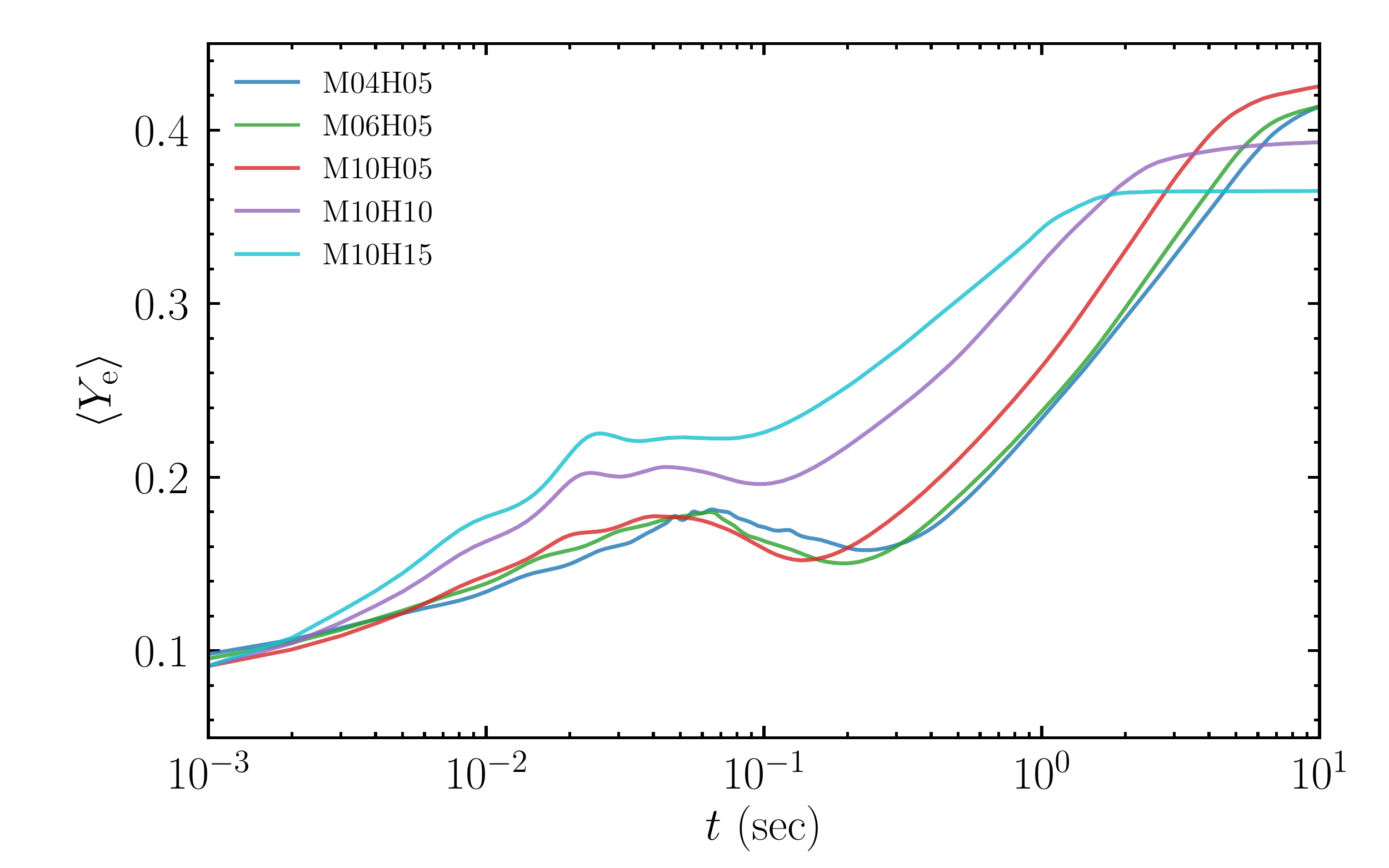} 
\includegraphics[width=85mm]{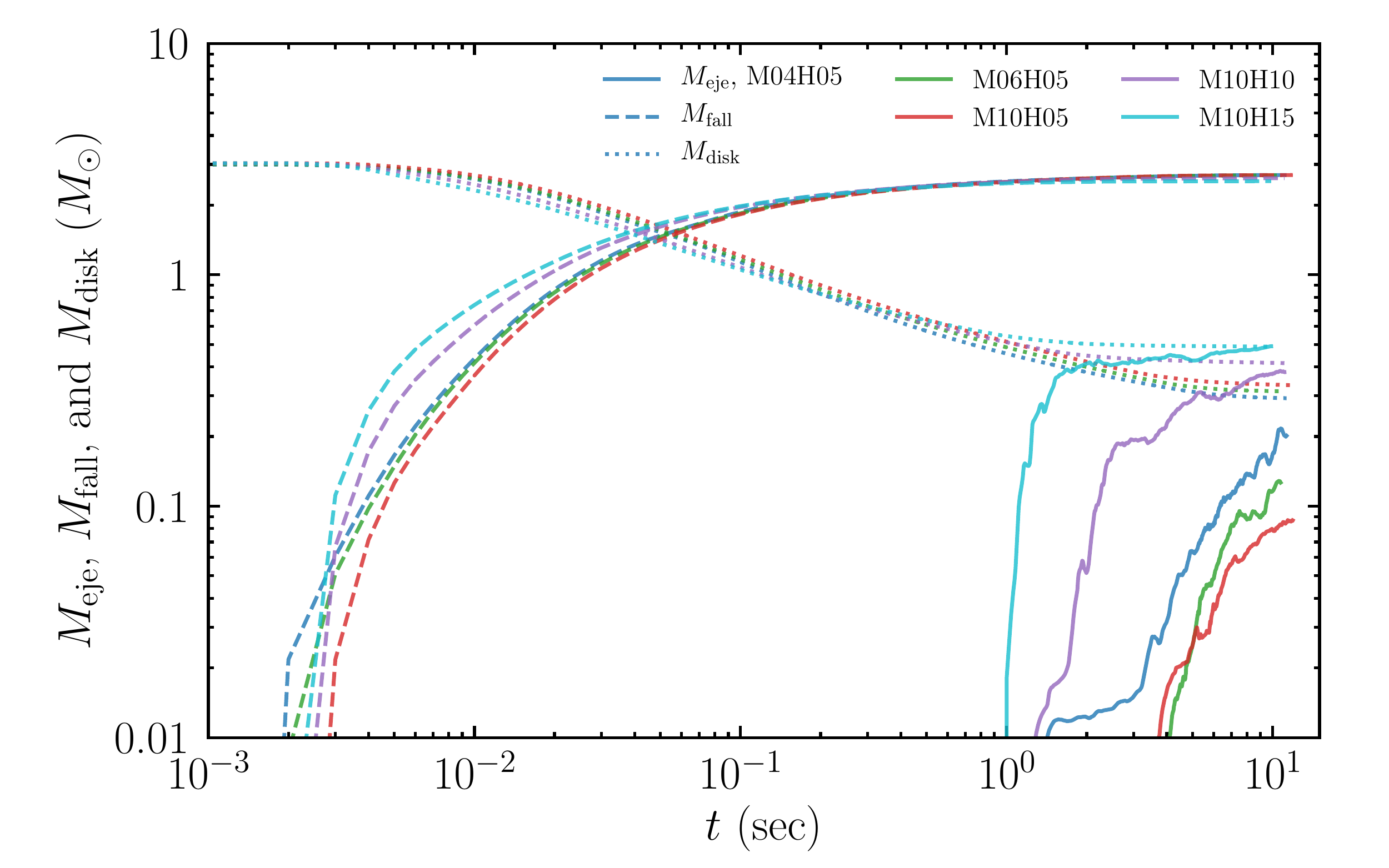} 
\caption{The same as Fig.~\ref{fig2} but for the system of $M_{\rm disk} \approx 3M_\odot$ and $M_{\rm BH,0}=4$, 6, and $10M_\odot$ with $\alpha_\nu=0.05$ (models M04H05, M06H05, and M10H05) and for $M_{\rm BH,0}=10M_\odot$ with $\alpha_\nu=0.10$ and 0.15 (models M10H10 and M10H15).
We note that for models M04H05 and M06H05, the density and velocity profiles are highly disturbed in the initial transition stage of $t \alt 0.2$\,s, and as a result, the average velocity and electron fraction oscillate due to an artifact of the initial condition.
\label{fig6}}
\end{figure*}

For these high-mass disk cases, $\approx 70\%$--90\% of the initial disk matter falls eventually into the black hole, and as a result, the black-hole mass increases approximately to $M_{\rm BH,0}+\zeta M_{\rm disk}$ with $\zeta \approx 0.7$--0.9.
For the initially compact disks with $\alpha_\nu=0.05$ (models M04H05,
M06H05, M10H05, and M10H05n), $\approx 90$\% of the initial disk matter eventually falls into the black hole irrespective of the models (cf.~the bottom-right panel of Fig.~\ref{fig6}), and thus, the final black-hole mass approaches approximately $M_{\rm BH,0}+0.9M_{\rm disk}$.
The rest-mass accretion rate is quite high ($\sim 50M_\odot$/s for $\alpha_\nu=0.05$) in the early stage for these models (see the bottom panels of Fig.~\ref{fig4}).
In this high-accretion rate stage, the neutrino emission efficiency is low because of the very high mass accretion rate (cf.~the bottom-right panel of Fig.~\ref{fig5}), and hence, the advection of the matter into the black hole is the major dissipation process of the matter energy.
We note that the short-term very high accretion rate in $t \leq 10$\,ms for the larger values of $\alpha_\nu (\geq 0.1)$ is achieved during the relaxation stage of the accretion disk in which the viscous tensor changes from zero to the quasi-steady state, and thus, it is an artifact due to the setting of the initial condition.

The rest-mass accretion rate, $dM_{\rm acc}/dt$, decreases steeply with time
after the peak is reached, and then, the growth of the black hole is
essentially stopped for $t \agt 5$\,s irrespective of the initial
conditions. Here, when the matter infall rate into the black hole
drops steeply, the neutrino luminosity also drops, and as a result,
the mass ejection sets in.  An interesting finding is that the rest-mass
accretion rate is approximately and universally proportional to
$t^{-3/2}$ after the peak is reached. As already mentioned, before the
mass ejection is activated, the dominant dissipation process of the
disk is the neutrino cooling, and thus, in this stage, the disk is in
a neutrino-dominate accretion flow (NDAF) state~\cite{NDAF} (see
Sec.~\ref{sec3-4} for more details).

For given values of $M_{\rm BH,0}$ and $r_{\rm out}$, the fraction of
the disk matter that falls into the black hole (i.e., the value of
$\zeta$) is larger than that for the low-mass disk models (cf.~Sec.~\ref{sec3-2}). Our interpretation for this is that the
self-gravity of the disk plays a role for confining the matter in the
central region of the system.

The values of $\zeta$ naturally decrease with the increase of the initial disk radius $r_{\rm out}$ (compare the results for models M10H05, M10H05w, and M10H05x): For M10H05w and M10H05x, only $\sim 80$\% and 70\% of $M_{\rm disk}$ falls into the black hole, respectively (cf.~the bottom-right panel of Fig.~\ref{fig9}).
It is also found that the peak rest-mass accretion rate is significantly reduced with the increase of $r_{\rm out}$, indicating that the value of $M_{\rm fall}$ is decreased.
Thus, the fraction of the disk mass that falls into the black hole, and hence, the final black-hole mass depend strongly on the initial disk radius.
This fraction also depends on the magnitude of the viscous coefficient (cf.~the bottom-right panel of Fig.~\ref{fig6}): For the larger viscous coefficients, the values of $\zeta$ are slightly smaller, and as a result, the final black-hole mass is slightly smaller.
For these models, the rest-mass accretion rate drops steeply for an earlier time of $t \agt 1$\,s at which the mass ejection sets in (see Sec.~\ref{sec3-4}).  This leads to the consequence that the ejecta mass increases with the increase of $\alpha_\nu$. 

Irrespective of the initial black-hole mass, the dimensionless spin of the black hole, $\chi$, monotonically increases toward an asymptotic value of $\chi_{\rm BH} \agt 0.9$ for $\alpha_\nu=0.05$.
This asymptotic value depends on the initial black-hole mass because the fraction of the accreted angular momentum to that of the initial black hole, which is proportional to $M_{\rm disk}/M_{\rm BH,0}$, is higher for the smaller-mass black hole in our setting.
However, the asymptotic values are always smaller than unity.
We note that because of the presence of the disk, the dimensionless angular momentum of the entire system defined by $\chi_{\rm tot}:=cJ_0/(GM_0^2)$, where $J_0$ and $M_0$ are the total angular momentum and initial gravitational mass, respectively, exceeds unity.
Specifically, $\chi_{\rm tot} \approx 1.20$, 1.15, 1.06, 1.18, and 1.28 for models M04H05, M06H05, M10H05, M10H05w, and M10H05x, respectively. Reflecting these initial conditions, for the relatively low-mass black holes (M04H05 and M06H05), the value of $\chi_{\rm BH}$ exceeds 0.97, but the further increase is halted for the chosen viscous coefficient.

The upper left and right panels of Fig.~\ref{fig4}, respectively, show that for the larger viscous coefficient or for the larger initial extent of the disk, the value of $\chi_{\rm BH}$ is smaller.
Both results are reasonable: For the larger viscous coefficient, the outward angular momentum transport becomes more efficient, and as a result, the angular momentum accretion into the black hole is suppressed.
For the larger initial extent of the disk, the fraction of the matter that falls into the black hole becomes smaller as the bottom-right panel of Fig.~\ref{fig4} shows.
This also suppresses the angular momentum gain by the black hole, leading to the smaller value of $\chi_{\rm BH}$.

It is interesting to note that in the absence of the neutrino absorption/irradiation effect, the final values of $M_{\rm BH}$ and $\chi_{\rm BH}$ are slightly smaller and larger than those in the presence of this effect, respectively.
The smaller value of $M_{\rm BH}$ may be interpreted as a consequence of less absorbed neutrino energy in the absence of neutrino absorption by the infalling matter.
Indeed, the neutrino luminosity (i.e., the fraction of neutrinos that escape to infinity) is slightly higher in the absence of the neutrino absorption/irradiation (cf. the bottom-left panel of Fig.~\ref{fig8}).
By contrast, the reason for the larger value of $\chi_{\rm BH}$ in the absence of the neutrino absorption/irradiation is not clear.
This result indicates that in the presence of the neutrino absorption/irradiation the angular momentum is more efficiently radiated away by neutrinos.
Thus, the possible interpretation for this is that in the presence of the neutrino absorption/irradiation, neutrinos are frequently reprocessed in the central region of the disk and the final emission toward infinity occurs in a relatively far region from the black hole for which the specific angular momentum of the disk matter is larger than that in the central region, and hence, the rotation effect (a relativistic effect) increases the angular momentum loss by the emitted neutrinos.

\subsection{Evolution of disks for high-mass disk models}\label{sec3-4}

\begin{figure*}[t]
\includegraphics[width=85mm]{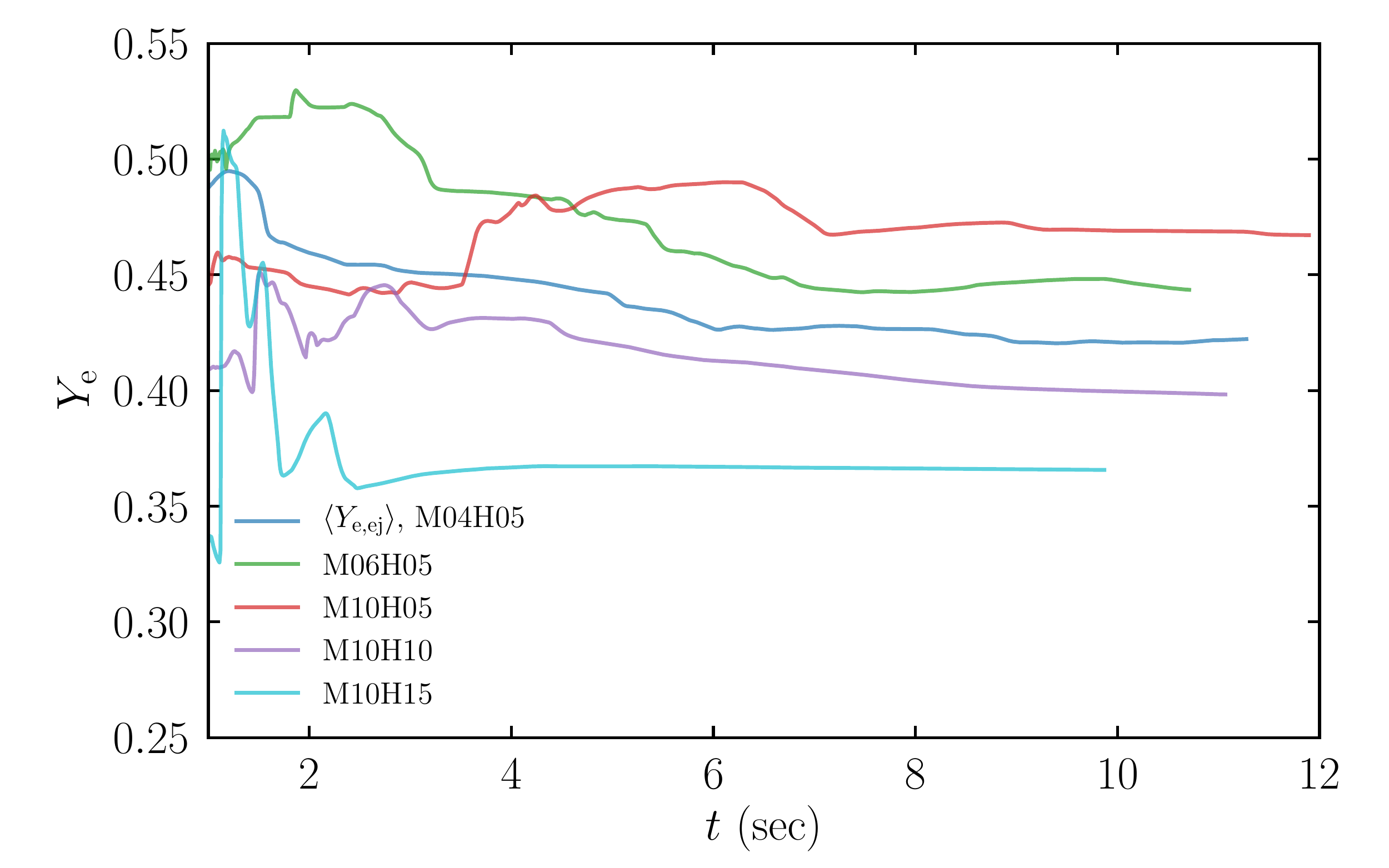} 
\includegraphics[width=85mm]{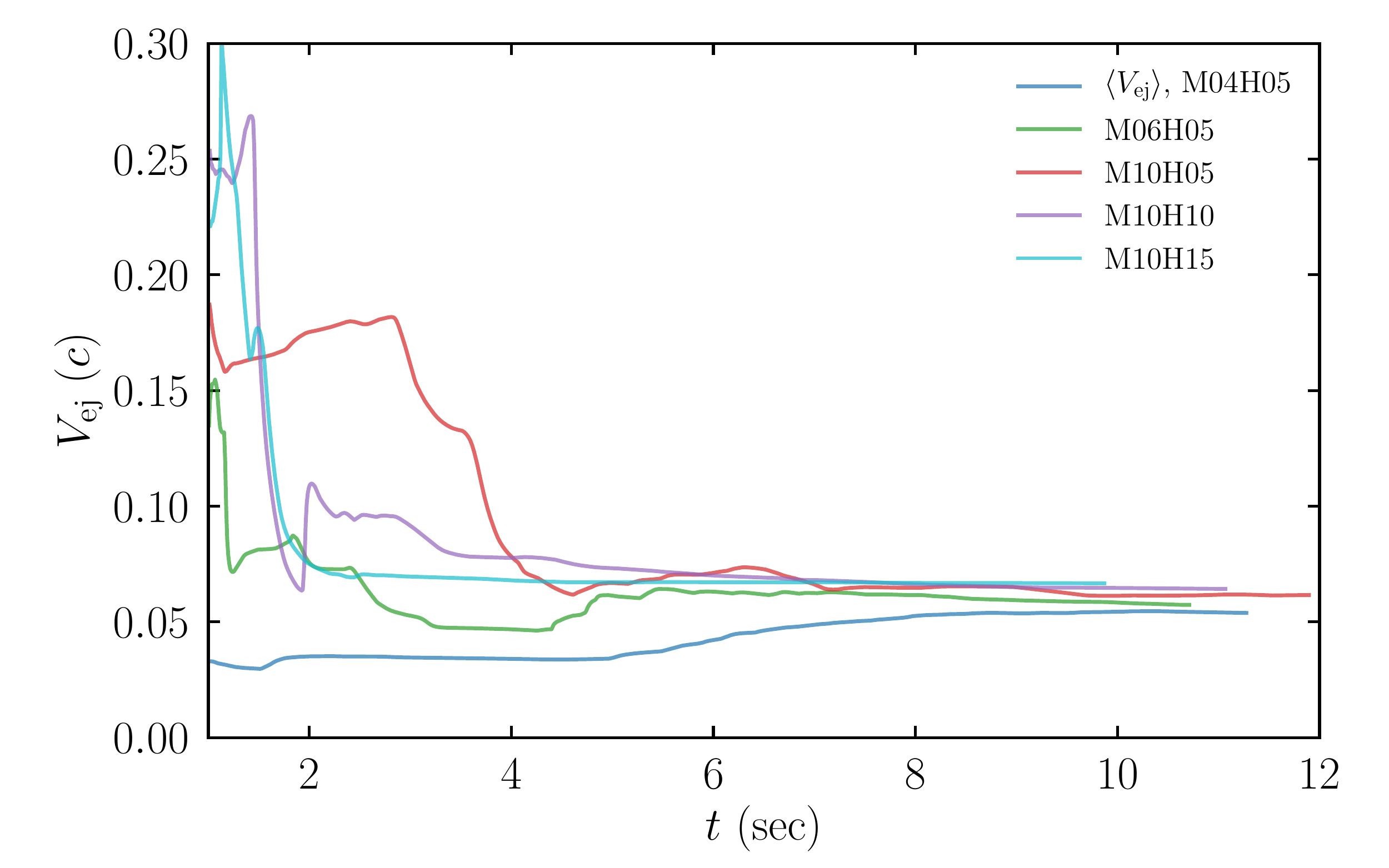}
\caption{The average values of $Y_e$ and velocity of the ejecta component for the models of $M_{\rm disk} \approx 3M_\odot$ and $M_{\rm BH,0}=4$, 6, and $10M_\odot$ with $\alpha_\nu=0.05$ (models M04H05, M06H05, and M10H05) and for $M_{\rm BH,0}=10M_\odot$ with $\alpha_\nu=0.10$ and 0.15 (models M10H10 and M10H15).
\label{fig7}}
\end{figure*}

\begin{figure*}[t]
\includegraphics[width=85mm]{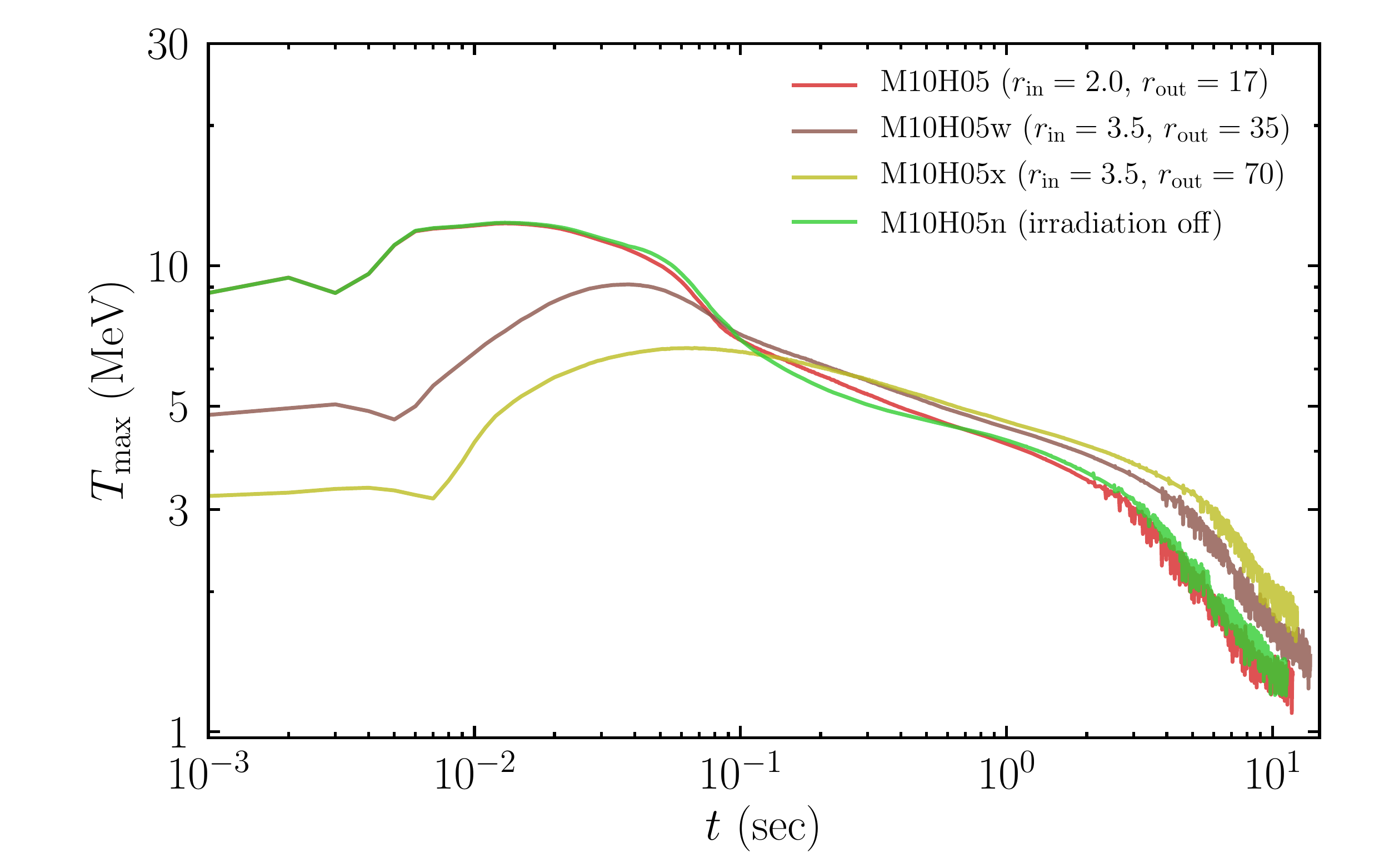} 
\includegraphics[width=85mm]{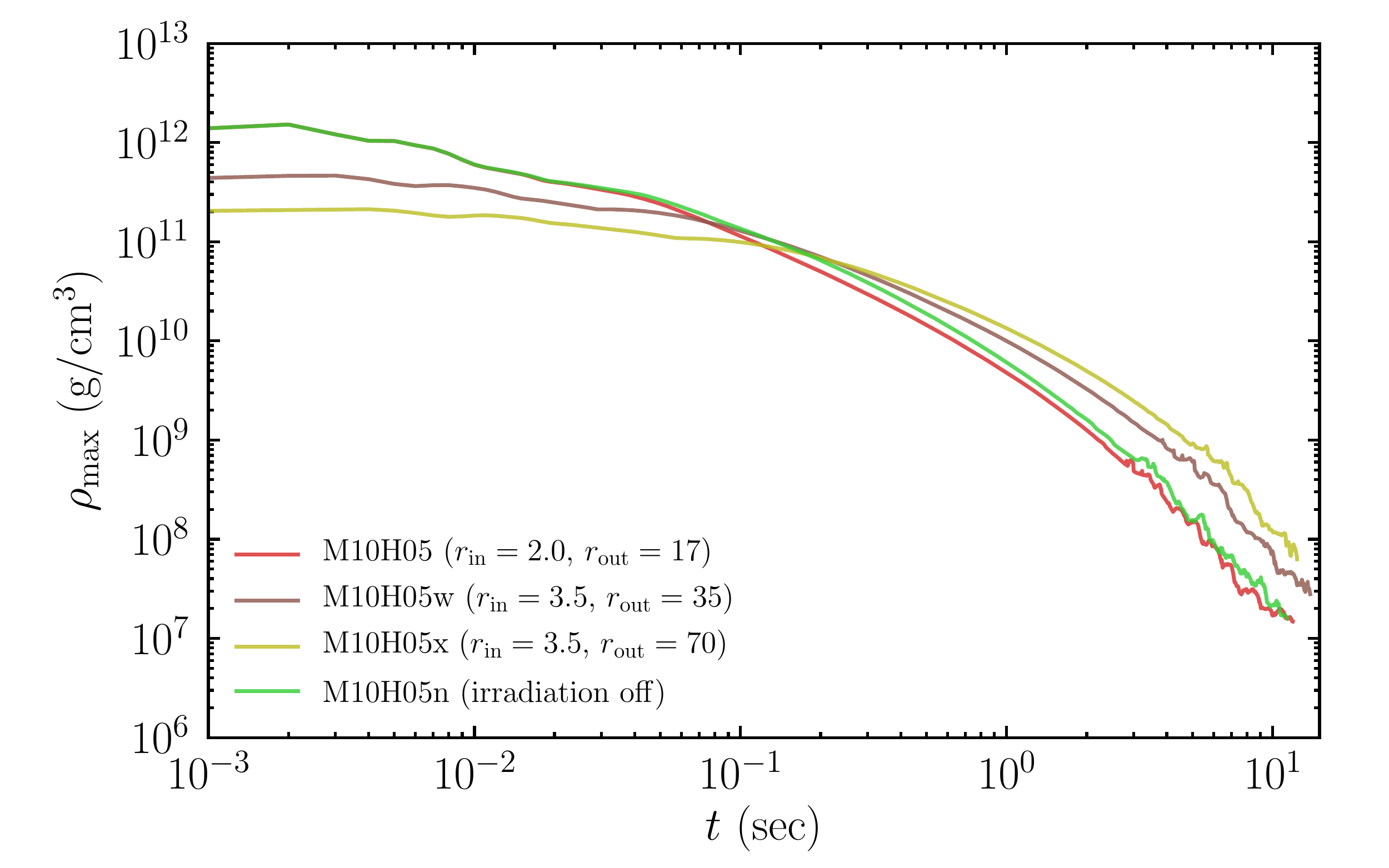} \\
\includegraphics[width=85mm]{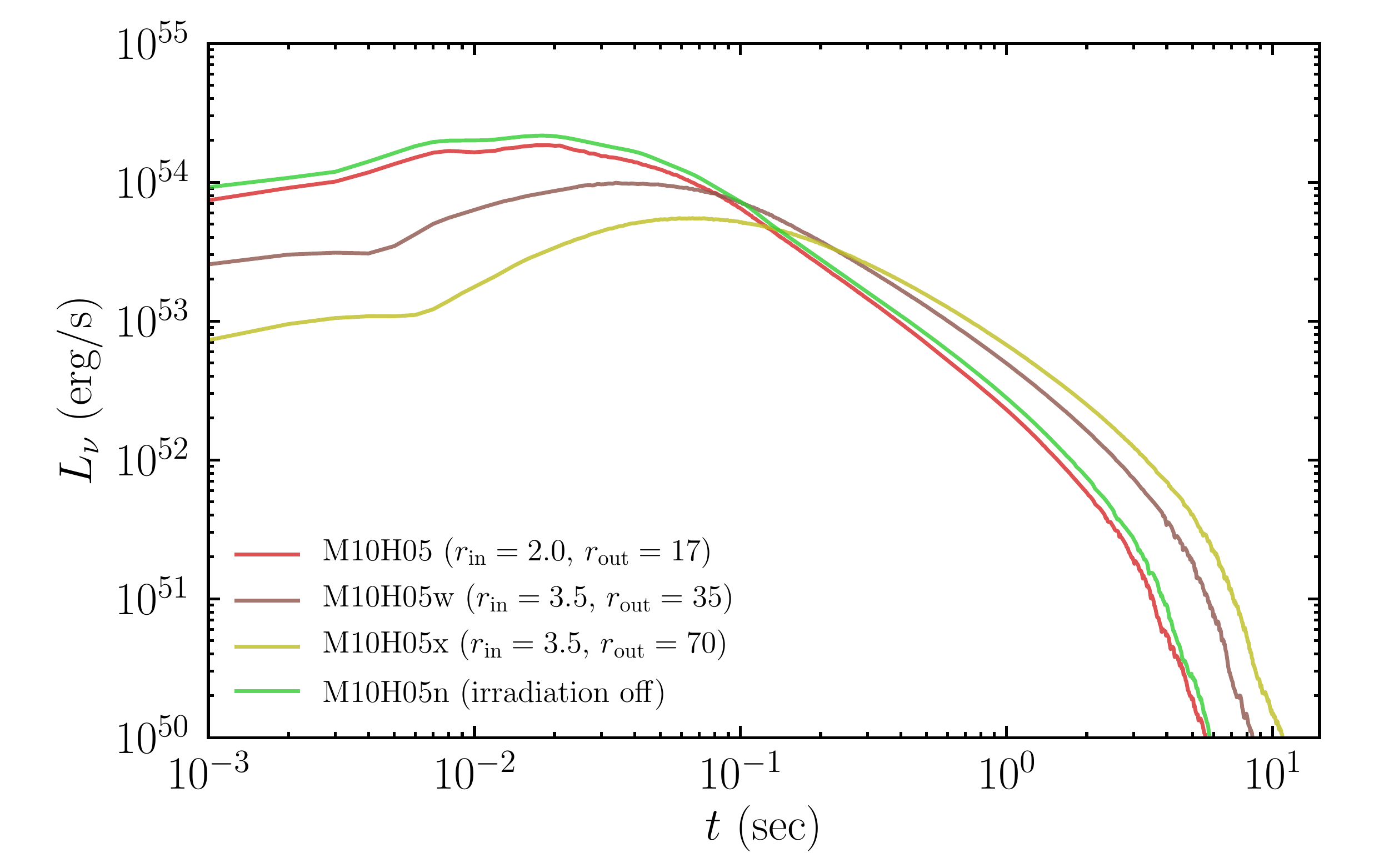} 
\includegraphics[width=85mm]{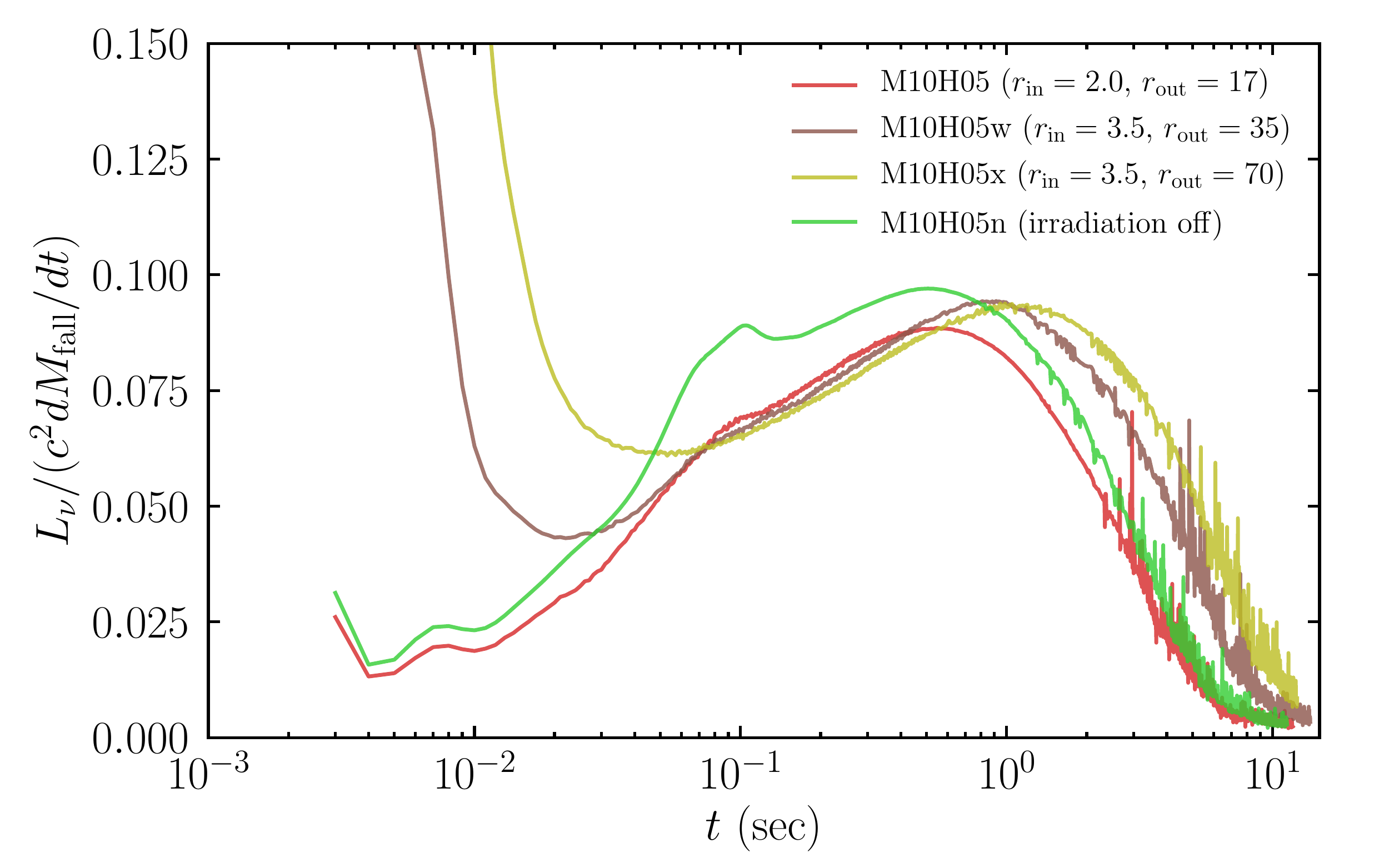} 
\caption{The same as Fig.~\ref{fig5} but for the system of $M_{\rm
    disk}=3M_\odot$ and $M_{\rm BH}=10M_\odot$ with and without 
neutrino irradiation (models M10H05 and M10H05n) and for
initially large disk radii (models M10H05w and M10H05x). 
\label{fig8}}
\end{figure*}

\begin{figure*}[t]
\includegraphics[width=85mm]{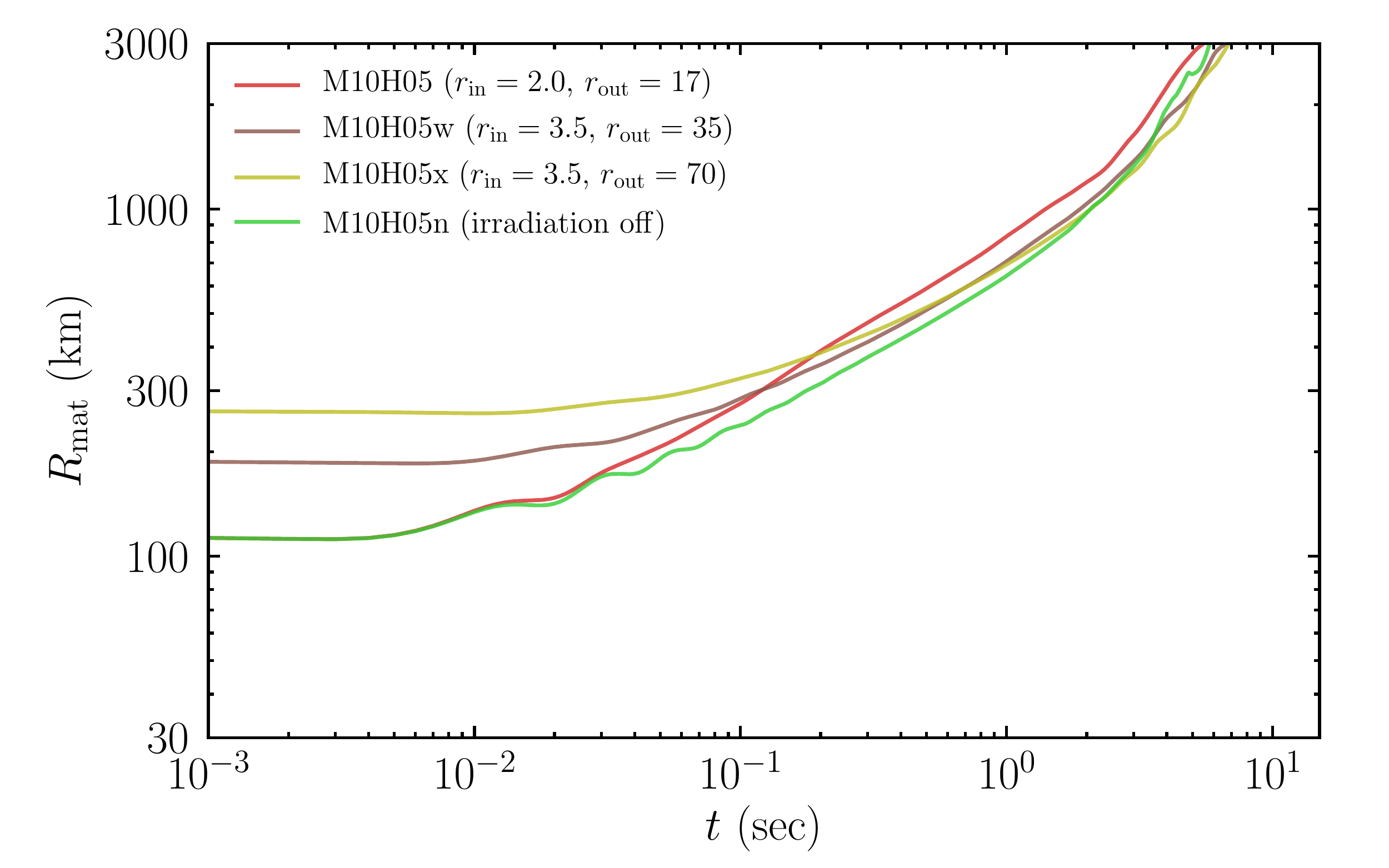} 
\includegraphics[width=85mm]{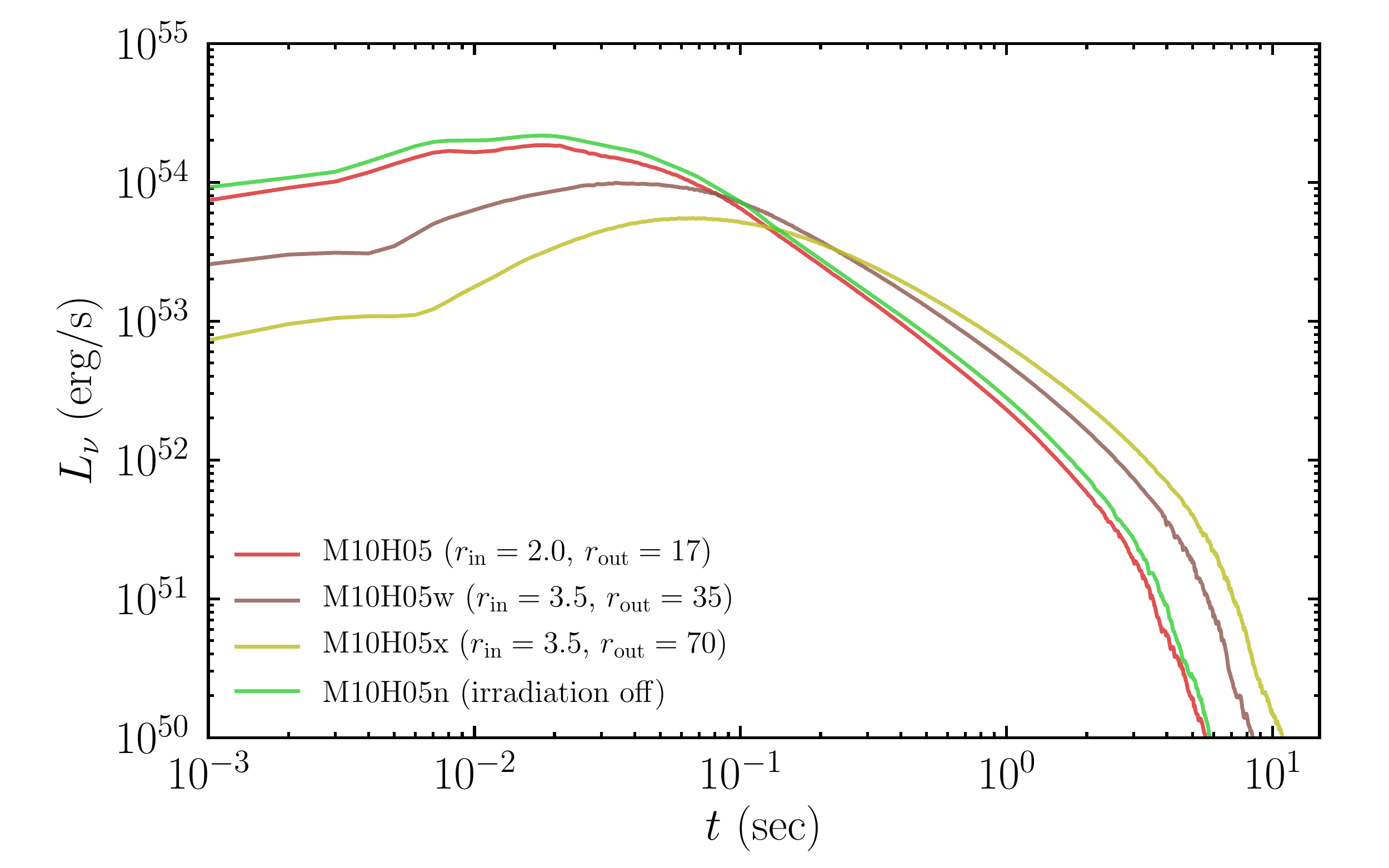} \\
\includegraphics[width=85mm]{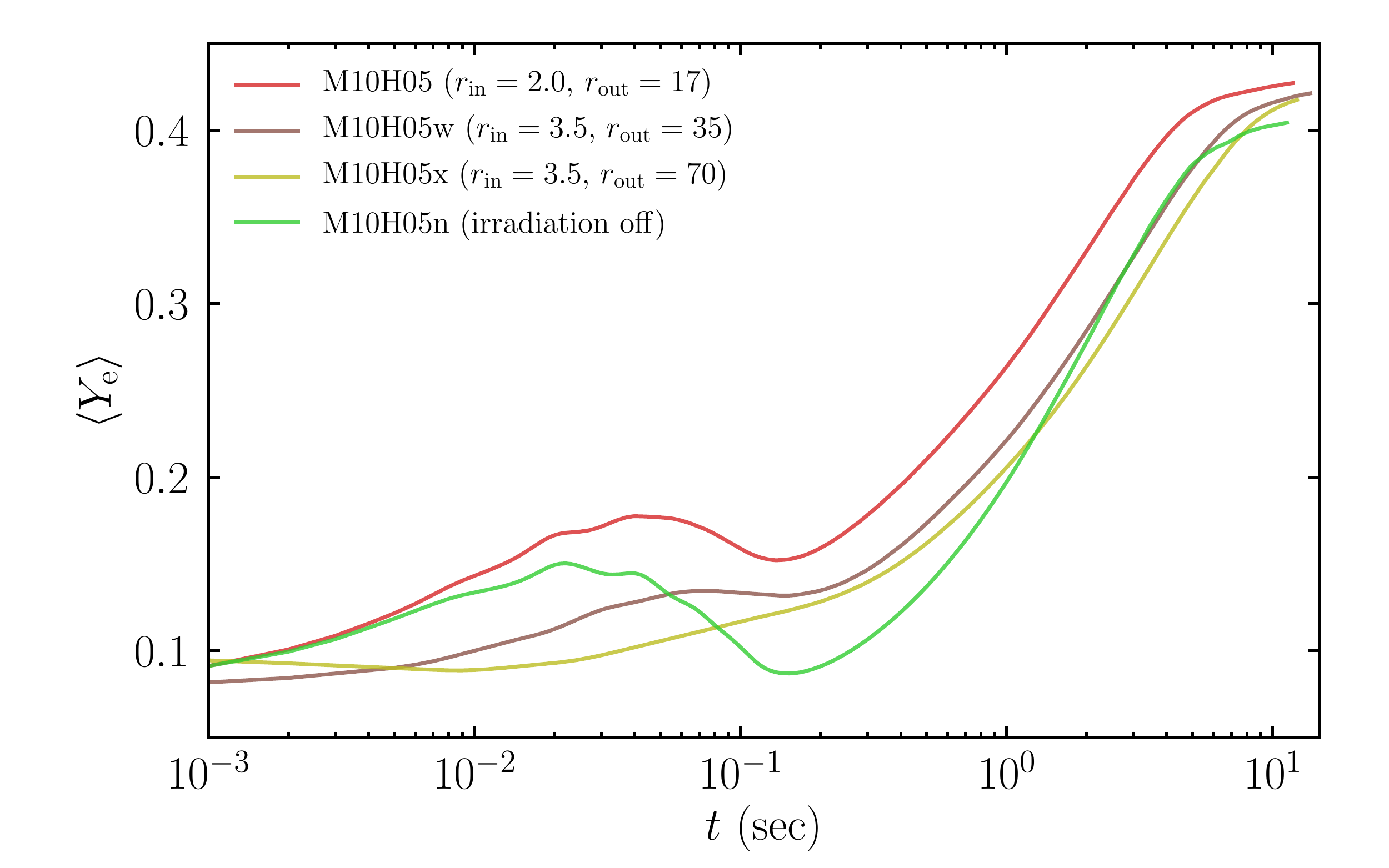} 
\includegraphics[width=85mm]{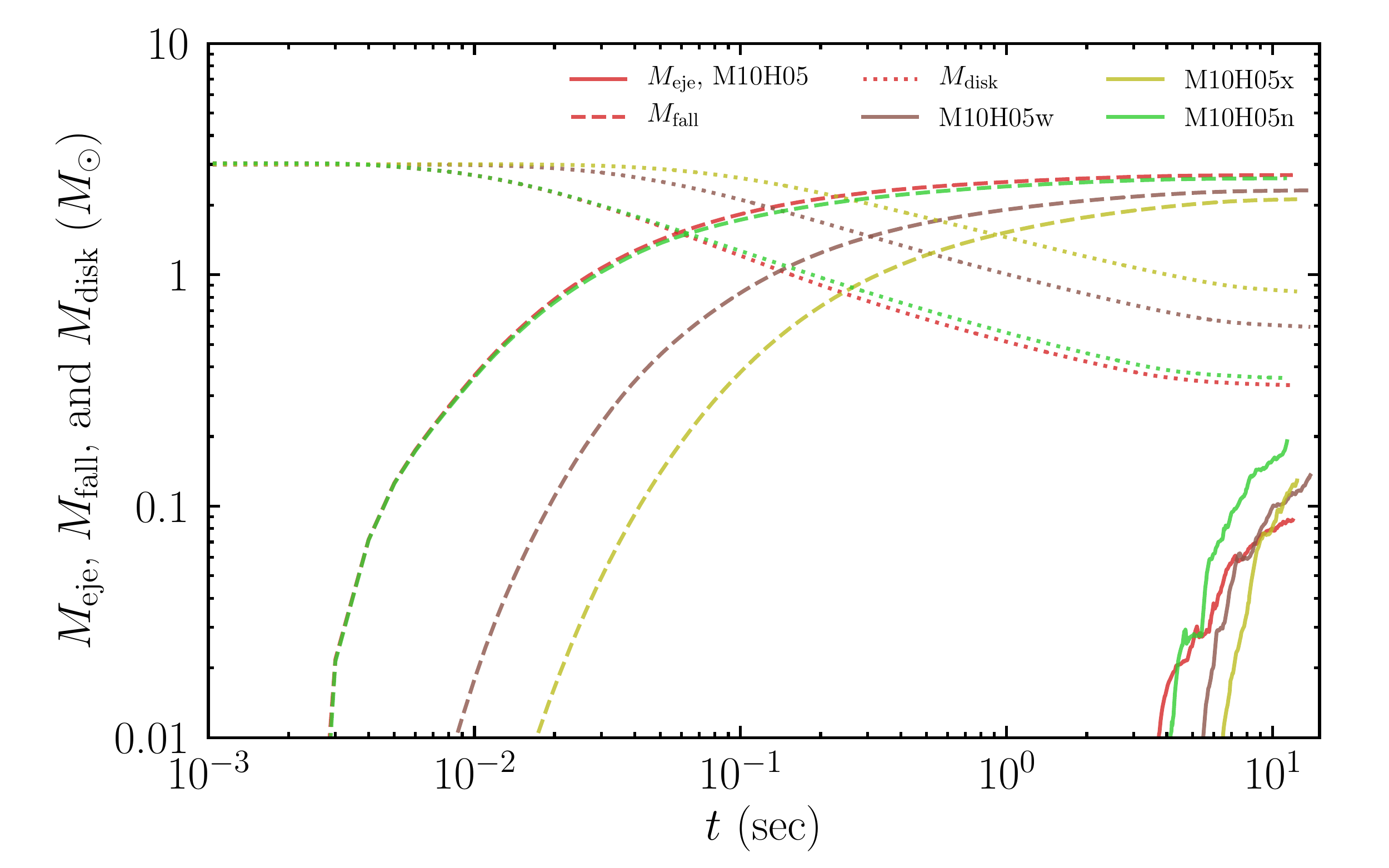} 
\caption{The same as Fig.~\ref{fig6} but for the system with $M_{\rm
    disk}=3M_\odot$ and $M_{\rm BH}=10M_\odot$ with and without 
neutrino irradiation (models M10H05 and M10H05n) and for an 
initially large disk radius (models M10H05w and M10H05x). 
\label{fig9}}
\end{figure*}

Next we summarize the viscous evolution of the disk for the case of
$M_{\rm disk} \approx 3M_\odot$. In this subsection, we focus on
models M04H05, M06H05, M10H05, M10H10, and M10H15. 

As in the low-mass disk cases, in the early stage of the evolution
of the system (in $t \alt 100$\,ms; see the bottom panels of
Figs.~\ref{fig4} and \ref{fig6}), $\sim 60$\% of the rest mass of the
disk falls into the black hole due to the viscous angular momentum
transport processes, and subsequently, the remaining part of the disk
component is evolved by the viscous heating and angular momentum
transport processes.\footnote{We note that for the high-mass disk 
  models, the density and velocity profiles are highly disturbed in
  the initial transition stage as mentioned in Sec.~\ref{sec3-1}. As a
  result, the average velocity and electron fraction oscillate due to
  an artifact of the initial condition. This effect is in particular
  conspicuous for $M_{\rm BH}=4$ and $6M_\odot$ for which the
  self-gravity of the disk plays a significant role for the evolution
  of the system (see Fig.~\ref{fig6}). However, after this transition,
  the disk is evolved quasi-steadily by viscous-hydrodynamics
  process.}  Quantitatively, the evolution process depends on the
black-hole mass, viscous coefficient, and initial extent of the disk.
This subsection focuses on the dependence on the black-hole mass and
viscous coefficient.

Figures~\ref{fig5} and \ref{fig6} display the evolution of several
quantities of the disk for models M04H05, M06H05, M10H05, M10H10, and
M10H15, i.e., for $\alpha_{\nu}$ $= 0.05$ with $M_{\rm BH,0}=4$, 6, and
$10M_\odot$ and for $M_{\rm BH,0}=10M_\odot$ with $\alpha_\nu=0.10$
and 0.15.  For these cases, the maximum temperature and density are by
a factor of $\sim 3$ higher than those for the low-mass disk cases
(compare Figs.~\ref{fig1} and \ref{fig5}). Associated with this
situation, the neutrino luminosity reaches $\agt 10^{54}\,{\rm
  erg/s}$, which is more than 10 times as high as that for the low-mass 
disk cases. We note that the emissivity of neutrinos via the
capture process of electrons and positrons by nucleons is
approximately proportional to $T^6$.  Thus, the neutrino emissivity is
not directly reflected in the luminosity (this is also observed from a
low-efficiency of the neutrino luminosity in the early stage of $t
\alt 10$\,ms: see the bottom-right panel of Fig.~\ref{fig5}).  The
reason for this is that the disk is so dense and hot that the emitted
neutrinos are trapped~\cite{NDAF}. However, the total energy carried
away by neutrinos is still high, $\sim 10^{53}$\,erg\,($\approx
0.05M_\odot c^2$), irrespective of the black-hole mass. Thus, more
than 1\% of the rest-mass energy of the disk is carried away by
neutrinos. All these properties are found irrespective of the
black-hole mass and the viscous coefficient.

The bottom-right panel of Fig.~\ref{fig5} shows that the efficiency of
the neutrino emission depends strongly on the black-hole mass.  As we
discussed in Sec.~\ref{sec3-2}, the primary reason for this may be
that for the lower-mass black-hole cases, the temperature and density
of the disk are higher than for the higher-mass cases, and as a
result, the neutrino luminosity and the efficiency of the neutrino
emission could be also higher. However, Fig.~\ref{fig5} shows that the
neutrino luminosity does not depend strongly on the black-hole mass,
due to the trapping of neutrinos. Our alternative interpretation for
this dependence of the efficiency on the black-hole mass is that the
black-hole spin is reflected. As we showed in Fig.~\ref{fig4}, the
dimensionless spin of the black holes, $\chi$, is higher for the
lower-mass black holes and close to unity for $M_{\rm BH}=4$ and
$6M_\odot$.  For such rapidly spinning black holes, the horizon radius
steeply decreases as $\chi$ approaches unity, and thus, the rest-mass
accretion rates decrease as a result of the decrease in the horizon
area. Simultaneously, as $\chi$ approaches unity, the gravitational
potential around the black hole becomes deeper, and hence, the efficiency
for releasing the gravitational potential energy is steeply
enhanced~\cite{ST83}. Thus, it is reasonable that the neutrino
emission efficiency steeply increases with the decrease of the
black-hole mass (i.e., with the increase of the resulting
dimensionless spin of the black hole).


The maximum neutrino emission efficiency depends very weakly on the
viscous coefficient (compare the results of M10H05, M10H10, and
M10H15). This agrees with the result in our previous
paper~\cite{Fujiba20}.  The timescale to reach the maximum is
approximately proportional to $\alpha_\nu^{-1}$, reflecting the
viscous timescale.  Overall, the viscous coefficient basically changes
the timescale for the evolution of the disk. The curves of the disk
quantities like the neutrino luminosity, maximum temperature, and
maximum density become similar if we plot these quantities as
functions of $\alpha_\nu t$.

The maximum temperature and the neutrino luminosity are maintained to
be higher than 2\,MeV$/k$ and $10^{51}\,{\rm erg/s}$ until $t \sim
2$--10\,s. Here, this timescale depends on the black-hole mass and
viscous coefficient. The neutrino cooling timescale is maintained to
be as long as the viscous heating timescale for $t \sim
1$--5\,s. During this NDAF stage~\cite{NDAF}, the thermal energy
generated by the viscous heating is consumed by the neutrino cooling,
and after this time, the mass ejection sets in. The onset of the mass
ejection is signaled again clearly by the accelerated increase of
$\langle s/k \rangle$ (see the top-right panel of Fig.~\ref{fig6}).

As found from the curves of the ejecta mass $M_{\rm eje}$ (see the bottom-right panel of Fig.~\ref{fig6}) as well as of $\langle s/k \rangle$, the mass ejection sets in at $t \sim 1$--5\,s, which is later than in the low-mass disk case.
This is due to the higher disk mass and resulting longer-term high neutrino-emissivity stage (compare Figs.~\ref{fig1} and \ref{fig5}).
Associated with the delay of the mass ejection, the electron fraction is enhanced during the longer-term disk evolution for the higher-mass disk case. The bottom-left panel of Fig.~\ref{fig6} shows that irrespective of the black-hole mass, the final average value of $Y_e$ exceeds $0.4$ for $\alpha_\nu=0.05$, which is higher than that for the low-mass disk case.
Therefore, we conclude that the ejecta from the system of a stellar-mass black hole and a high-mass disk with mass $\agt M_\odot$ has high values of the electron fraction with its average $\sim 0.4$, if the mass ejection proceeds via the viscous effects.

For the larger viscous coefficient, the average value of $Y_e$ is higher before the mass ejection sets in ($t\lesssim 1$\,s) because of the stronger viscous effect that results in the faster decrease of the electron degeneracy and higher equilibrium value of $Y_e$ by the electron/positron capture. 
However, its final average value is lower for the larger viscous coefficients. 
The reason for this is that for the larger viscous coefficients, 
the viscous timescale of the disk is shorter, and thus, 
the freezeout of the electron/positron capture occurs earlier (see also Ref.~\cite{Fujiba20}).
For a very high value of $\alpha_\nu$ as 0.15, the final average value of $Y_e$ can be as low as $\sim 0.35$.
Thus, in the presence of an efficient mass-ejection mechanism which we do not take into account in the present work, the value of $Y_e$ could be also decreased.


Figure~\ref{fig7} displays the evolution for the average values of $Y_e$ and velocity of the ejecta component.
Reflecting the evolution of $Y_e$ in the disk during its viscous expansion, the average value of $Y_e$ for the ejecta becomes higher than that for the low-mass disk case: approximately 0.35--0.45.
Here, the value depends on the black-hole mass and viscous coefficient.
As in the low-mass disk cases, the final average value of $Y_e$ is higher for the higher black-hole mass while it becomes lower for the larger viscous coefficients.

The final average velocity of the ejecta, $V_\mathrm{ej}$, is $\approx 0.05$--$0.07c$ 
and, for a given value of $\alpha_\nu$, it 
depends very weakly on the black-hole mass as in the low-mass disk case 
(compare the right panels of Figs~\ref{fig3} and \ref{fig7}). 
The value of $V_\mathrm{ej}$ is larger for larger viscous parameters 
because of the stronger viscous effect (angular-momentum transport and 
viscous heating effects).


As in the low-mass disk case, after the viscous evolution of the disk and subsequent mass ejection, the system relaxes to a state composed of a spinning black hole surrounded by a geometrically-thick torus and a narrow funnel in the vicinity of the rotation axis.
In particular, for the high-mass disk case, the resulting black hole is rapidly spinning with the dimensionless spin $\agt 0.9$.
Such an outcome would be quite suitable for generating gamma-ray bursts, i.e., for driving a collimated jet accelerated by thermal energy generated from the deep gravitational potential near the spinning black hole (see the discussion in Sec.~\ref{sec3-7}).

\subsection{Dependence on the initial disk model and neutrino irradiation}\label{sec3-5}

This subsection briefly shows the dependence of the disk evolution on the initial disk extent and neutrino irradiation referring to the models with $M_{\rm disk} \approx 3M_\odot$ and $M_{\rm BH}=10M_\odot$.
Figures~\ref{fig8} and \ref{fig9} compare the results for models M10H05, M10H05w, M10H05x, 
and M10H05n.
For the larger initial disk radii (compare the results of models M10H05w and M10H05x with those of M10H05), the evolution process is obviously delayed because the viscous timescale becomes longer for the larger initial disk radii (see Eq.~\eqref{tvis}).
Specifically, for the larger initial disk radii, the peak maximum temperature and neutrino luminosity are reached later (at $t\sim 50$--100\,ms) and associated with this, the neutrino luminosity remains higher for the later stage.
This fact is also inferred from the bottom-right panel of Fig.~\ref{fig4}, which shows that the peak of the rest-mass accretion is reached at $t \sim 30$\,ms and 60\,ms for M10H05w and M10H05x, respectively.
As a result of the delay in the evolution, the disk expansion timescale becomes longer and the mass ejection sets in slightly later.
In spite of the difference in the onset time of the mass ejection, the final average value of $Y_e$ depends only weakly on the initial disk radius.
This would be due to the fact that the physical condition at the onset of the mass ejection such as temperature, density, and neutrino luminosity depends only weakly on the initial disk radius.
One significant difference among the models is found in the total amount of the disk mass that swallowed into the black hole.
For the larger initial disk radii, the fraction of the disk matter that falls into the black hole is smaller naturally, and as a result, the fraction of the ejecta can be increased.
However, besides this quantitative difference, the initial disk radius does not lead to the remarkable difference in the disk evolution and the property of the ejecta; e.g., for models M10H05w and M10H05x, the final average values of $Y_e$ and velocity are as high as those for model M10H05.

By comparing the results for models M10H05 and M10H05n, we find that the effect of the neutrino irradiation modifies the disk evolution process quantitatively.
For example, in the absence of the neutrino irradiation, the neutrino emissivity is slightly enhanced and as a result, the onset of the disk expansion is delayed while the final average value of $Y_e$ is decreased due to the omission of the irradiation.
However, these effects do not substantially change the disk evolution.

The properties of the ejecta reflect the disk evolution.  
In the absence of the neutrino absorption/irradiation, the asymptotic average value of $Y_e$ is $\approx 0.41$ which is lower than that in the presence the neutrino absorption/irradiation (see Fig.~\ref{fig9}).
Thus, the neutrino absorption effect plays a role for quantitatively determining the electron fraction of the ejecta.
In the absence of this neutrino effect, the asymptotic ejecta velocity is slightly smaller ($\approx 0.05c$) than in its presence ($\approx 0.06c$), reflecting the absence of the neutrino absorption/irradiation (i.e., neutrino radiation pressure).

\begin{table}
\caption{Average quantities for tracer particles.  $\langle Y_e\rangle$, $\langle s/k\rangle$, $\langle t_\mathrm{exp}\rangle$, and $\langle C_r\rangle$ are the mass-weighted average values of the electron fraction and entropy per baryon at $T=5$\,GK, expansion timescale defined by $t_\mathrm{exp} = t(T=2.5\, \mathrm{GK})-t(T=5\, \mathrm{GK})$, and $C_r$ defined by Eq.~\eqref{eq:cr}, respectively.}
\label{tab:average}
\begin{ruledtabular}
\begin{tabular}{lcccc}
Model    & $\langle Y_e\rangle$ &$\langle s/k\rangle$& $\langle t_\mathrm{exp}\rangle$ (s) & $\langle C_r\rangle$\\
\hline
K8       &  0.343 & 18.6  & 0.16 &  0.063\\
\hline
M04L05   &  0.354 & 21.0  & 0.12 &  0.069\\
M06L05   &  0.379 & 27.5  & 0.27 &  0.081\\
M10L05   &  0.369 & 35.3  & 0.22 &  0.085\\
\hline
M04H05   &  0.439 & 19.8  & 0.74 &  0.027\\
M06H05   &  0.445 & 20.6  & 0.75 &  0.024\\
M10H05   &  0.453 & 23.5  & 0.78 &  0.029\\
M10H10   &  0.404 & 22.4  & 1.02 &  0.034\\
M10H15   &  0.377 & 25.6  & 0.64 &  0.043\\
\hline
M10H05w  &  0.445 & 19.1  & 0.93 &  0.021\\
M10H05x  &  0.441 & 19.7  & 0.81 &  0.021\\
M10H05n  &  0.425 & 21.1  & 0.79 &  0.030\\
\end{tabular}
\end{ruledtabular}
\end{table}

\begin{table*}
\caption{\label{tab:nucleosynthesis}
Nucleosynthesis results. Described are the model name, ejecta mass at the end of simulation, estimated total ejecta mass, mass of $^{48}$Ca, mass of $^{56}$Ni, lanthanide mass fraction, logarithmic value of $p(Z, A)$ defined by Eq.~(\ref{eq:opro}) at the maximum, isotope name with the maximum of $p(Z, A)$, maximum fraction of events represented by the model with respect to that of core-collapse supernovae, and maximum Galactic rate of events represented by the model.}
\begin{ruledtabular}
\begin{tabular}{lccccccccc}
Model   & $M_\mathrm{ej}$ ($M_\odot$) & $M_\mathrm{ej,tot}$ ($M_\odot$) & $^{48}$Ca ($M_\odot$) & $^{56}$Ni ($M_\odot$) & $X_\mathrm{lan}$ & $\log p_\mathrm{max}$ 
& Isotope & $f_\mathrm{max}$ ($10^{-3}$) & $R_\mathrm{max}$ (Myr$^{-1}$) \\
\hline
K8      & 0.0088 & 0.015 & 0.00012  & 0.0000052 & $1.9\times 10^{-5}$  & 7.0 & $^{87}$Rb & 1.7  & 39 \\
\hline
M04L05  & 0.011  & 0.016 & 0.00012  & 0.000050 & $6.8\times 10^{-5}$  & 7.0 & $^{87}$Rb  & 1.6   & 37  \\
M06L05  & 0.0087 & 0.014 & 0.00016  & 0.000028 & $2.4\times 10^{-4}$  & 6.9 & $^{87}$Rb  & 1.9   & 45  \\
M10L05  & 0.0093 & 0.017 & 0.000097 & 0.000095 & $3.4\times 10^{-4}$  & 7.9 & $^{87}$Rb  & 1.5   & 34  \\
\hline
M04H05  & 0.13   & 0.23  & 0.014    & 0.019    & $1.1\times 10^{-10}$ & 6.7 & $^{82}$Se  & 0.20  & 4.5 \\
M06H05  & 0.11   & 0.13  & 0.0080   & 0.013    & $2.4\times 10^{-10}$ & 6.6 & $^{82}$Se  & 0.39  & 9.0 \\
M10H05  & 0.081  & 0.090 & 0.0029   & 0.011    & $2.0\times 10^{-8}$  & 6.4 & $^{82}$Se  & 0.94  & 22  \\
M10H10  & 0.33   & 0.42  & 0.021    & 0.012    & $3.5\times 10^{-5}$  & 6.8 & $^{82}$Se  & 0.050 & 1.2 \\
M10H15  & 0.47   & 0.49  & 0.013    & 0.014    & $7.1\times 10^{-5}$  & 6.9 & $^{81}$Br  & 0.054 & 1.3 \\
\hline
M10H05w & 0.11   & 0.12  & 0.0051   & 0.016    & $2.0\times 10^{-9}$  & 6.5 & $^{82}$Se  & 0.55  & 13  \\
M10H05x & 0.10   & 0.13  & 0.0075   & 0.018    & $2.2\times 10^{-11}$ & 6.6 & $^{82}$Se  & 0.44  & 10  \\
M10H05n & 0.14   & 0.19  & 0.011    & 0.021    & $5.0\times 10^{-4}$  & 6.6 & $^{132}$Xe & 0.32  & 7.4 \\
\end{tabular}
\end{ruledtabular}
\end{table*}

\subsection{Nucleosynthesis}\label{sec3-6}

\begin{figure}
\includegraphics[width=0.46\textwidth]{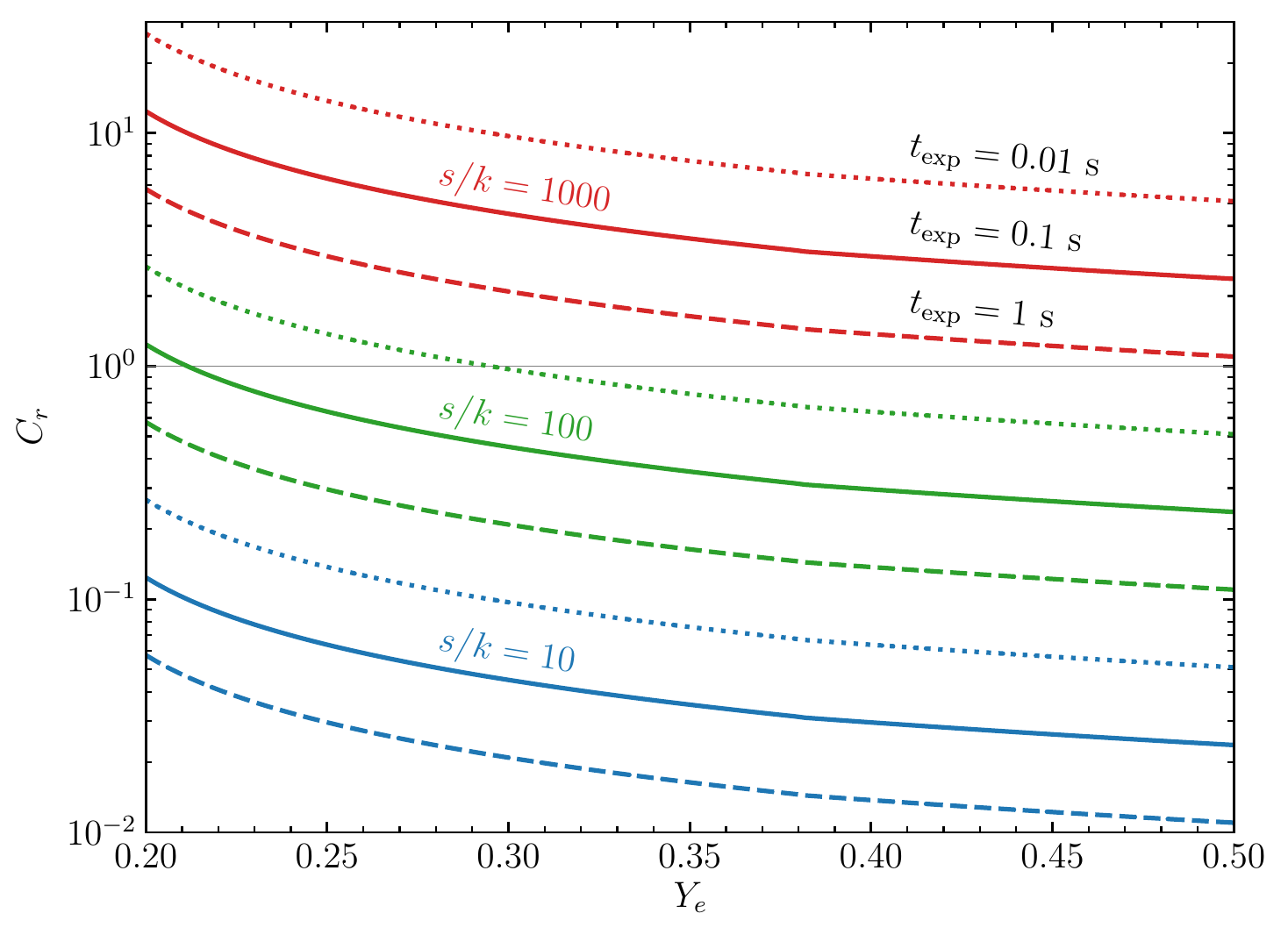}
\caption{Dependence of $C_r$ defined by Eqs.~(\ref{eq:cr})--(\ref{eq:cr2}) on $Y_e$ for fixed values of $s/k = 10$ (blue), 100 (green), and 1000 (red) and $t_\mathrm{exp} = 0.01$ (dotted curves), 0.1 (solid curves), and 1\,s (dashed curves). The gray horizontal line marks $C_r = 1$, above which the $r$-process nuclei with $A = 200$ are expected to be abundantly produced.
\label{fig10}}
\end{figure}

\begin{figure*}
\includegraphics[width=0.46\textwidth]{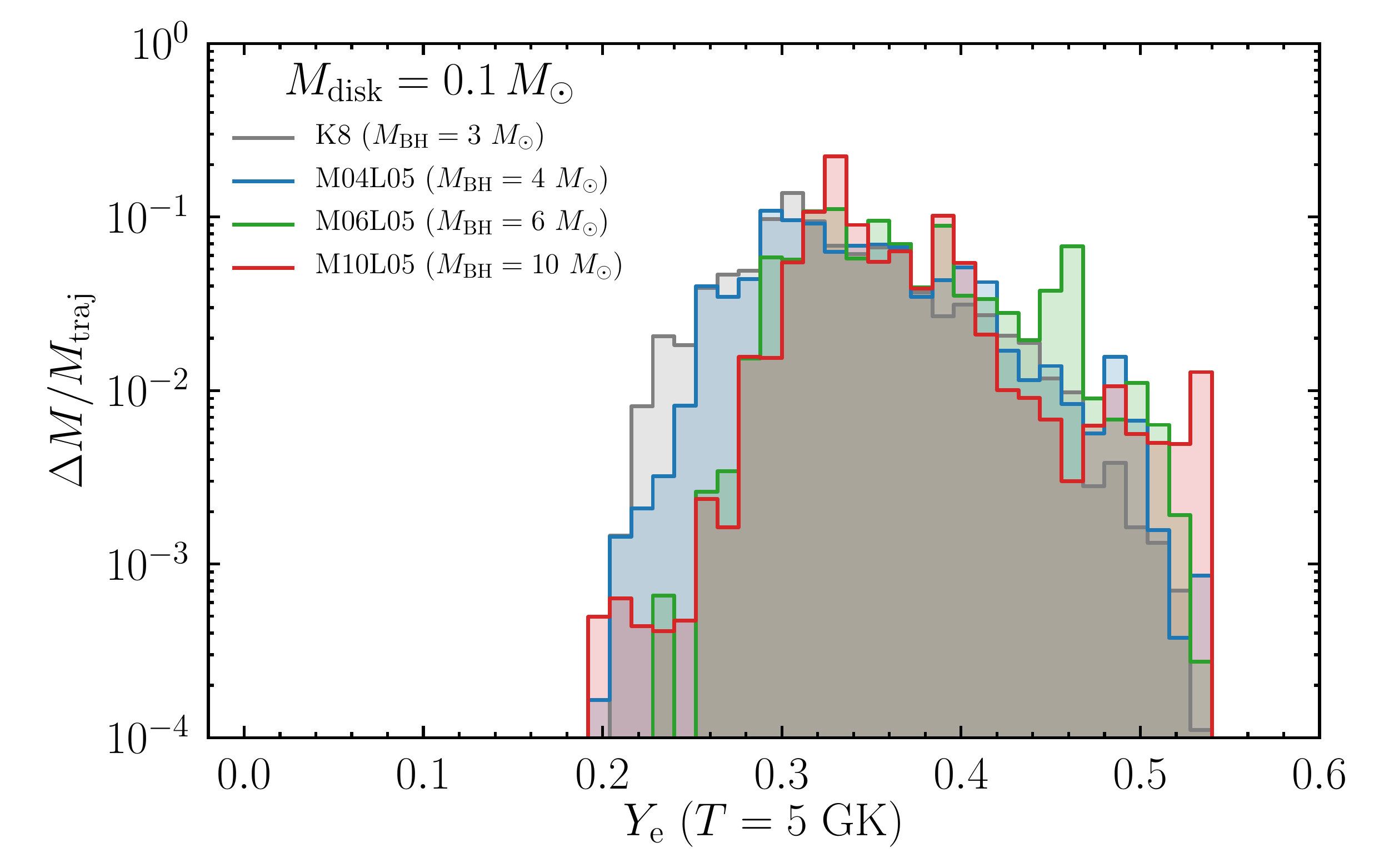}
\includegraphics[width=0.46\textwidth]{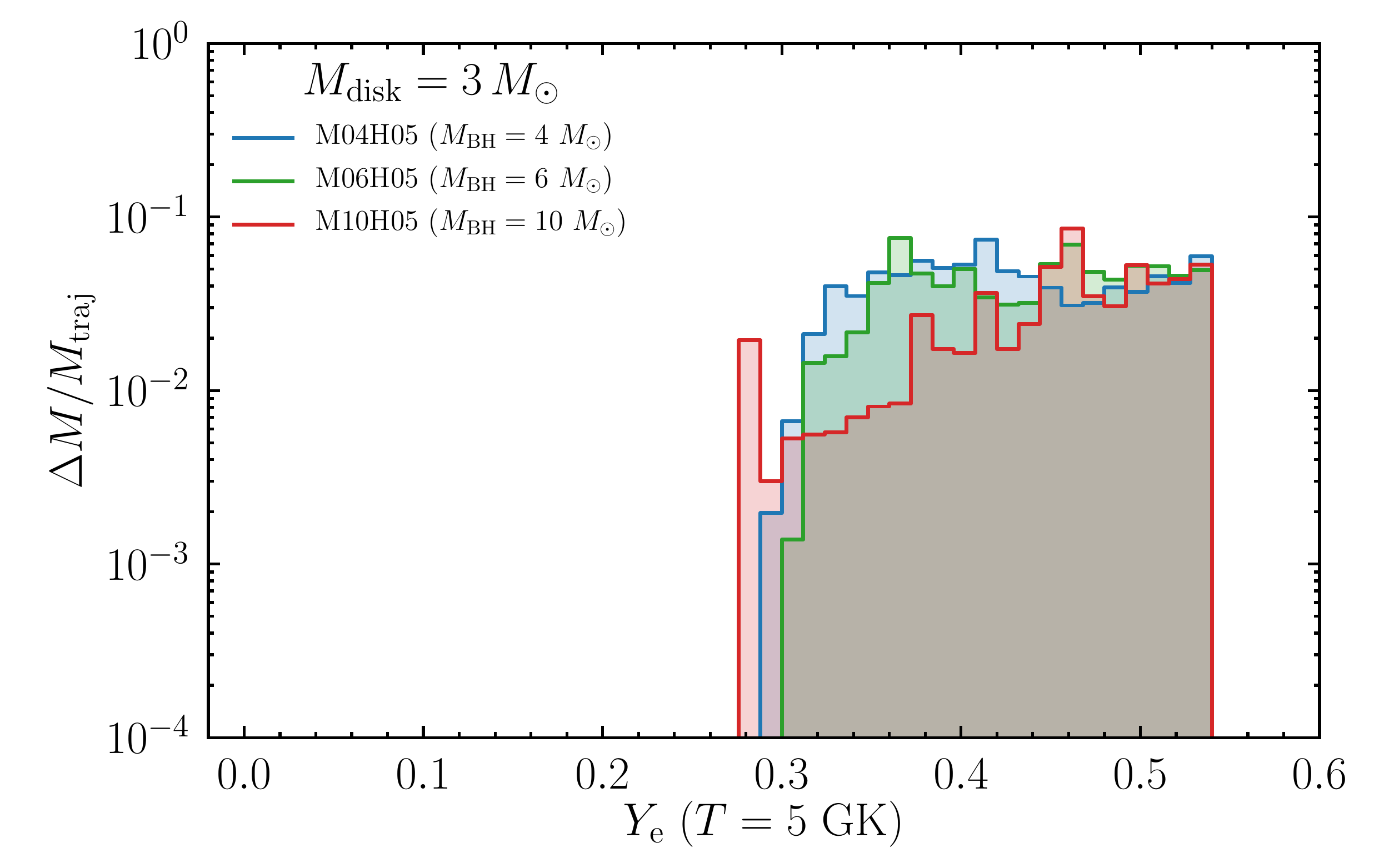}
\includegraphics[width=0.46\textwidth]{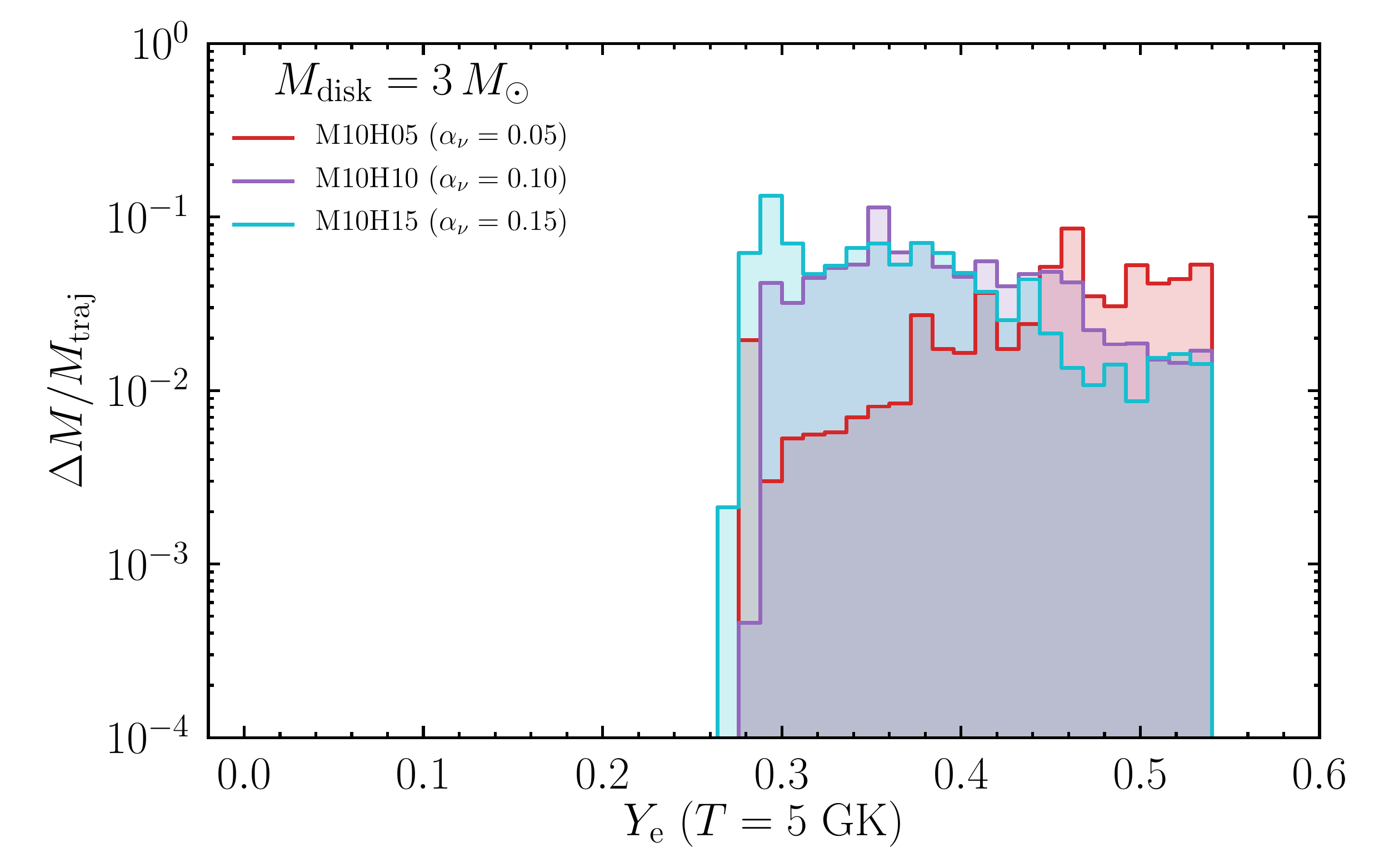}
\includegraphics[width=0.46\textwidth]{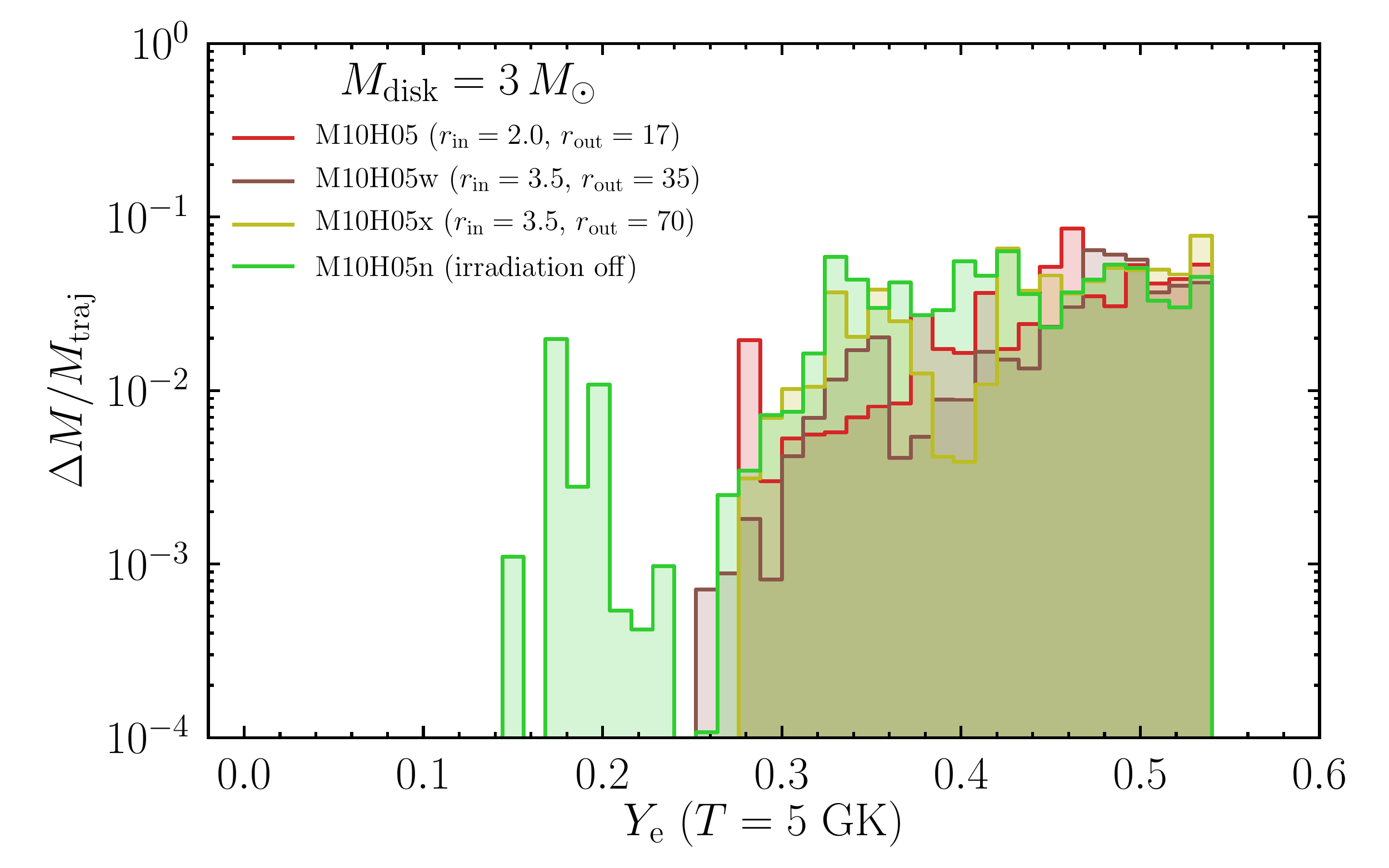}
\caption{Mass histogram (mass fraction) of the ejecta as a function of
  $Y_e$ for models M04L05, M06L05, and M10L05 (top-left), M04H05, M06H05,
  and M10H05 (top-right), M10H05, M10H10, and M10H15 (bottom-left), and  M10H05, M10H05w, and
  M10H05n (bottom-right). Here, the value of $Y_e$ is determined at the time when the
  temperature of each ejecta component decreases to $5 \times 10^9$\,K
  (referred to as 5\,GK). The cutoff of the distribution at the high-$Y_e$ end is due to the fact that we limit the range being $Y_e \le 0.55$ in our simulations.
\label{fig11}}
\end{figure*}

\begin{figure*}
\includegraphics[width=0.46\textwidth]{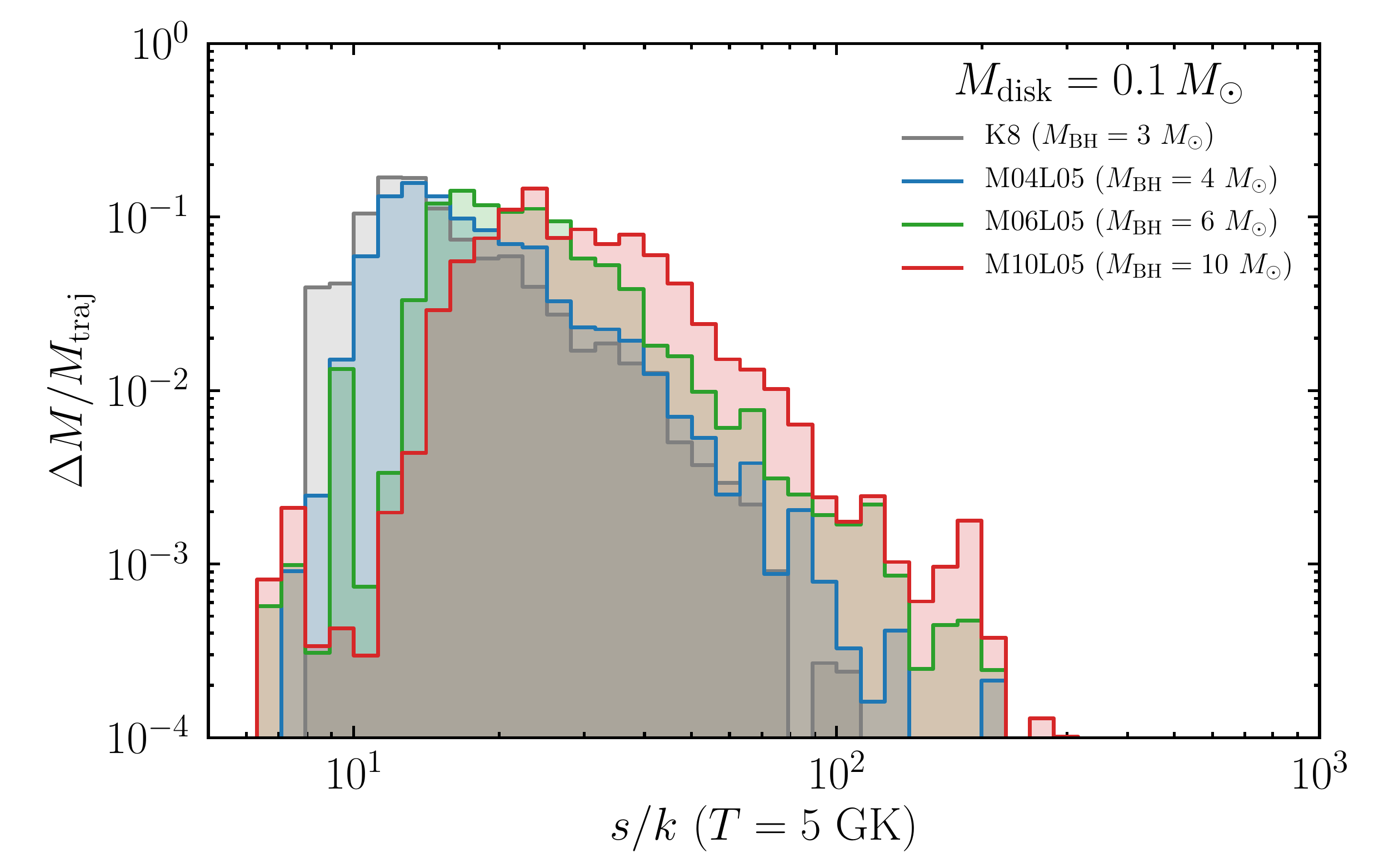}
\includegraphics[width=0.46\textwidth]{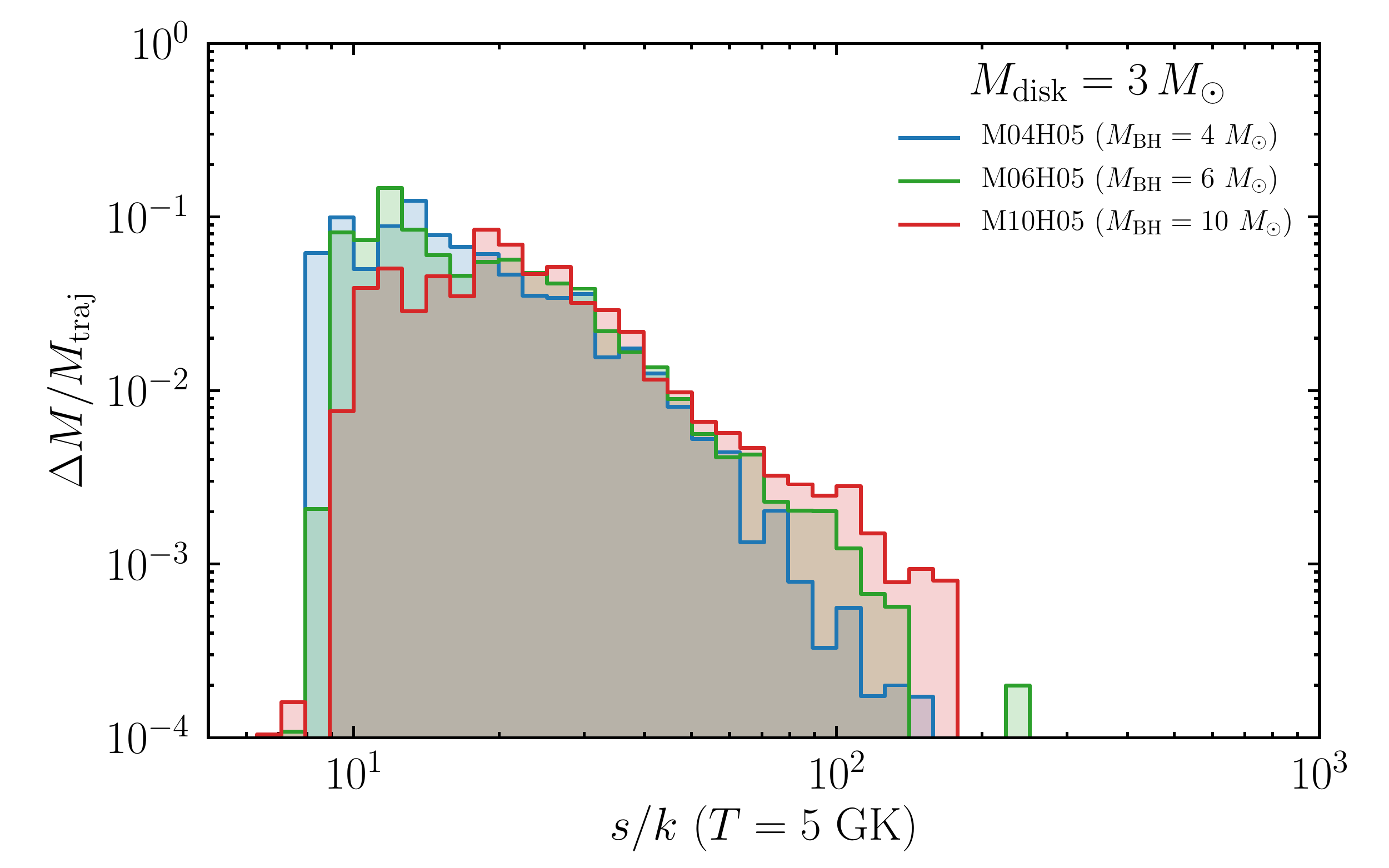}
\includegraphics[width=0.46\textwidth]{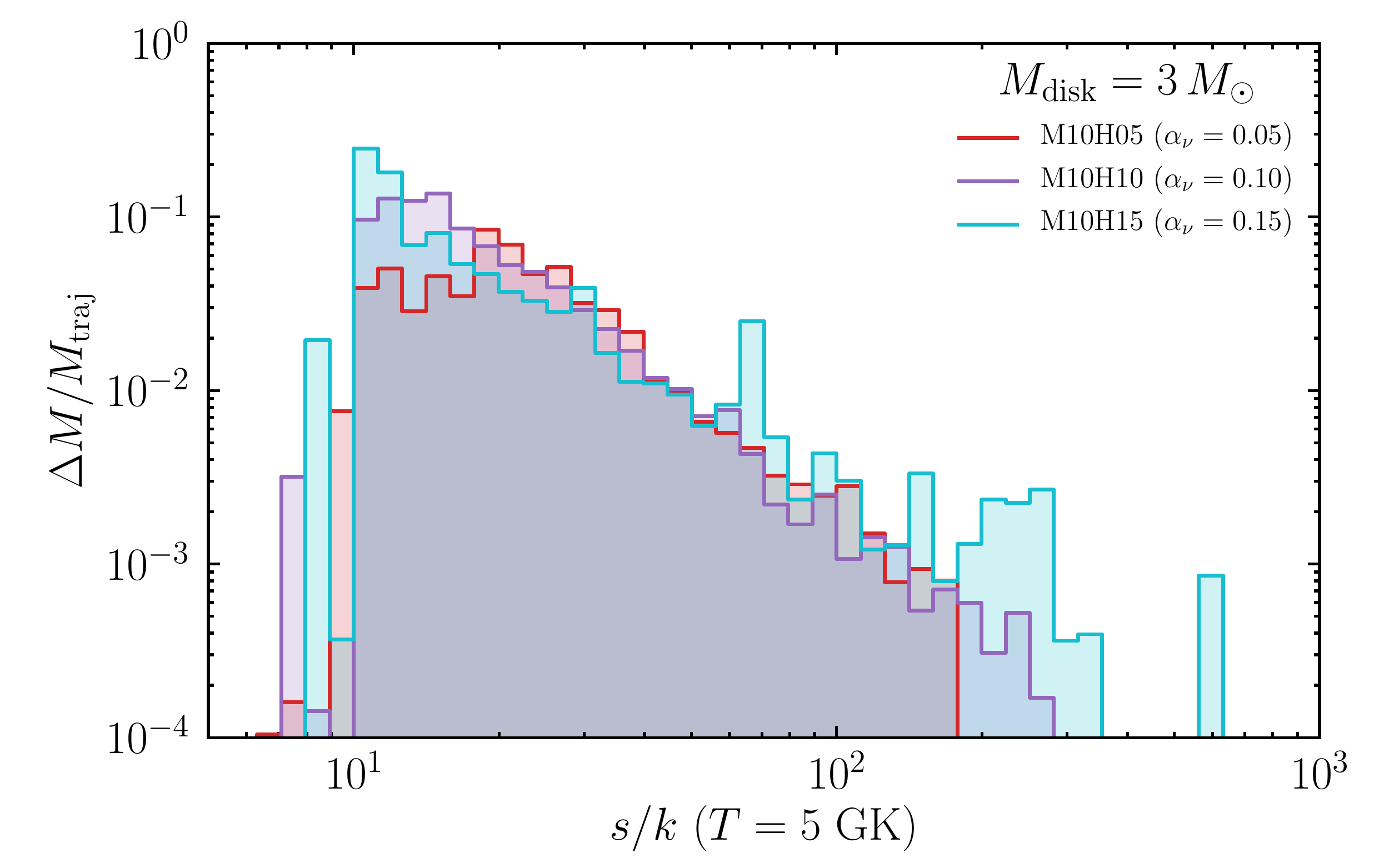}
\includegraphics[width=0.46\textwidth]{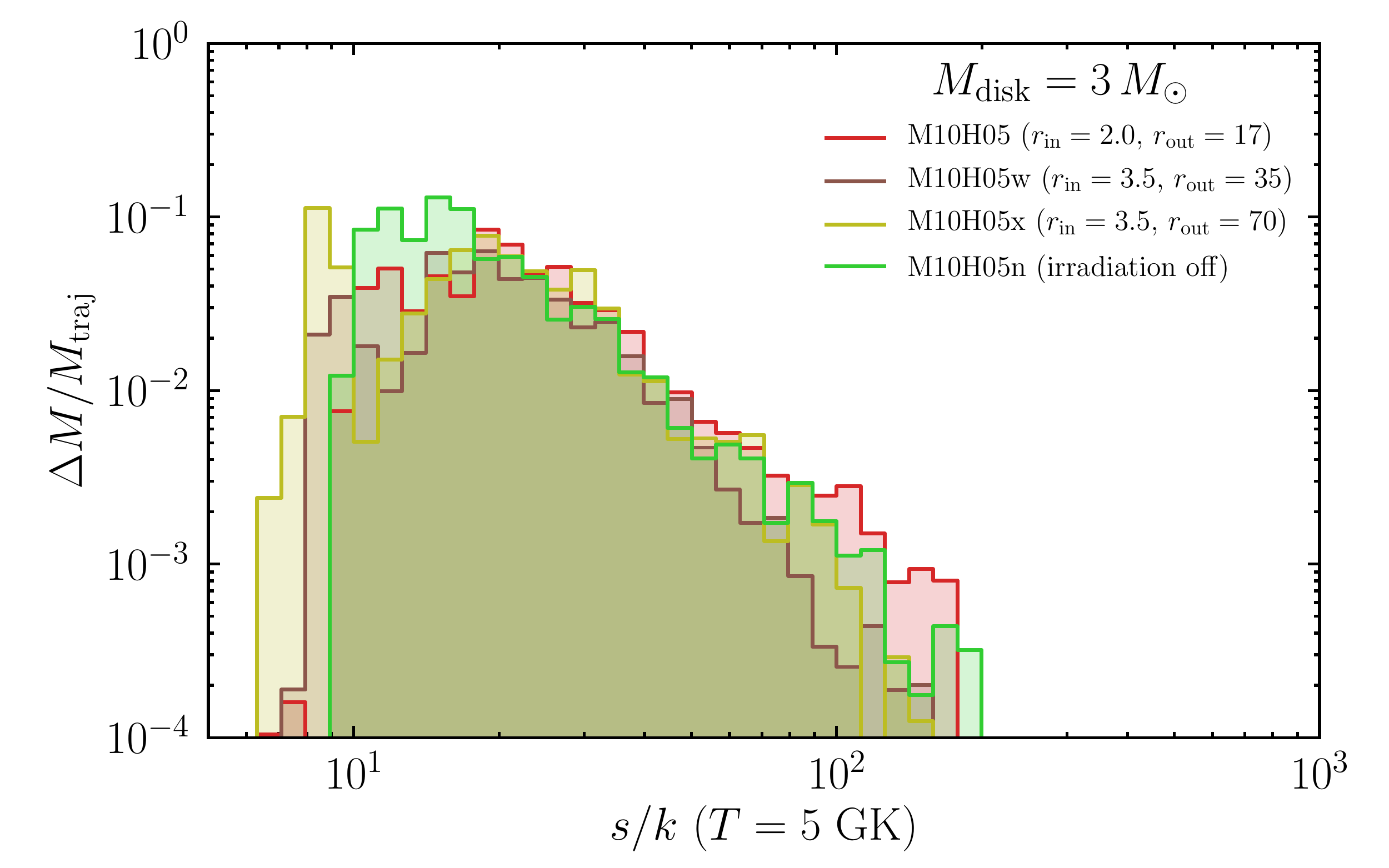}
\caption{The same as Fig.~\ref{fig11} but as a function of the specific entropy of the ejecta. 
\label{fig12}}
\end{figure*}

\begin{figure*}[t]
\includegraphics[width=0.46\textwidth]{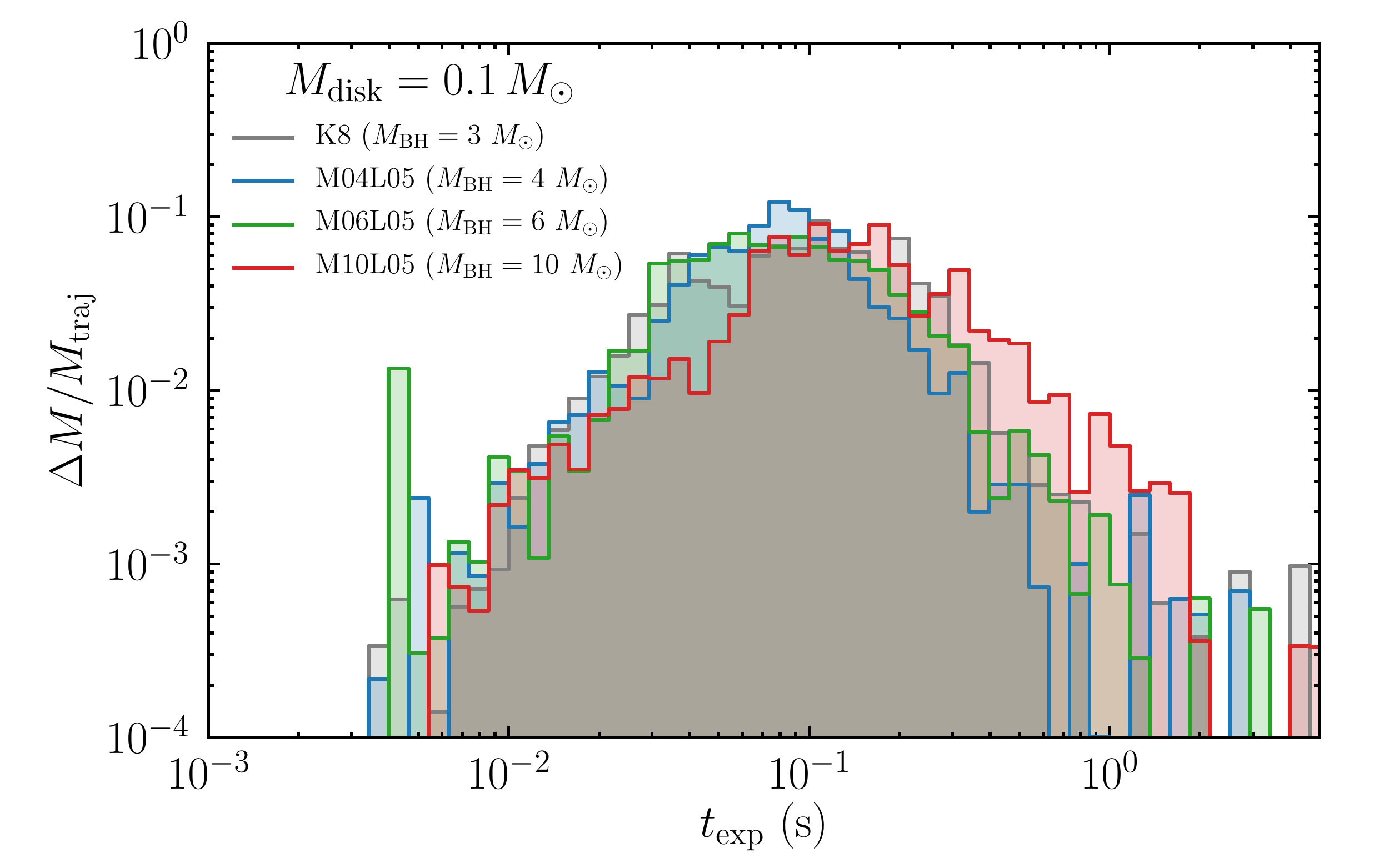}
\includegraphics[width=0.46\textwidth]{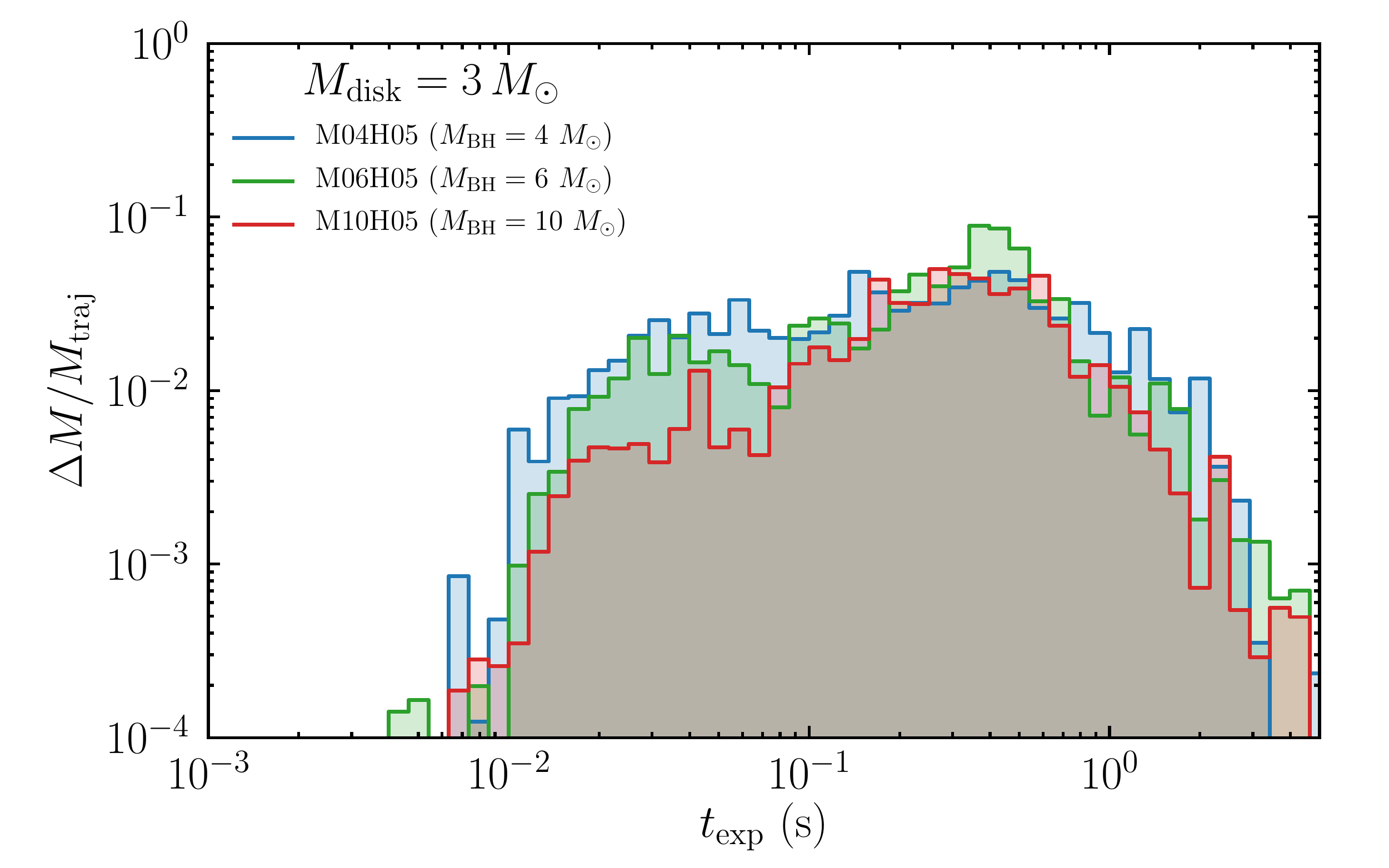}
\includegraphics[width=0.46\textwidth]{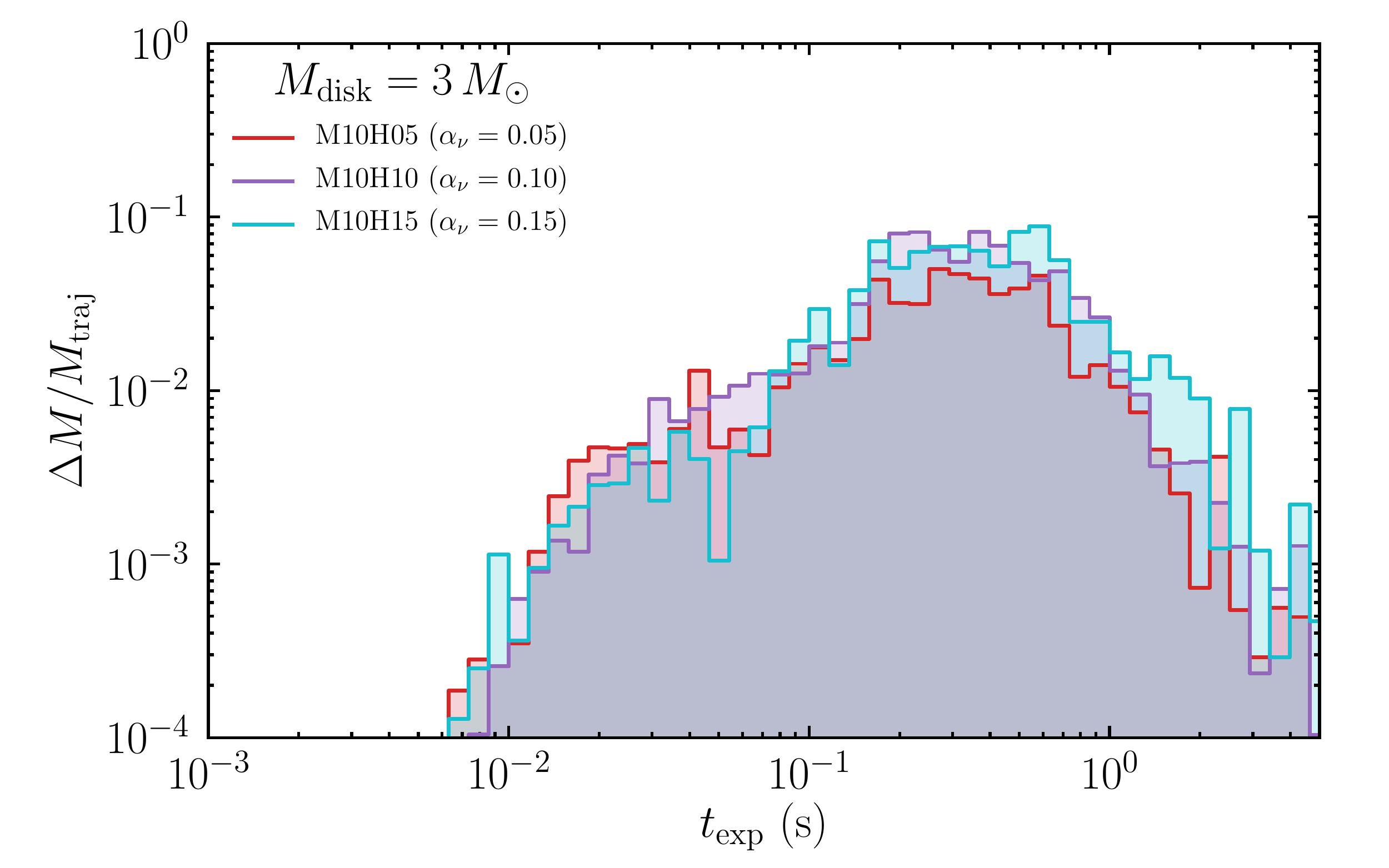}
\includegraphics[width=0.46\textwidth]{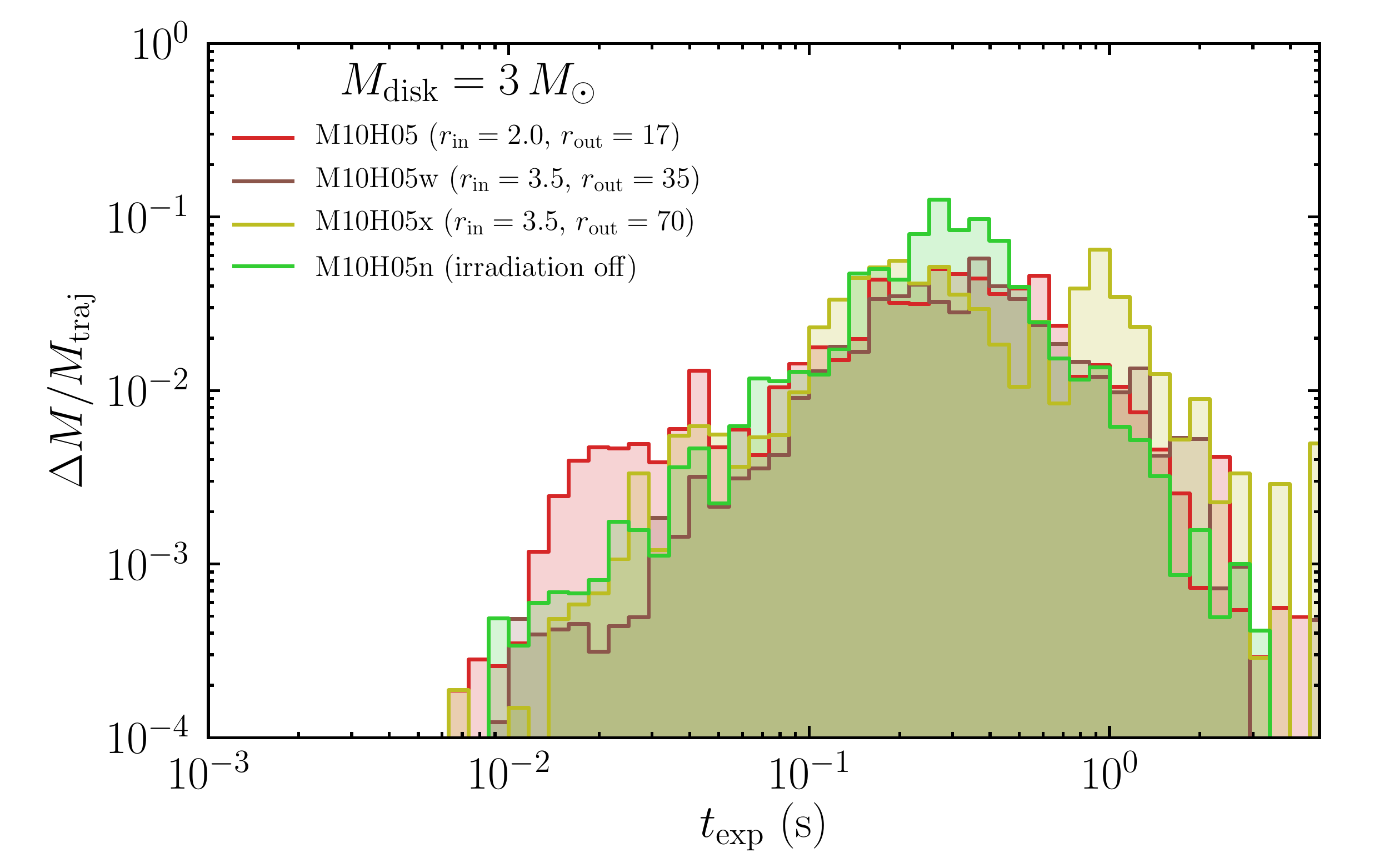}
\caption{The same as Fig.~\ref{fig11} but as a function of the expansion timescale from 5~GK to 2.5~GK.
\label{fig13}}
\end{figure*}

\begin{figure*}[t]
\includegraphics[width=0.46\textwidth]{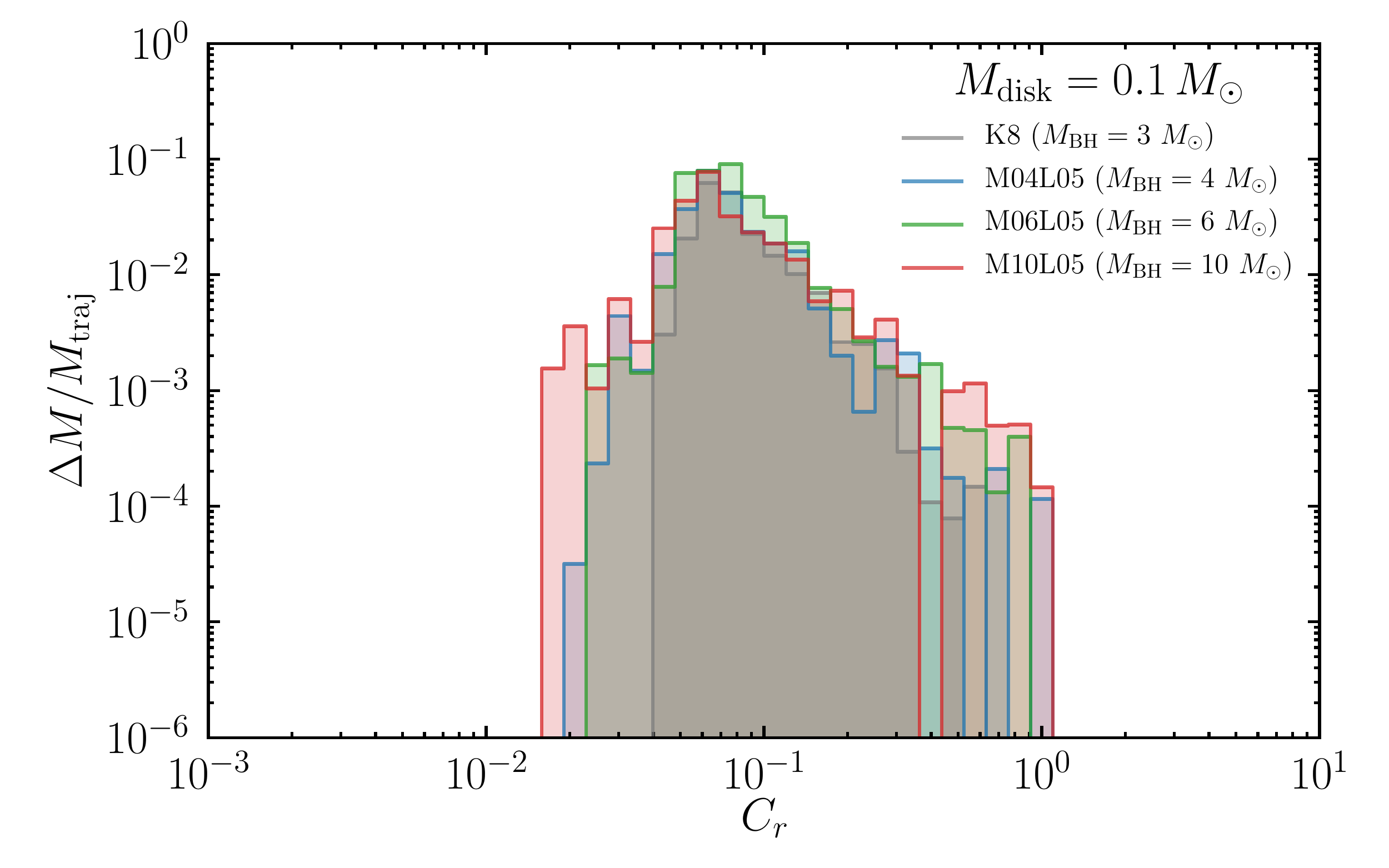}
\includegraphics[width=0.46\textwidth]{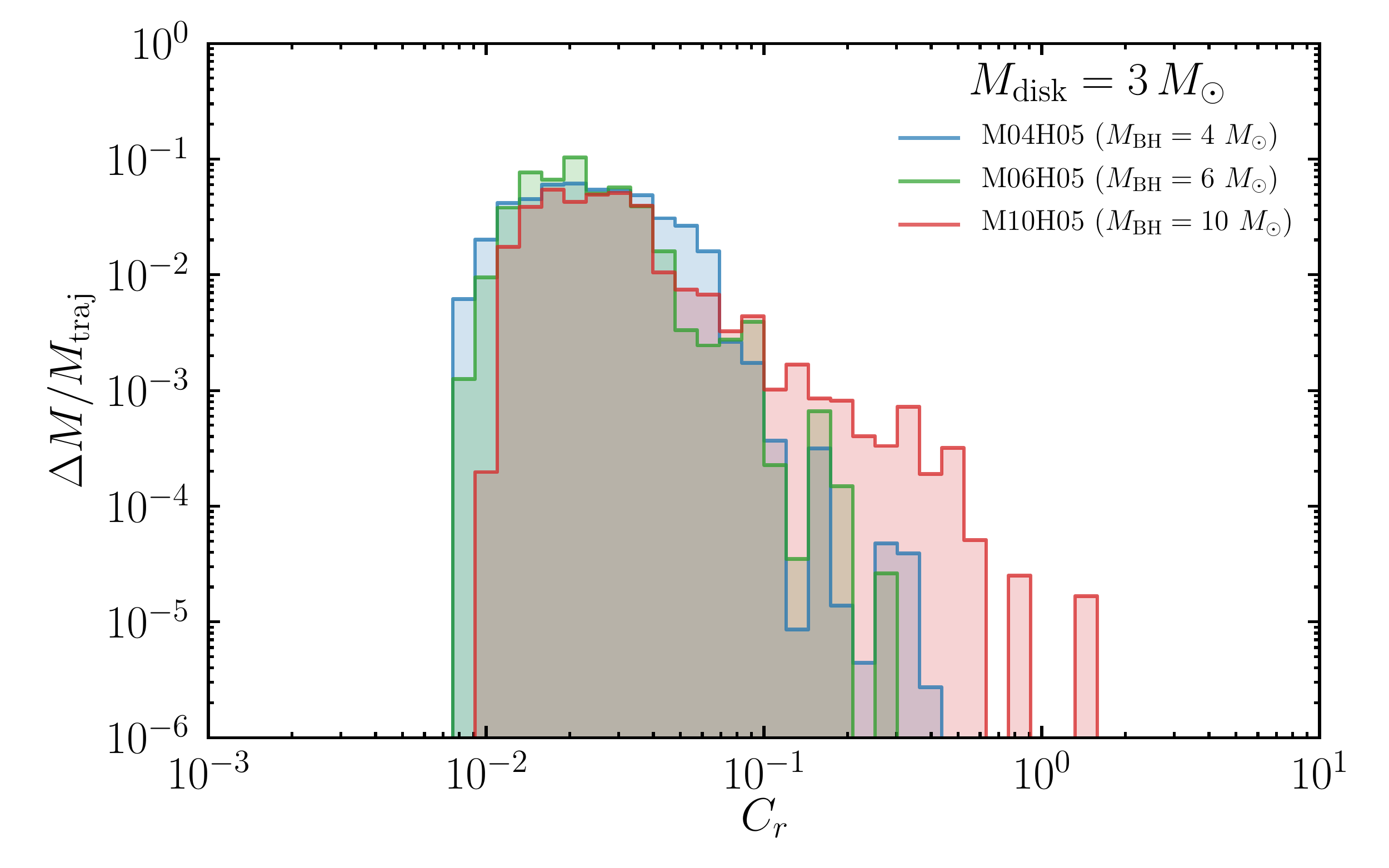}
\includegraphics[width=0.46\textwidth]{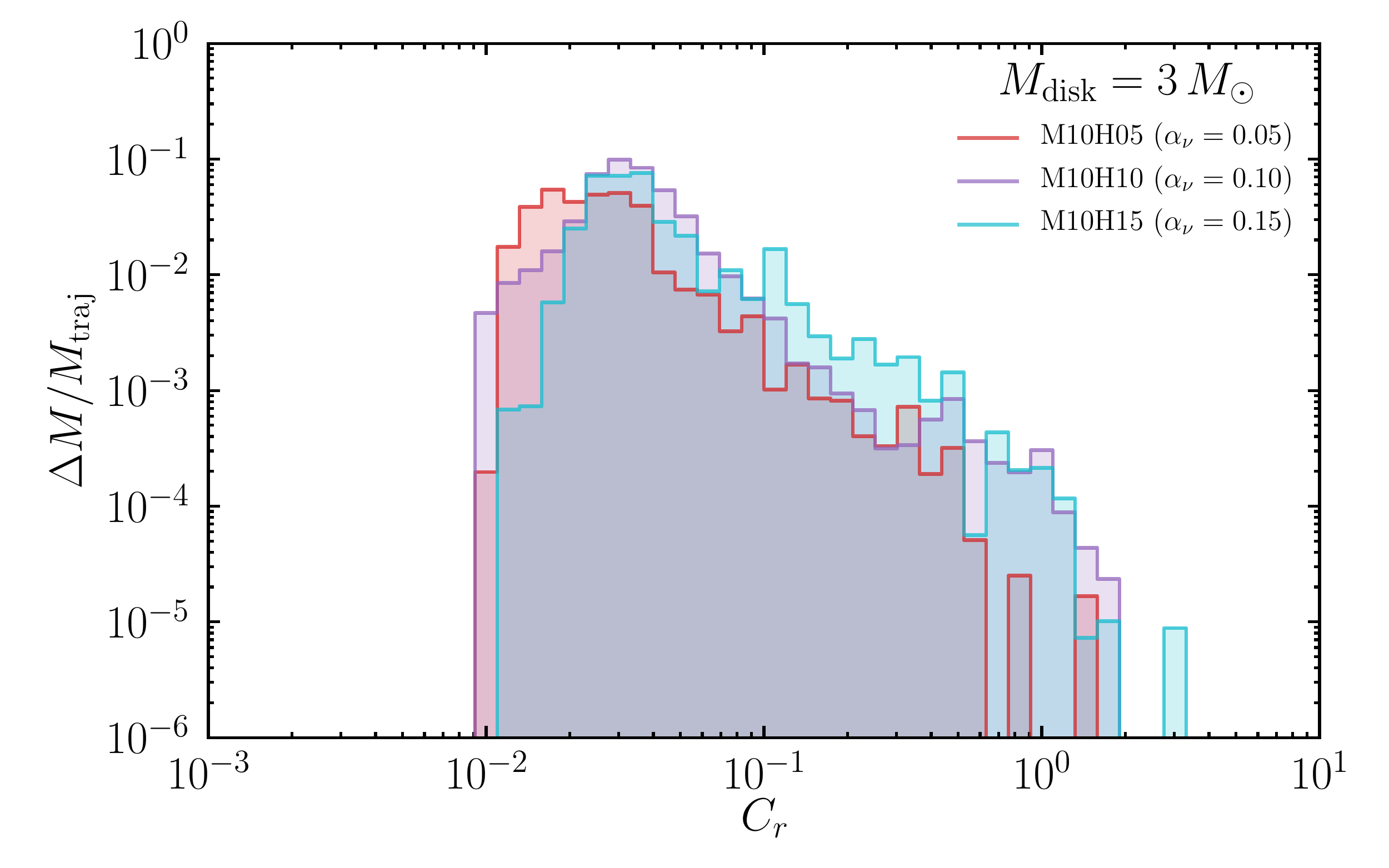}
\includegraphics[width=0.46\textwidth]{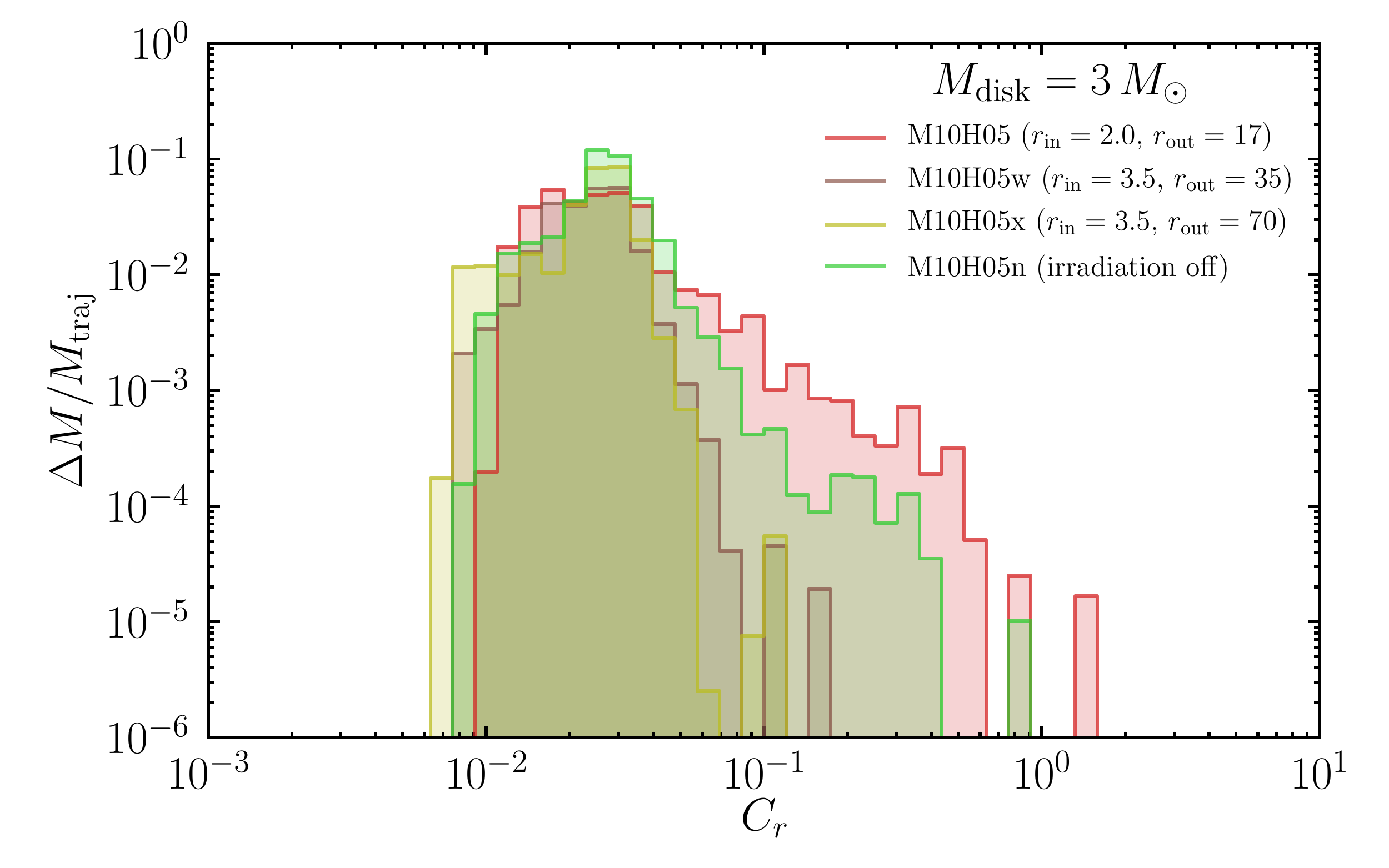}

\caption{The same as Fig.~\ref{fig11} but as a function of $C_r$ defined by Eq.~(\ref{eq:cr}). The tracer particles with $Y_e < 0.2$ or $Y_e > 0.5$ (at 5~GK) are excluded from the analysis, which are out of the range for the use of Eq.~(\ref{eq:cr}).
\label{fig14}}
\end{figure*}

\begin{figure*}
\includegraphics[width=0.47\textwidth]{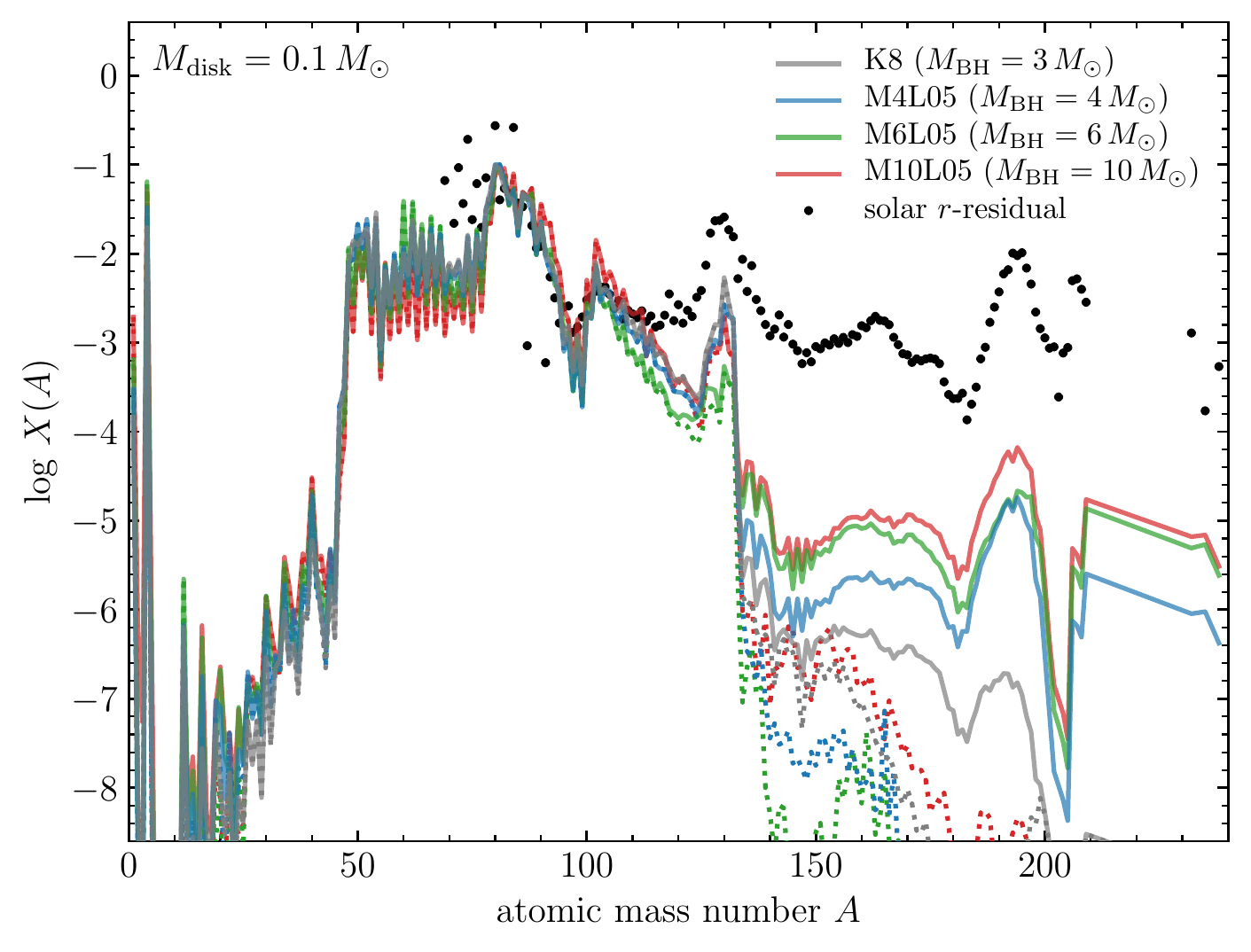}
\includegraphics[width=0.47\textwidth]{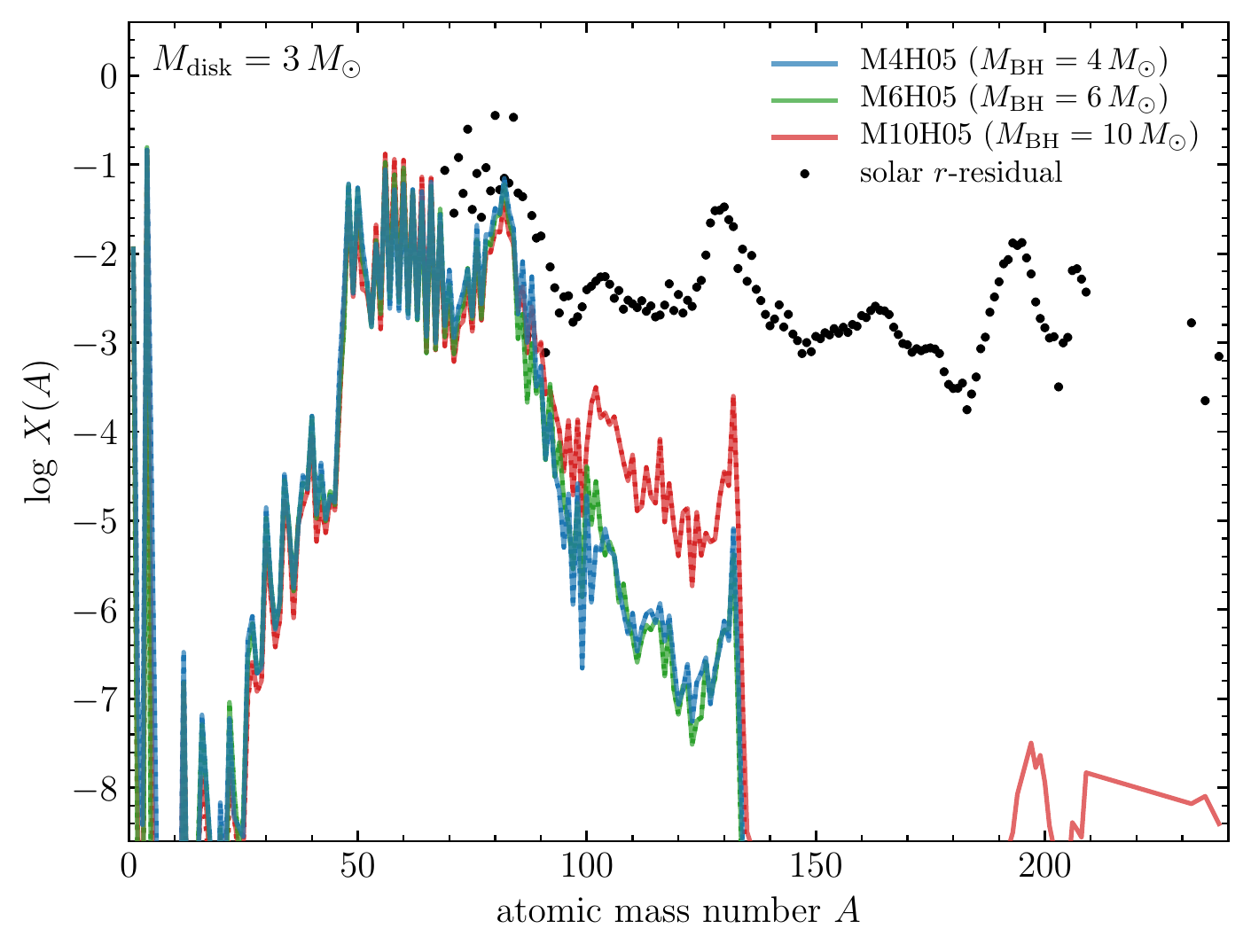}
\includegraphics[width=0.47\textwidth]{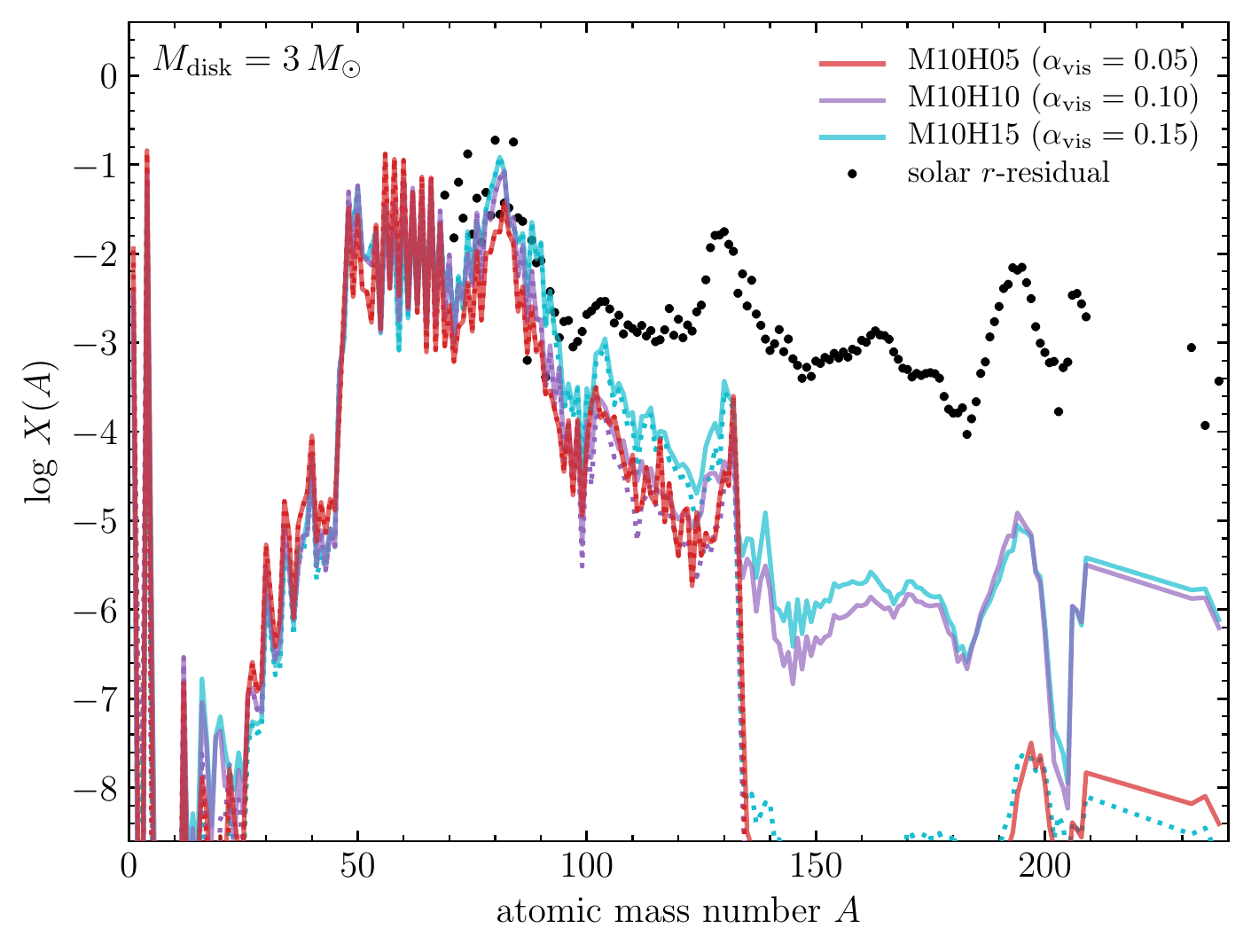}
\includegraphics[width=0.47\textwidth]{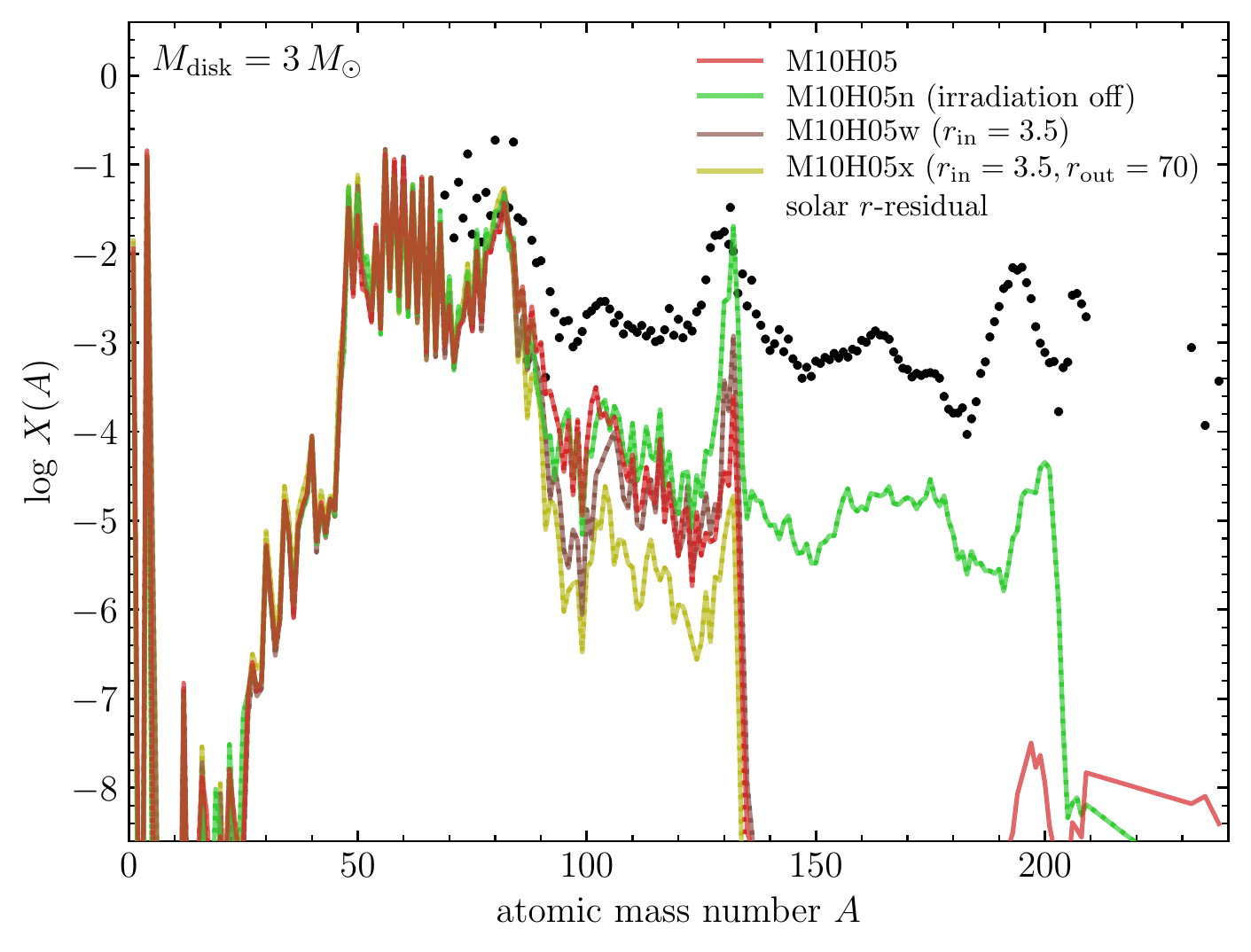}
\caption{Mass fraction of isobaric product (solid lines) for models
  M04L05, M06L05, and M10L05 (top-left), M04H05, M06H05, and M10H05 (top-right),
  M10H05, M10H10, and M10H15 (bottom-left), and M10H05, M10H05w, M10H05x, and
  M10H05n (bottom-right). The dotted lines indicate those excluding the tracer particles with $s/k > 100$ at 5~GK. The filled circles denote the $r$-process residuals to the solar system abundances for $A \ge 69$ \cite{Prantzos2020a}, which
  are shifted to match the calculated mass fractions of $X(82)$ for M10L05 and M10H05 in the top-left and the other panels, respectively.}
\label{fig16}
\end{figure*}

\begin{figure*}
\includegraphics[width=0.46\textwidth]{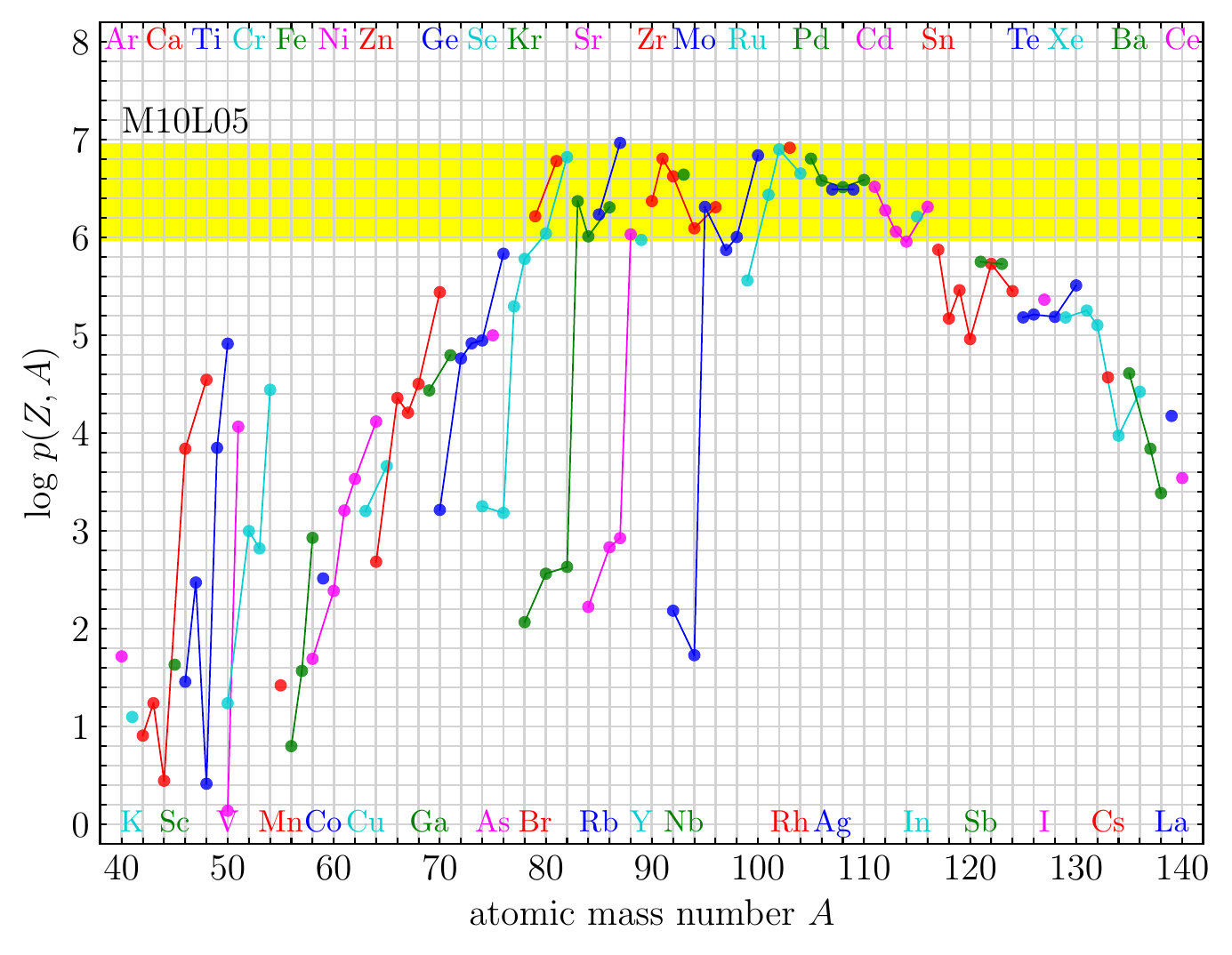}
\includegraphics[width=0.46\textwidth]{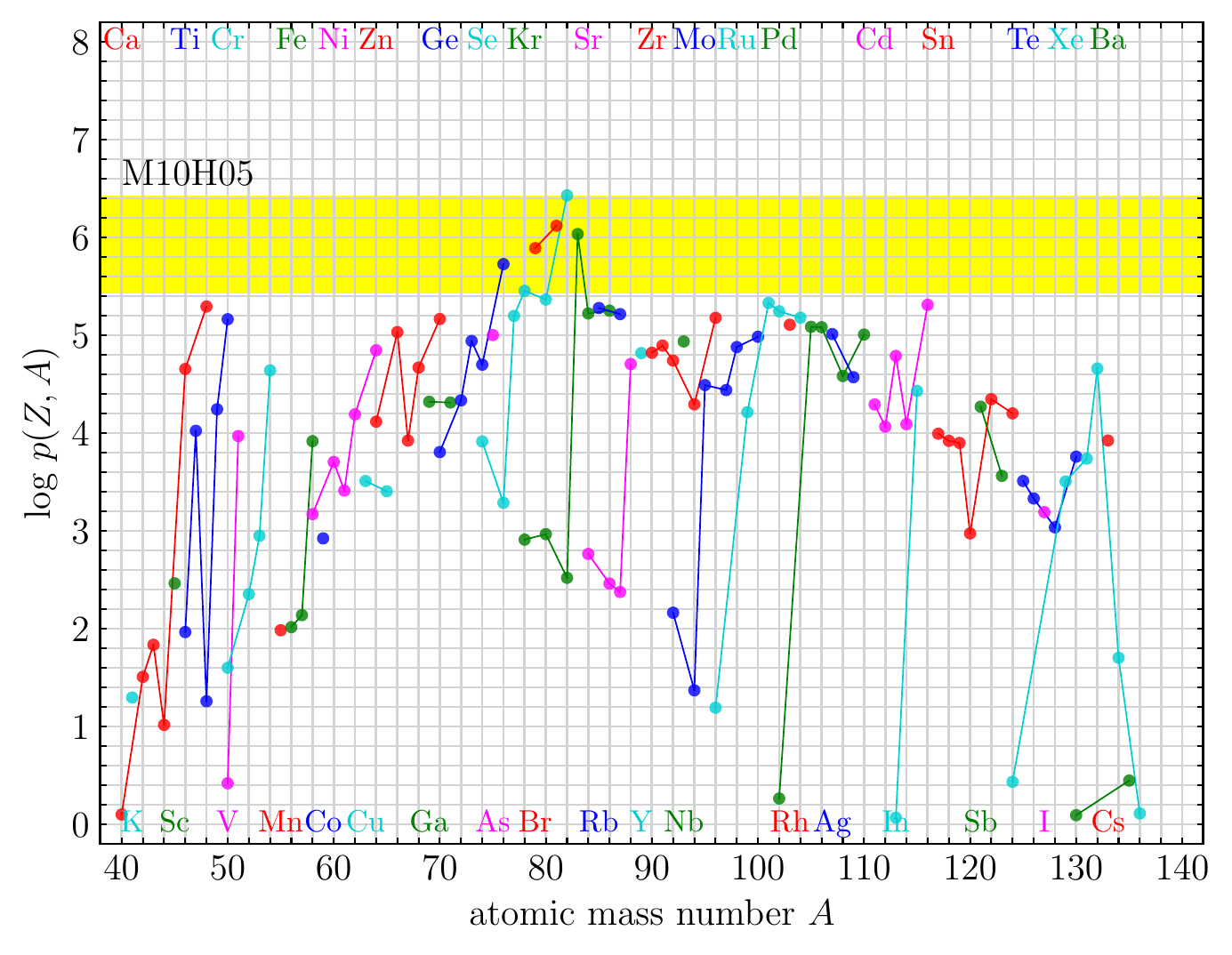}
\caption{Overproduction factors defined by Eq.~(\ref{eq:opro}) for the isotopes synthesized in models M10L05 (left) and M10H05 (right) as representatives of low-mass and high-mass disk models, respectively. The yellow band for each panel marks the range between the maximum of $p(Z, A) \equiv p_\mathrm{max}$ and $p_\mathrm{max}/10$. We regard the isotopes that reside on the yellow band potentially originating from the remnant of black hole-neutron star mergers (left) or collapsars (right) represented by our models. The isotopes of each element displayed at the top (even $Z$) or bottom (odd $Z$) of the panel are connected by the line with the same color.
}
\label{fig18}
\end{figure*}

Nucleosynthetic abundances are calculated for all the models in post-processing steps by using the reaction-network code, \texttt{rNET} \cite{Wanajo2018b}, consisting of 6300 isotopes with the range of atomic number, $Z = 1$--110.
The adopted theoretical rates for both light-particle (nucleon and $\alpha$-particle) capture (TALYS \cite{Goriely2008a}) and $\beta$-decay (GT2 \cite{Tachibana1990a}) are based on the microscopic mass prediction of HFB-21 \cite{Goriely2010a}.
The experimental rates, if available, are taken from REACLIB V2.0 \cite{Cyburt2010a}.
Neutrino-induced reactions are not included in the present nucleosynthesis calculations.
The thermodynamic histories of $\sim 5000$ tracer particles for each model are deduced as described in our previous work \cite{Fujiba2019}.
Each nucleosynthesis calculation is started with the initial mass fractions of protons and neutrons being $Y_e$ and $1-Y_e$, respectively, at which the temperature in a given tracer particle decreases to $10^{10}$\,K (=10~GK; here 1\,GK$=10^9$\,K).
At such high temperature, nuclear statistical equilibrium (NSE) immediately establishes and thus any initial compositions with the same net charge, i.e., $\sum XZ/A = Y_e$ (here $X$, $Z$, and $A$ indicate the mass fraction, atomic number, and atomic mass number of a given isotope, respectively) should give the same result.
We note that the $Y_e$ values at 5~GK are adopted for the initial composition.

As all nucleosynthesis-relevant quantities such as $Y_e$, $s$, and $V_\mathrm{ej}$ (or expansion timescale) span wide ranges across our explored models, the trend of resulting abundance distributions cannot be interpreted based on solely a single quantity (e.g., $Y_e$).
Before presenting the nucleosynthesis results, therefore, we analyse these relevant quantities in some detail by utilizing the criterion for the production of heavy $r$-process nuclei described in Ref.~\cite{Hoffman1997a},
\begin{equation}
    C_r = f(Y_e)\, s\, {t_\mathrm{exp}}^{-1/3} > 1,
    \label{eq:cr}
\end{equation}
where
\begin{equation}
    f(Y_e) = 10^{-3} \left\{\frac{1-2Y_e}{[0.33/(Y_e-0.17)]^2-(0.5/Y_e)^2}\right\}^{-1/3}
    \label{eq:cr1}
\end{equation}
for $0.2 < Y_e < 0.38$ (neutron-rich condition) and,
\begin{equation}
    f(Y_e) = 5.5\times 10^{-4}\, {Y_e}^{-1}
    \label{eq:cr2}
\end{equation}
for $0.38 < Y_e < 0.5$ (neutron-deficient condition) with $t_\mathrm{exp}$ being the duration of the expansion from 5\,GK (the end of NSE) to 2.5\,GK (the beginning of neutron capture) in units of second.
Note that the original value of the coefficient in Eq.~(\ref{eq:cr2}), $5\times 10^{-4}$ in Ref. \cite{Hoffman1997a}, is replaced by $5.5\times 10^{-4}$ such that $f(Y_e)$ becomes continuous at $Y_e = 0.38$.
Here, the proton-to-nucleon ratio of the seed nuclei synthesized in nuclear quasi-statistical equilibrium (QSE) is assumed to be 0.38.

According to Ref.~\cite{Hoffman1997a}, the $r$-process nuclei with $A = 200$ are expected to be abundantly produced for the case that the condition of $C_r > 1$ is satisfied.
This criterion also can be used to inspect the productivity of nuclei beyond the seeds ($A \sim 80$--90) even for $C_r < 1$.
Figure~\ref{fig10} shows the dependence of $C_r$ on $Y_e$ for fixed values of $s$ and $t_\mathrm{exp}$.
We find that a combination of higher $s$, lower $Y_e$, and shorter $t_\mathrm{exp}$ gives a larger value of $C_r$, i.e., a condition for synthesizing heavier nuclei.
Among these quantities, a change of $t_\mathrm{exp}$ has a relatively small impact on $C_r$ as evident from Eq.~(\ref{eq:cr}) and Fig.~\ref{fig10}.
As pointed out in Ref.~\cite{Hoffman1997a}, for $Y_e < 0.2$ (out of the range for the use of Eq.~(\ref{eq:cr})), the heavy $r$-process nuclei are synthesized less dependently on $s$ and $t_\mathrm{exp}$.
However, such a condition is not met in our explored models except for M10H05n (without neutrino irradiation).

Figures~\ref{fig11}--\ref{fig14} plot the mass histograms of the ejecta as a function of $Y_e$, $s/k$, $t_\mathrm{exp}$, and $C_r$ for all the models employed in this paper.
The mass-averaged values of these quantities are also presented in Table~\ref{tab:average}.
Here, $Y_e (T=5\,{\rm GK})$ is the electron fraction of each tracer particle determined at the time when its
temperature decreases to 5\,GK.  For the low-mass disk cases, the
electron fraction is distributed between $\sim 0.2$ and 0.5, and for
the high-mass disk cases, it is between $\sim 0.3$ and 0.5 (except for M10H05n, for which the heating/irradiation is absent).  As we
showed in Secs.~\ref{sec3-2} and \ref{sec3-4}, the average values of
$Y_e$ for the ejecta is between $\approx 0.30$ and 0.35 for low-mass
disk models and between $\approx 0.35$ and 0.50 for high-mass disk
models (see also $\langle Y_e \rangle$ in Table~\ref{tab:average}). Thus the distribution is consistent with the average. Note that the cutoff of the distribution at the high-$Y_e$ end in each panel of Fig.~\ref{fig11} is a consequence of the fact that we limit the range as $Y_e \le 0.55$ in our simulations.

The distribution of the electron fraction depends quantitatively on the black-hole mass, disk mass, and the values of $\alpha_\nu$.
Figure~\ref{fig11} shows that the electron fraction is higher for the higher-mass black-hole models.
The bottom-left panel of Fig.~\ref{fig11} also shows that the electron fraction is lower for the larger viscous coefficient case (i.e., for the case that the mass ejection timescale is shorter).
However, even in the extremely high value of $\alpha_\nu=0.15$, the value of $Y_e$ is always higher than 0.25; ejecta is not extremely neutron-rich for the high-mass disk case.
The bottom-right panel of Fig.~\ref{fig11} indicates that the changes of $r_\mathrm{in}$ or $r_\mathrm{out}$ only slightly modify the $Y_e$ distribution.
This panel also confirms that the inclusion of neutrino irradiation is necessary to obtain a reliable $Y_e$ distribution. 

Figure~\ref{fig12} shows that the typical specific entropy of the
ejecta is 10--$30k$ irrespective of the black-hole mass, the disk mass,
the value of $\alpha_\nu$, and the disk size. We find, however, that the distribution extends to higher entropy as $dM/ds \propto s^{-3}$ (note that the bin spacing in Fig.~\ref{fig12} is proportional to $\log s$). As a result, non-negligible amounts of ejecta have the entropy values exceeding $s/k = 100$ in many of our explored models. Figure~\ref{fig13} shows the
expansion timescales defined by $t_\mathrm{exp} = t(T=2.5\, \mathrm{GK})-t(T=5\, \mathrm{GK})$. For the 
low-mass disk cases (see the top-left panel of Fig.~\ref{fig13}), the typical expansion timescale is $\sim 100$\,ms 
irrespective of the black-hole mass.
For the high-mass disk cases,  this timescale is longer, typically $\sim 500$\,ms  (see Fig.~\ref{fig13}).
A reason for this trend is that the radius, at which the temperature of the fluid becomes $\sim 5$\,GK, is larger for higher-mass disks due to their higher temperature. 
This makes the timescale for the decrease of the temperature longer, because the temperature in radiation-dominated material drops approximately as $\propto r^{-1}$.

From the distributions of $Y_e$, $s/k$, and $t_\mathrm{exp}$ presented in Figs.~\ref{fig11}--\ref{fig13}, the distribution of $C_r$ is derived by using Eq.~(\ref{eq:cr}). Here, the tracer particles with electron fractions only in the range of $0.2 < Y_e < 0.5$, for which Eq.~(\ref{eq:cr}) is applicable, are considered. We find that the ejecta satisfying $C_r > 1$ are subdominant or absent in our explored models. It is anticipated from Fig.~\ref{fig10} that the component with $s/k > 100$ is necessary to meet the criterion $C_r > 1$ for $0.2 < Y_e < 0.5$ and $0.01 < t_\mathrm{exp}/\mathrm{s} < 1$. Therefore, the heavy $r$-process nuclei
synthesized in our models should originate from high entropy ejecta with $s/k > 100$. Such ejecta cannot be the dominant component in the present cases because of the entropy distribution scaled approximately as $dM/ds \propto s^{-3}$ (Fig.~\ref{fig12}) with the lowest value of $s/k \sim 10$. This is evident also from the average values of $C_r$, $\langle C_r \rangle < 0.1$, in Table~\ref{tab:average}. We note that the high entropy ejecta with $s/k > 100$ can be only mildly neutron-rich.
The viscous and neutrino heating inevitably induces weak interaction, which leads to an increasing trend of the specific entropy with $Y_e$. As a result, the electron fractions in the ejecta with $s/k > 100$ become $Y_e \gtrsim 0.35$.

The high-entropy component of the ejecta originates from the innermost part of the disk,
for which the specific entropy of the matter is increased efficiently by the viscous heating. The material is ejected intermittently by the buoyancy force after the weak interaction in the disk freezes out.
The high-entropy component with $s/k \gtrsim 100$ has a moderate electron fraction of $Y_e \sim 0.35$ for the low-mass disk models (e.g., M10L05), while the high-entropy component has a high electron fraction of $Y_e \gtrsim 0.45$ for the high-mass disk models (e.g., M10H05).
This is because the temperature of the disk is lower and the weak interaction timescale is longer for the low-mass disk models, and thus, the initial moderate-value electron fraction in the outer part of the disk is preserved more easily than that for the high-mass disk models. 
For higher viscosity models, M10H10 and M10H15, a part of the high-entropy component has somewhat low electron fraction of $Y_e\sim 0.35$--0.45 because of the lower freeze-out values of $Y_e$ for the higher-viscosity cases.


The mass fraction of nucleosynthetic products as a function of atomic
mass number, $X(A)$, is shown in Fig.~\ref{fig16} for all the models (solid lines). The dotted lines indicate the results excluding the tracer particles with $s/k > 100$. The $r$-process residuals to the solar system abundances
\cite{Prantzos2020a} (filled circles; hereafter referred to as the solar $r$-residuals) are also shown, which are vertically shifted to match the value of $X(82)$ for M10L05 (the top-left panel of Fig.~\ref{fig16}) or M10H05 (the other panels of Figs.~\ref{fig16}). We find that the heavy $r$-process nuclei with $A > 132$ (beyond the neutron shell closure of $N = 82$) are exclusively synthesized in the ejecta with $s/k > 100$ as anticipated from the analysis of $C_r$ value (except for M10H05n).

For the low-mass disk models ($M_\mathrm{disk} \approx 0.10\,M_\odot$), the abundance distributions follow the low-$A$ ($= 80$--110) side of the solar $r$-residual pattern (see the top-left panel of Fig.~\ref{fig16}). However, the heavy $r$-process abundances with $A > 132$ are deficient compared to the scaled solar $r$-residuals by more than three orders of magnitude. The result is consistent with the small values of $\langle C_r \rangle$ ($\sim 0.06$--0.09; Table~\ref{tab:average}). This is a weak $r$-process signature that is also found for model K8 (gray line)~\cite{Fujiba20} as well as for the post-merger ejecta from massive neutron stars studied in our previous work \cite{Fujiba2019}. Although the amounts are small, we find that these nuclei ($A > 132$) are more abundant for higher $M_\mathrm{BH}$ models, despite a tendency of their higher values of $Y_e$ (Fig.~\ref{fig11} and Table~\ref{tab:average}). This is due to the fact that the ejecta for higher $M_\mathrm{BH}$ models contain larger amounts of matter with higher values of $s/k$ ($> 100$; the top-left panel of Fig.~\ref{fig12}) and thus $C_r$ ($\gtrsim 1$; the top-left panel of Fig.~\ref{fig14}).

For the high-mass disk models ($M_\mathrm{disk} \approx 3.0\,M_\odot$), the abundance distribution is characterized by a sharp peak at $A \approx 82$ (see Fig.~\ref{fig16}), which is formed in modestly neutron-rich ($Y_e=0.3$--0.4) and low-entropy ($s/k = 10$--20) ejecta.
Compared to the low-mass disk models, the overall electron fraction (Fig.~\ref{fig11}) and expansion timescale (Fig.~\ref{fig13}) for the high-mass disk models are higher and longer, respectively, while those of entropy are similar (Fig.~\ref{fig12}). Such a physical condition, i.e., the modest
neutron-richness, long expansion timescale, and relatively low entropy leads to
nucleosynthesis in QSE rather than by
neutron capture~\cite{Hoffman1997a,Wanajo2018a}. This also can be anticipated from the smaller values of $\langle C_r \rangle$ ($= 0.02$--0.04; Table~\ref{tab:average}) in the high-mass disk models, which indicate weaker productivity of heavy nuclei beyond the seeds ($A=80$--90). As a result, the
abundance pattern exhibits a sharp peak at $^{82}$Se (synthesized as $^{82}$Ge that has
the proton-to-nucleon ratio of 0.390) with negligible mass fractions
of the nuclei beyond $A \approx 132$.

The change of $M_\mathrm{BH}$ modifies the abundance
patterns in the range of $A = 100$--130 (see the top-right panel of Fig.~\ref{fig16}) for the high-mass disk models, reflecting the difference of the entropy distribution on the high-$s$ side ($> 50k$; see the top-right panel of Fig.~\ref{fig12}). This is also evident from the top-right panel of Fig.~\ref{fig14}, which shows the larger ejecta mass with $C_r > 0.1$ for model M10H05 than those for other models. 

Higher values of $\alpha_\nu$ ($=0.10$ and 0.15) lead to the production of the
nuclei with $A > 132$ (see the bottom-left panel of Fig.~\ref{fig16}) 
because of larger amounts of high-$s$
($> 100k$) components (see the bottom-left panel of Fig.~\ref{fig12}) and thus of $C_r > 1$ 
(see the bottom-left panel of Fig.~\ref{fig14}). 
For model M10H10 (and M10H15), the heavy $r$-process nuclei with $A > 140$ are synthesized entirely in the high-entropy ejecta with $s/k > 100$. A similar trend has 
also been found in our previous work for models with $M_\mathrm{BH} = 3\, M_\odot$ and $M_\mathrm{disk} = 0.1\, M_\odot$
(K8h and K8s in Ref.~\cite{Fujiba20}, although the heavy $r$-process nuclei originate also from the lowest-$Y_e$\,($\lesssim 0.2$) components for the small disk mass). In fact, the exclusion of ejecta with $s/k > 100$ (dotted lines in the bottom-left panel of Fig.~\ref{fig16}) diminishes the abundances for $A > 132$ (near or below the lower-bound of the plot), which confirms the origin of heavy $r$-process nuclei being from the 
high-entropy ejecta. 
However, these nuclei are under-abundant by about 3 orders of magnitude compared to
the solar $r$-residuals (scaled at $A = 82$). 

The change of the disk size slightly modifies the abundance patterns for $A > 90$ as found in the bottom-right panel of Fig.~\ref{fig16} that compares models M10H05, M10H05w, and M10H05x.
This is due to the overall lower entropy for the wider disk models (see the bottom-right panel of Fig.~\ref{fig12}) because of longer viscous timescale due to larger initial radius of the disk.
This leads to the weaker productivity of nuclei beyond the seeds (i.e., smaller $C_r$; see the bottom-right panel of Fig.~\ref{fig14} and Table~\ref{tab:average}).  We
note that, for the high-mass disk models, the neutrino absorption on
free nucleons plays a significant role for shaping the $Y_e$
distribution by diminishing the low-$Y_e$ ($\lesssim 0.3)$
component (M10H05 and M10H05n in the bottom-right panel of Fig.~\ref{fig11}), which suppresses the production of nuclei with $A > 132$ (see the bottom-right panel of Fig.~\ref{fig16}). This is due to the high neutrino luminosity ($\gtrsim 10^{54}$
erg/s at the peak; cf.~Fig.~\ref{fig5}), which are about 10 times
higher than those for the low-mass disk models (cf.~Fig.~\ref{fig1}). Note that, for model M10H05n (for which the neutrino irradiation is absence), 
the production of nuclei with $A > 132$ is due to the presence of low-$Y_e \sim 0.2$ ejecta (the solid and dotted lines in the bottom-right panel of Fig.~\ref{fig16} are overlapped).

\subsection{Implications from the Nucleosynthesis Calculation} \label{subsec:nucl}

We presume that our low-mass and high-mass disk models represent the remnants for the merger of black hole-neutron star and the core-collapse of rapidly rotating massive stars (collapsars), respectively. In this respect, the upper limits for the frequencies of such events can be obtained from the nucleosynthesis results as what follows. Here, we do not consider the contribution from the earlier ejecta (if any) of black hole-neutron star mergers or collapsar.
By including this contribution, the constraint on the frequency will be even stronger.
Table~\ref{tab:nucleosynthesis} (seventh and eighth columns) presents the maximum overproduction factor ($p_\mathrm{max}$) and the relevant isotope for each model. The overproduction factor is defined by 
\begin{equation}
p(Z, A) = X(Z, A)/X_\odot(Z, A), \label{eq:opro}
\end{equation}
where $X(Z, A)$ and
$X_\odot(Z, A)$ denote the mass fractions of the isotope with $Z$ and $A$ for each model and in the solar system \cite{Lodders2009a},
respectively. In Fig.~\ref{fig18}, the overproduction factors for M10L05 (left) and M10H05 (right) are plotted as representatives of the low-mass and high-mass disk models, respectively. The maximum Galactic fraction of black hole-neutron star mergers or collapsars represented
by each model with respect to that of core-collapse supernovae (CCSNe), $f_\mathrm{max}$ ($\ll 1$, Table~\ref{tab:nucleosynthesis};
ninth column), can be estimated as \cite{Wanajo2018a}
\begin{equation}
    f_\mathrm{max} \approx \frac{f_\mathrm{max}}{1-f_\mathrm{max}} 
= \frac{\langle M_\mathrm{CCSN}(^{16}\mathrm{O})\rangle/X_\odot(^{16}\mathrm{O})}{M_\mathrm{ej, tot}\, p(Z, A)},
\end{equation}
where $\langle M_\mathrm{CCSN}(^{16}\mathrm{O})\rangle = 1.5\,
M_\odot$ is the average ejecta mass of $^{16}$O per CCSN and $X_\odot(^{16}$O$) = 6.60\times 10^{-3}$ is the mass fraction of $^{16}$O in the solar system~\cite{Lodders2009a}. 
Here, the total ejecta mass for each model ($M_\mathrm{ej, tot}$; third column in Table~\ref{tab:nucleosynthesis}) estimated as in our previous study \cite{Fujiba2019} 
is adopted instead of the ejecta mass at the end of simulation ($M_\mathrm{ej}$; 
second column in Table~\ref{tab:nucleosynthesis}) that is still increasing as found in Figs.~\ref{fig2}, \ref{fig6}, and \ref{fig9}. The fraction
$f_\mathrm{max}$ can be translated to the maximum rate of black hole-neutron star mergers or collapsars
represented by each model in the Galaxy, $R_\mathrm{max}$
(Table~\ref{tab:nucleosynthesis}; last column), by adopting the
inferred Galactic CCSN rate $\approx 2.30\times 10^{-2}$ yr$^{-1}$
\cite{Li2011a}.

In Table~\ref{tab:average}, we find $f_\mathrm{max} \sim 2\times 10^{-3}$ and $R_\mathrm{max} \sim 40$ Myr$^{-1}$ for the low-mass disk models. These values are similar to what are expected for binary neutron star mergers to be the dominant $r$-process site in the Galaxy and in the solar system \cite{Ishimaru2015a,Hotokezaka2018a}. However, as shown in the top-left panels of Figs.~\ref{fig16} and \ref{fig18}, the low-mass disk models account for the origin of isotopes in the range $A=80$--110 only. In fact, spectroscopic \cite{McWilliam1998a,Johnson2002a} and galactic chemical evolution studies \cite{Ojima2018a,Hirai2019a} imply the presence of the production site of light trans-iron elements, such as Sr, in addition to the main $r$-process site. Our results indicate that the black hole-accretion disks followed by black hole-neutron star mergers can be the second sources of such elements. Spectroscopic studies have also revealed the presence of metal-poor stars that exhibit a descending trend of trans-iron abundance patterns \cite{Honda2006a,Aoki2017a}, which is presumed to be a signature of weak $r$-processing \cite{Wanajo2006a}. Such a weak $r$-process-like abundance patterns may be explained by our low-mass disk models (see also a similar result for the post-merger ejecta from massive neutron star-accretion disks in our recent work~\cite{Fujiba2019}). It should be stressed, however, that the contribution from the early dynamical ejecta should be added to assess the full nucleosynthetic outcomes from black hole-neutron star mergers.

For the high-mass disk models, we find $f_\mathrm{max} \sim (0.02$--$0.9)\times 10^{-3}$ and $R_\mathrm{max} \sim 5$--$20$ Myr$^{-1}$ (see Table~\ref{tab:average}).
Provided that collapsars are the dominant sources of long duration
gamma-ray bursts, the local volumetric rate of collapsars is estimated to be
$R_\mathrm{GRB}/f_\mathrm{b} \approx 260$ Gpc$^{-3}$ yr$^{-1}$, where
$R_\mathrm{GRB} \approx 1.3$ Gpc$^{-3}$ yr$^{-1}$ is the local volumetric rate of long gamma-ray bursts pointing toward the Earth \cite{Wanderman2010a} and $f_\mathrm{b} \approx 5\times 10^{-3}$ is the beaming factor \cite{Goldstein2016a}. This can be translated
to the Galactic rate of long gamma-ray bursts $\approx 26$ Myr$^{-1}$, i.e., about
0.1\% of the Galactic CCSN rate, by using the number of Milky Way
analogous galaxies $\approx 0.01$ Mpc$^{-3}$ yr$^{-1}$
\citep{Kopparapu2008a}. These values are in good agreement with the upper bounds for the high-mass disk models. Thus, our models, in particular M10H05 ($f_\mathrm{max} = 0.94\times 10^{-3}$ and $R_\mathrm{max} =22$ Myr$^{-1}$; Table~\ref{tab:average}), may reasonably represent the black hole-accretion disks formed in collapsars.

As the sources of heavy nuclei, the black hole-accretion disks formed in collapsars can contribute only to the light trans-iron elements with $A \sim 80$ according to the right panel of Fig.~\ref{fig18}. It is noteworthy, however, that the overproduction factor of the neutron-rich isotope $^{48}$Ca accounts for $\sim 0.1$ of $p_\mathrm{max} = p(^{82}$Se). This implies that collapsars can be the sources of $^{48}$Ca up to $\sim 10\%$ of its total amount in the solar system. In Table~\ref{tab:nucleosynthesis} (fourth column), the ejecta mass of $^{48}$Ca calculated as $M_\mathrm{ej, tot} X(^{48}$Ca) is presented for all the models. Currently, the astrophysical sources of $^{48}$Ca are unknown, for which the suggested sites include high-density type-Ia supernovae \cite{Meyer1996a,Woosley1997a} and electron-capture supernovae (both core-collapse \cite{Wanajo2013a} and thermonuclear types \cite{Jones2019a}).
It should be noted that neutrino-induced reactions are not included in the present nucleosynthesis calculations. The high neutrino luminosity for the high-mass disk models (Fig.~\ref{fig5}) may lead to a $\nu p$-process \cite{Froehlich2006a,Froehlich2006b,Pruet2006a,Wanajo2006b,Wanajo2011a} that produces some proton-rich isotopes for which the astrophysical origins are uncertain (e.g., $^{92}$Mo). 

The type-Ic supernovae accompanying with long gamma-ray bursts are
considered to be powered by a large amount (0.1--$0.6\, M_\odot$
\cite{Cano2016a}) of $^{56}$Ni. Indeed, the pioneering work of
collapsars \cite{MacFadyen1999a} has predicted the mass ejection of
$^{56}$Ni as large as $\sim 1\, M_\odot$, assuming $Y_e = 0.5$ in the
accretion disk. However, the ejecta from our high-mass disk models
contain only $\sim 0.01$--$0.02\, M_\odot$ of $^{56}$Ni (calculated as $M_\mathrm{ej, tot} X(^{56}$Ni); fifth column in Table~\ref{tab:nucleosynthesis}). This is due to
the fact that the ejecta in our models have relatively small amounts
of matter with $Y_e > 0.49$ (see Fig.~\ref{fig11}), for which $^{56}$Ni is
efficiently produced \cite{Wanajo2018a}. 
This may indicate that the bulk of $^{56}$Ni in this type of supernovae comes from the early ejecta \cite{Tominaga2007}, for which the
driving mechanism is unsettled ( neutrino heating or
magnetic pressure). It is also suggested that the detonation of the late-time accreted material can be an additional source of $^{56}$Ni \cite{Zenati2020}.

\subsection{Relation with gamma-ray bursts}\label{sec3-7}

\begin{figure*}[t]
\includegraphics[width=85mm]{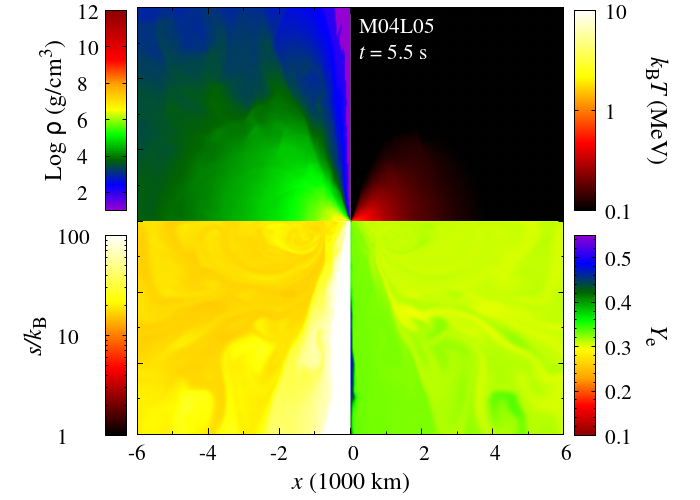} 
\includegraphics[width=85mm]{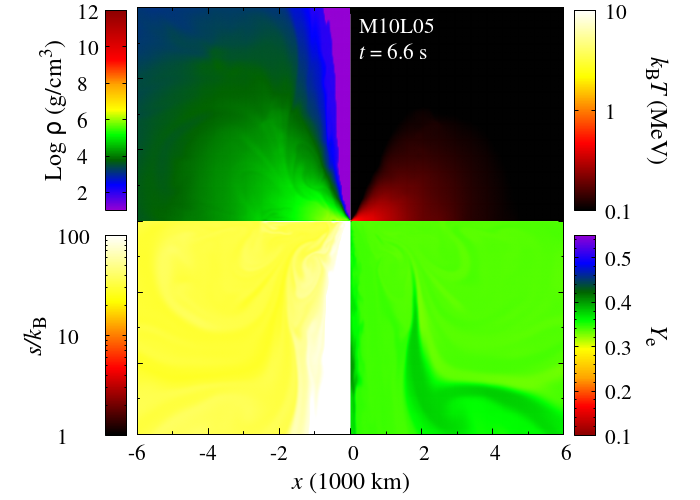} \\
\includegraphics[width=85mm]{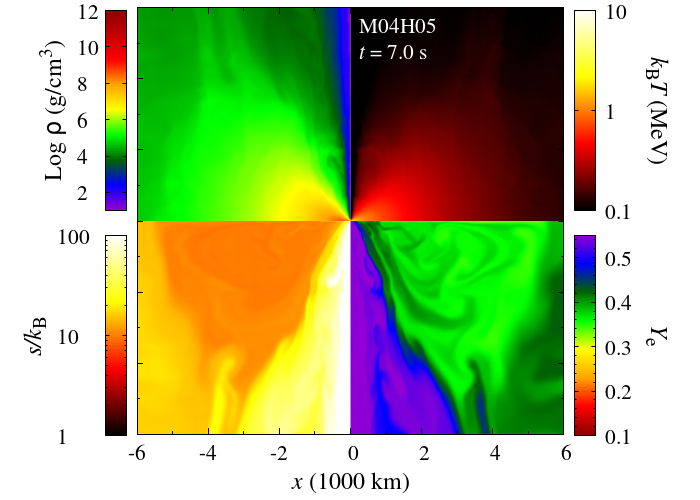} 
\includegraphics[width=85mm]{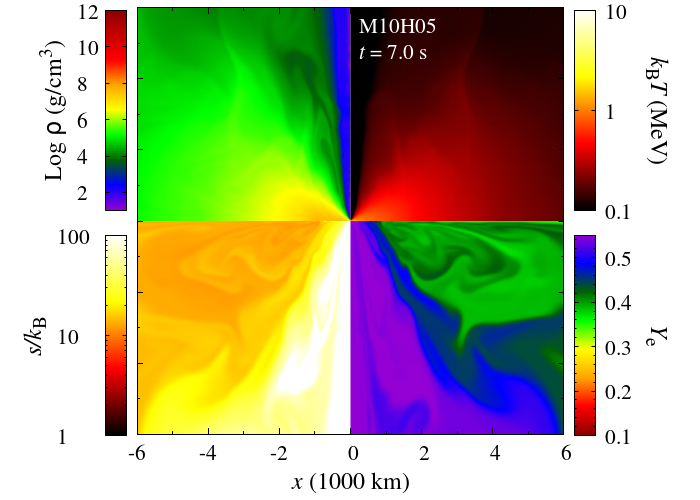} \\
\includegraphics[width=85mm]{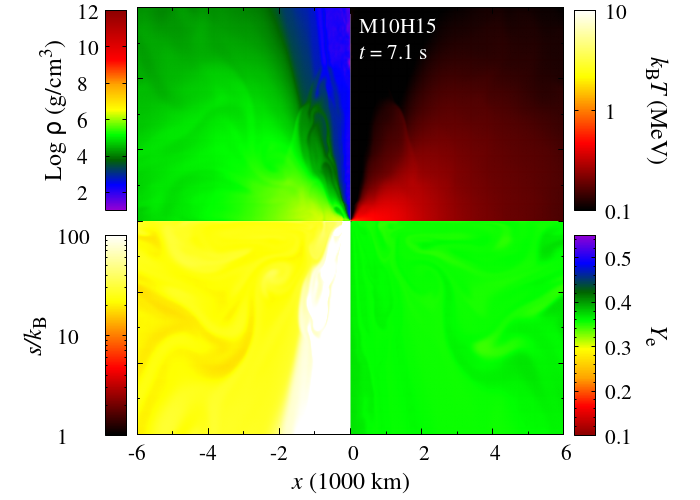} 
\includegraphics[width=85mm]{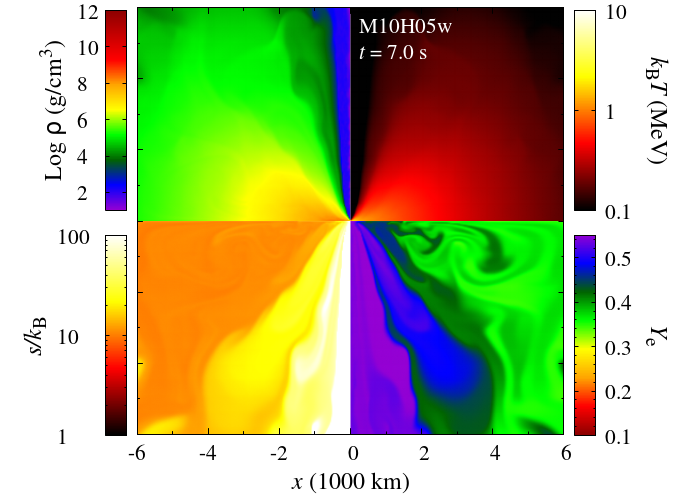} 
\caption{Snapshots for the rest-mass density in units of ${\rm
    g/cm^3}$, temperature ($k T$) in units of MeV, specific entropy
  per baryon in units of $k$, and electron fraction $Y_e$ at late times 
  for models M04L05 (top left), M10L05 (top right), M04H05 (middle left), 
  M10H05 (middle right), M10H15 (bottom left), and M10H05w (bottom right).
\label{fig20}}
\end{figure*}

Figure~\ref{fig20} displays snapshots for the profiles of the rest-mass density, temperature, specific entropy, and electron fraction for models M04L05, M10L05, M04H05, M10H05, M10H15, and M10H05w at late times ($t\approx 5.5$--7.1\,s). 
By these times, mass ejection has already set in and the disk has relaxed to a quasi-steady state. These figures show that the outcomes of the evolution of the system are composed of a rapidly spinning black hole surrounded by a torus and a narrow funnel in the vicinity of the rotation axis, irrespective of the black-hole mass and initial disk mass.
The dimensionless spin is larger for the larger initial disk mass.
The rest-mass density in the funnel region is $\alt 10^2\,{\rm g/cm^3}$ for the initially low-mass disk models and $\alt 10^{2.5}\,{\rm g/cm^3}$ for the initially high-mass disk models.
The density is close to the artificial atmosphere density, and thus, in reality, the density might be even lower.
The specific entropy in the funnel region is always very high, $s/k \agt 100$, and this indicates that the radiation pressure dominates over the gas pressure.
By contrast, the region outside the funnel, the specific entropy is of the order of $10k$. 

The half opening angle of the funnel region, which we define as the
region with the density less than $10^3\,{\rm g/cm^3}$, is $\sim
0.1$\,rad and is narrower for the smaller-mass black holes and for the
initially larger-mass disks. This angle depends weakly on the initial
disk radius although it is wider for the large viscous coefficient. For
all the models that we studied, the total rest mass in the funnel
region is of the order of $10^{-6}M_\odot$ in the computational
region, which may be suitable for a high-energy relativistic jet passing
through avoiding the baryon loading. All these facts indicate that the
outcome of the viscous evolution of the system of a rapidly spinning
black hole and a massive disk is suitable for generating gamma-ray
bursts~\cite{Meszaros2004,Piran2005}.


In the late-evolution stage of the disk with $t > 1$\,s for which the
matter temperature satisfies $kT < 3$\,MeV, the neutrino emissivity is
much smaller than the viscous heating. In such a stage, viscous heating
is fully used not only for the disk expansion and resulting steady mass
ejection but also for driving a high-velocity outflow in the vicinity
of the rotation axis and torus surface. In particular, near the
innermost region of the disk, the viscous heating rate is high and the
high velocity outflow is generated.  Since the matter density is low
above such a region, the high velocity is naturally realized. 
The order of the viscous heating rate is estimated by 
\beqn
L_{\rm vis} &\sim& \nu M_{\rm disk,inn} \Omega^2 \nonumber \\
&=& 6 \times 10^{49} \,{\rm erg/s}
\left({\alpha_\nu \over 0.05}\right)
\left({H \over 20\,{\rm km}}\right) \nonumber \\
&& \times \left({c_s \over 0.1c}\right)
\left({M_{\rm disk,inn} \over 10^{-4}M_\odot}\right)
\left({\Omega \over 10^3\,{\rm rad}}\right)^2,
\label{Lvis}
\eeqn
where $M_{\rm disk,inn}$ denotes the rest mass of the inner region of
the disk which contributes to the efficient viscous heating, and we
supposed that the black-hole mass would be 5--$7M_\odot$.  Note that
around rapidly spinning black holes with $\chi \agt 0.95$, a small
orbital radius of $\sim 2GM_{\rm BH}/c^2$ is possible and in such a
case, $\Omega$ can be $\approx 10^4$\,rad/s for $M_{\rm BH}=7M_\odot$.

The isotropic luminosity of typical long gamma-ray bursts is
$10^{51}\,{\rm erg/s}$~\cite{Meszaros2004,Piran2005}. If the beaming
effect (by which the total luminosity can be smaller) and energy
conversion efficiency to gamma-rays (by which higher luminosity is
necessary) are taken into account, the viscous heating rate shown in
Eq.~(\ref{Lvis}) is a substantial fraction of the long gamma-ray burst
luminosity.  Unfortunately, for the viscous heating, the matter is
always accompanied, and hence, it is not possible to drive the
outflow of the Lorentz factor to $\sim 100$ by this heating effect due
to the baryon loading problem.  However, an outflow by this heating
effect can play an important role for cleaning up the region in the
vicinity of the rotation axis to reduce the density there.


In the presence of electromagnetic fields, the Blandford-Znajek
effect~\cite{BZ77} could provide another energy injection mechanism
by extracting the huge rotational kinetic energy of the remnant black
hole. In the presence of a radial magnetic field on the black-hole
horizon $B^r$, the outward Poynting flux from the horizon is written
as (e.g., Ref.~\cite{MG04})
\beq
F_{\rm BZ}=2 {(GM_{\rm BH})^2 \over c^3}(B^r)^2 \omega 
\hat r_+ (\Omega_H-\omega)\sin^2\theta,
\eeq
where $\hat r_+=1+\sqrt{1-\chi^2}$, $\Omega_H=c^3\chi/(2GM_{\rm
  BH}\hat r_+)$, and $\omega$ is a rotational frequency of the
electromagnetic field on the horizon for which one often assumes
$\omega=\Omega_H/2$. Then, assuming that $B^r$ and $\omega$ are
uniform on the horizon, we obtain the total luminosity
\beqn 
L_{\rm BZ}&=&{2 \pi \over 3} {(GM_{\rm BH})^2 \over c^3}(B^r)^2
\hat r_+ \chi^2 \nonumber \\ & \approx & 7 \times 10^{50}\,{\rm
  erg/s}\, \left({M_{\rm BH} \over 7M_\odot}\right)^2 \left({B^r \over
  10^{14}\,{\rm G}}\right)^2 \hat r_+ \chi^2. \nonumber \\
\label{BZL}
\eeqn
Here, a strong magnetic field of the order of $10^{14}$\,G would be
achieved if the magnetorotational instability efficiently enhances the
magnetic-field strength, $B$, until the equi-partition is satisfied in
the disk, i.e., $B^2/(4\pi) \sim \rho_{\rm disk}c_s^2$ with $\rho_{\rm
  disk}$ being the typical rest-mass density of the disk at $t \agt
1$\,s which is $\sim 10^8\,{\rm g/cm^3}$ and with $c_s \sim 0.1c$. The
luminosity, $L_{\rm BZ}$, is comparable to the isotropic luminosity of
long gamma-ray bursts, and thus, the Blandford-Znajek effect could
also be a reasonable energy source. 

The funnel structure shown in Fig.~\ref{fig20} looks suitable for
confining a jet launched from the central part. In the funnel region,
the rest-mass density of the matter is likely to be much lower than
the electromagnetic pressure $B^2/8\pi$, and hence, a force-free
magnetosphere would be established. By contrast in the geometrically
thick torus next to the funnel, the rest-mass density increases
steeply to $\agt 10^6\,{\rm g/cm^3}$ for which the rest-mass energy
density of the matter exceeds the magnetic pressure. Therefore, only
in the narrow funnel, a clear spiral-shape magnetic field would be
established and an efficient particle acceleration would be achieved, 
leading to generating a narrow jet in the funnel. 

\section{Summary}\label{sec4}

As an extension of our previous study~\cite{Fujiba20}, we performed
viscous neutrino-radiation hydrodynamics simulations for accretion
disks surrounding a spinning black hole with $M_{\rm BH}=4$, 6,
and $10M_\odot$ and initial dimensionless spin $\chi \approx 0.8$. We
consider a compact disk with $M_{\rm disk} \approx 0.1$ or
$3M_\odot$ and with the outer edge located at $r_\mathrm{out}=
200$--1000\,km. For $M_{\rm disk} \approx 0.1M_\odot$, we find that
$\sim 20$\% of $M_{\rm disk}$ is ejected and the average electron
fraction of the ejecta is $\langle Y_e \rangle = 0.30$--0.35. Here,
$\langle Y_e \rangle$ is slightly higher for the larger values of
$M_{\rm BH}$. Nucleosynthesis calculation shows that only light
$r$-process elements with mass number of $80$--110 are
efficiently synthesized in such ejecta, because of the absence of the
low-$Y_e$ component with $Y_e \alt 0.2$. Thus, the results obtained
are qualitatively the same as those found in Ref.~\cite{Fujiba20} 
irrespective of the black-hole mass.

For high-mass disks with $M_{\rm disk} \approx 3M_\odot$, we found
that the luminosity of neutrinos exceeds $10^{54}$\,erg/s in the early
stage of the disk evolution and neutrino cooling is the dominant
cooling process for the first seconds. This timescale is
several times longer than that for $M_{\rm disk}\approx 0.1M_\odot$
for given black-hole mass. We also found that (i) $\sim 10$\% of
$M_{\rm disk}$ is ejected for the compact disk models with the disk
extent $\sim 200$\,km and (ii) the ejecta mass increases with the
increase of the initial disk radius.  Irrespective of the models,
$\langle Y_e \rangle$ of the ejecta is enhanced to be $\agt 0.35$
because the electron fraction is increased significantly during the
long-term viscous evolution of the disk until the neutrino cooling
timescale becomes longer than the viscous heating timescale. Our
nucleosynthesis calculation indicates that not $r$-process elements
but trans-iron elements with atomic mass number $\sim 80$ are predominately 
synthesized in the matter
ejected from a massive torus surrounding stellar-mass black holes.
This suggests that if the remnant of the collapsars is composed of a
black hole and a massive disk, such remnants may not be the sites for
the $r$-process nucleosynthesis. 

By the matter accretion, the dimensionless spin of the black hole,
$\chi$, increases.  For the initially high-mass disk case, the value
of $\chi$ exceeds 0.9. In particular for the case that the initial
ratio of the disk mass to the black-hole mass is larger than 1/2, it
exceeds 0.95. This value depends on the viscous coefficient, and for
the larger viscous coefficient, the resulting value of $\chi$ is
slightly smaller. 

For the high-mass disk case, the outcome after the viscous evolution
is the system composed of a rapidly spinning black hole with $\chi
\agt 0.9$, a geometrically thick torus, and a narrow funnel in the
vicinity of the rotation axis. In particular, due to the outflow
activity in a late stage of the disk evolution, the rest-mass density
in the rotation axis becomes quite low $\alt 10^2\,{\rm g/cm^3}$,
and the total rest mass in the funnel region becomes as small as
$10^{-6}M_\odot$. The outcome appears to be suitable for driving
gamma-ray bursts. Since the viscous heating is unlikely to be its
central engine, we cannot reproduce the gamma-ray bursts in our
simulation.  However, the final outcome that we find in this paper
indicates that in the presence of magnetic fields for which the
electromagnetic energy is comparable to the thermal energy of the disk
together with a rapidly spinning black hole, the Blandford-Znajek
mechanism is likely to provide the energy injection required for
launching the gamma-ray bursts.

\acknowledgments

This work was in part supported by Grant-in-Aid for Scientific
Research (Grant Nos.~JP16H02183, JP16K17706, JP18H01213, JP19K14720,
and JP20H00158) of Japanese MEXT/JSPS.  Numerical computations were
performed on Sakura and Cobra clusters at Max Planck Computing and 
Data Facility.


\end{document}